\documentclass[english,twocolumn,10pt]{article}

\usepackage[subpreambles=true]{standalone}
\usepackage{import}

\usepackage{braket}
\usepackage{times}
\usepackage{graphicx}
\usepackage[labelfont=bf]{caption}
\usepackage{siunitx}
\usepackage[caption=false]{subfig}

\usepackage[section]{placeins}
\usepackage{xspace} 
\usepackage{url}

\usepackage{breakurl}
\usepackage[breaklinks]{hyperref}

\usepackage{titling}  
\usepackage{mathtools}

\RequirePackage[T1]{fontenc}
\RequirePackage[utf8]{inputenc}
\RequirePackage{lmodern}

\usepackage[nomain, acronym]{glossaries}
\newacronym{efg}{EFG}{electric field gradient}
\newacronym{esr}{ESR}{electron spin resonance}
\newacronym{nmr}{NMR}{nuclear magnetic resonance}
\newacronym{ner}{NER}{nuclear electric resonance}
\newacronym{set}{SET}{single electron transistor}
\newacronym{awg}{AWG}{arbitrary waveform generator}
\newacronym{fpga}{FPGA}{field-programmable gate array}
\newacronym{dds}{DDS}{direct digital synthesis}
\newacronym{pcb}{PCB}{printed circuit board}
\newacronym{cpw}{CPW}{coplanar waveguide}
\newacronym{sem}{SEM}{scanning electron microscope}
\newacronym{sqt}{SQT}{single quantum transition}
\newacronym{dqt}{DQT}{double quantum transition}
\newacronym{lqse}{LQSE}{linear quadrupole stark effect}
\newacronym{dft}{DFT}{Density functional theory}
\newacronym{lpcvd}{LPCVD}{low pressure chemical vapor deposition}
\newacronym{pmma}{PMMA}{poly(methyl methacrylate)}
\newacronym{nmp}{NMP}{N-Methyl-2-pyrrolidone}
\newacronym{hf}{HF}{hydrogen fluoride}
\newacronym{ebl}{EBL}{electron beam lithography}
\newacronym{al}{Al}{aluminum}

\newacronym{srs}{SRS}{Stanford Research Systems}
\newacronym{ni}{NI}{National Instruments}
\newacronym{cst}{CST}{Computer Simulation Technology}

\newcommand{\Sb}{$^{123} \mathrm{Sb}$\xspace}

\renewcommand{\figurename}{Fig.}

\DeclareCaptionLabelSeparator{bar}{ | }
\title{Coherent electrical control of a single high-spin nucleus in silicon} 

\usepackage{authblk}

\author[1]{Serwan Asaad
\footnote{Contributed equally.}}
\newcommand\CoAuthorMark{\footnotemark[\arabic{footnote}]}
\author[1]{Vincent Mourik\protect\CoAuthorMark}
\author[1]{Benjamin Joecker}
\author[1]{Mark A. I. Johnson}
\author[2]{Andrew D. Baczewski}
\author[1]{Hannes R. Firgau}
\author[1]{Mateusz T. M\k{a}dzik}
\author[1]{Vivien Schmitt}
\author[3]{Jarryd J. Pla}
\author[1]{Fay E. Hudson}
\author[4]{Kohei M. Itoh}
\author[5]{Jeffrey C. McCallum}
\author[1]{Andrew S. Dzurak}
\author[1]{Arne Laucht}
\author[1]{Andrea Morello
\thanks{To whom correspondence should be addressed; E-mail: a.morello@unsw.edu.au}}

\affil[1]{Centre for Quantum Computation and Communication Technologies, School of
Electrical Engineering and Telecommunications, UNSW Sydney, Sydney, New
South Wales 2052, Australia}
\affil[2]{Center for Computing Research, Sandia National Laboratories, Albuquerque, NM 87185, USA}
\affil[3]{School of Electrical Engineering and Telecommunications, UNSW Sydney, Sydney, New South Wales 2052, Australia}
\affil[4]{School of Fundamental Science and Technology, Keio University, Kohoku-ku, Yokohama, Japan}
\affil[5]{Centre for Quantum Computation \& Communication Technology, School of Physics, University of Melbourne, Melbourne, VIC 3010, Australia}

\date{}

\begin{document} 

\maketitle 
\addcontentsline{toc}{section}{Main text}

\textbf{
Nuclear spins are highly coherent quantum objects. 
In large ensembles, their control and detection via magnetic resonance is widely exploited, e.g. in chemistry, medicine, materials science and mining. 
Nuclear spins also featured in early ideas \cite{Kane1998} and demonstrations \cite{Vandersypen2005} of quantum information processing. 
Scaling up these ideas requires controlling individual nuclei, which can be detected when coupled to an electron \cite{Jelezko2004,Pla2013,Willke2018}. 
However, the need to address the nuclei via oscillating magnetic fields complicates their integration in multi-spin nanoscale devices, because the field cannot be localized or screened. 
Control via electric fields would resolve this problem, but previous methods \cite{Thiele2014,Laucht2015,Sigillito2017} relied upon transducing electric signals into magnetic fields via the electron-nuclear hyperfine interaction, which severely affects the nuclear coherence. 
Here we demonstrate the coherent quantum control of a single antimony (spin-7/2) nucleus, using localized electric fields produced within a silicon nanoelectronic device. 
The method exploits an idea first proposed in 1961 \cite{Bloembergen1961} but never realized experimentally with a single nucleus. 
Our results are quantitatively supported by a microscopic theoretical model that reveals how the purely electrical modulation of the nuclear electric quadrupole interaction, in the presence of lattice strain, results in coherent nuclear spin transitions. 
The spin dephasing time, 0.1 seconds, surpasses by orders of magnitude those obtained via methods that require a coupled electron spin for electrical drive. 
These results show that high-spin quadrupolar nuclei could be deployed as chaotic models, strain sensors and hybrid spin-mechanical quantum systems using all-electrical controls.
Integrating electrically controllable nuclei with quantum dots \cite{Tosi2017,hensen2019silicon} could pave the way to scalable, nuclear- and electron-spin-based quantum computers in silicon that operate without the need for oscillating magnetic fields.
}

Nuclear Magnetic Resonance (NMR) relies on the presence of a static magnetic field, $B_0$, that separates the energy levels of the nuclear spins, and a radio-frequency (RF) oscillating magnetic field, $B_1$, that induces transitions between such levels. 
Magnetic fields cannot easily be confined or screened at the nanoscale. 
Therefore, identical nuclear spins within large regions would all respond to the same signal, preventing the spins from being individually addressed. 
Electric fields, instead, can be efficiently routed and confined within highly complex nanoscale devices, a prime example being the sophisticated interconnects found in modern silicon computer chips. 
These observations suggest that an ideal route to scale up nuclear spin based quantum devices would involve the use of RF electric fields for spin control.

A theoretical idea in this direction was proposed by Bloembergen as early as 1961 \cite{Bloembergen1961}: for nuclei with spin $I>1/2$ and nonzero electric quadrupole moment $q_n$, a resonant electric field induces nuclear spin transitions by modulating the nuclear quadrupole interaction, if the nuclei are placed in solids that lack point inversion symmetry at the lattice site. 
In bulk ensembles, the static shift of the NMR frequency by a DC electric field, named Linear Quadrupole Stark Effect (LQSE), was observed in the 1960s \cite{Dixon1964b}. 
The resonant version of LQSE, called Nuclear Electric Resonance (NER) was only demonstrated more recently \cite{Ono2013} in a bulk GaAs crystal.

We report here the demonstration of NER and coherent electrical control of a single $^{123}$Sb nucleus in silicon. 
The discovery that this nucleus could be electrically controlled was, in fact, fortuitous. The $^{123}$Sb atom possesses a nuclear spin $I=7/2$ with electric quadrupole moment $q_n = -0.69$~barn. 
Depending on its electrochemical potential relative to a nearby electron reservoir, an electron (with spin $S=1/2$) can optionally be bound to the nucleus. 
The atom was implanted in a metal-oxide-semiconductor nanostructure \cite{supp} (Fig. 1A) fabricated on isotopically-enriched $^{28}$Si, similar to those developed for $^{31}$P spin qubits \cite{Pla2013,Pla2012,Muhonen2014}. 
The structure comprises a single-electron transistor for electron spin readout \cite{Morello2010}, electrostatic gates to control the donor electrochemical potential, and a broadband on-chip microwave antenna \cite{Dehollain2013}. 
The antenna is nominally terminated by a short circuit, in order to obtain maximum current at its tip and produce strong oscillating magnetic fields to control both the electron (at $\approx 40$~GHz) and the nuclear (at $\approx 10$~MHz) spins of the donor. 
In this device, however, an electrostatic discharge damaged the short circuit termination (Fig.~1A). 
The small gap in the termination has a low enough impedance at 40 GHz to allow current flow for electron spin resonance, but at $\approx 10$~MHz it produces solely an RF electric field. 
Once we realized that NER was possible, we changed to using the electrostatic gates fabricated right above the donor, which have an even stronger effect.

\begin{figure*}
    \centering
    \makebox[\textwidth][c]{\includegraphics[width=183mm]{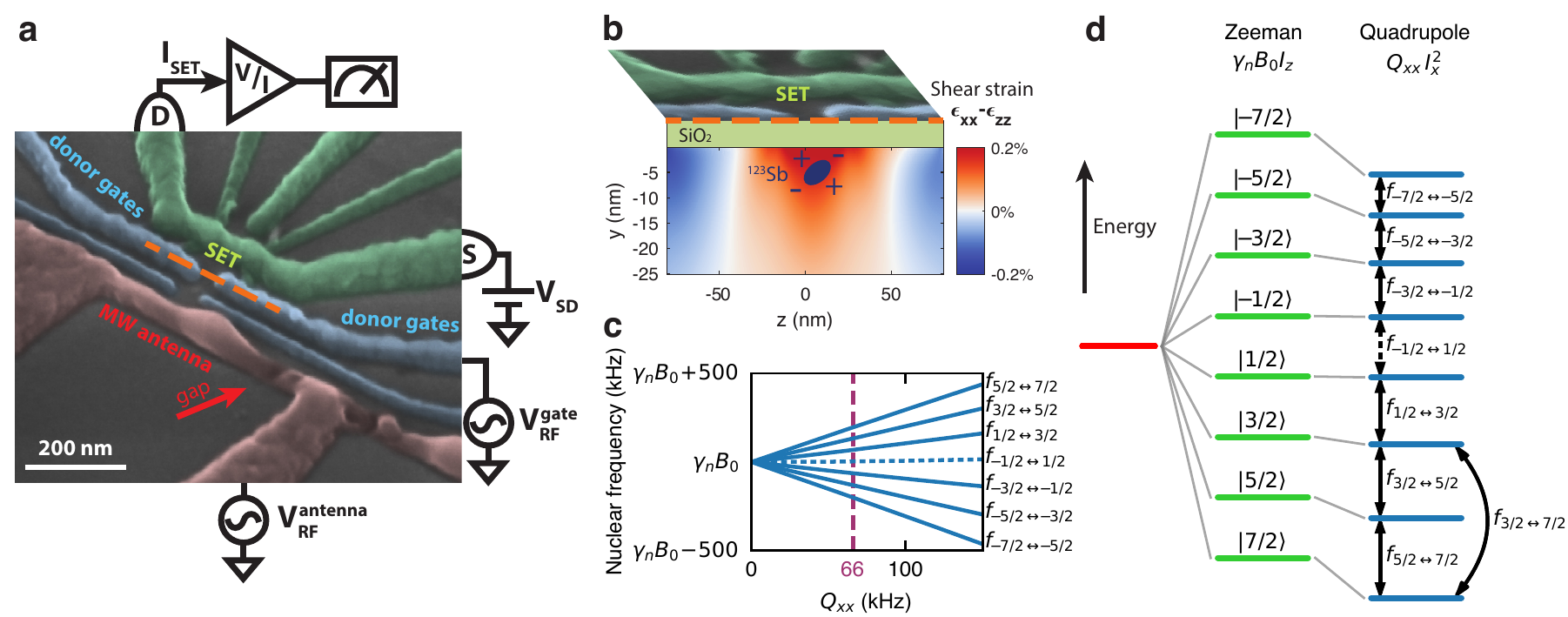}}%
    \caption{
    \textbf{$^{123}$Sb nuclear spin in a silicon device.}
    \textbf{a}, False-colored scanning electron micrograph of the silicon metal-oxide-semiconductor device used in the experiment.
    Note the gaps in the nominally short-circuited antenna terminations. 
    \textbf{b}, Shear strain in the silicon substrate, calculated on a vertical cross-section under the orange dashed line in \textbf{a}. 
    \textbf{c}, Energy level diagram of the spin-$7/2$ nucleus of an ionized $^{123}$Sb donor. 
    The electric quadrupole interaction $Q_{xx}$ shifts the Zeeman energies, resulting in \textbf{d} 7 individually addressable nuclear resonances. 
    The $m_I = -1/2 \leftrightarrow +1/2$ transition (dashed) is forbidden in NER.
    }
    \label{fig:figure1}
\end{figure*}

We focus here on the $^{123}$Sb donor in its ionized state; the removal of the donor-bound electron precludes any interpretation of the data involving modulation of hyperfine fields \cite{Thiele2014,Sigillito2017}.

In nanoscale Si devices, the aluminum gates can cause significant lattice strain at low temperatures, due to the different thermal contraction of Al and Si \cite{Thorbeck2015}. 
Lattice strain creates an electric field gradient (EFG) $\mathcal{V}_{\alpha\beta}=\partial^2 V/\partial \alpha \partial \beta$ ($V$ is the electric potential and $\alpha,\beta = x,y,z$) at the nuclear site \cite{Franke2015,Pla2018} (Fig. 1B), which produces a static nuclear quadrupole interaction $Q_{\alpha\beta} = e q_n \mathcal{V}_{\alpha\beta}/[2I(2I-1)]$, resulting in a quadrupole splitting $f_Q$ of the nuclear resonance frequencies (Fig. 1D), making all transitions individually addressable.

\begin{figure*}
    \centering
    \makebox[\textwidth][c]{\includegraphics[width=183mm]{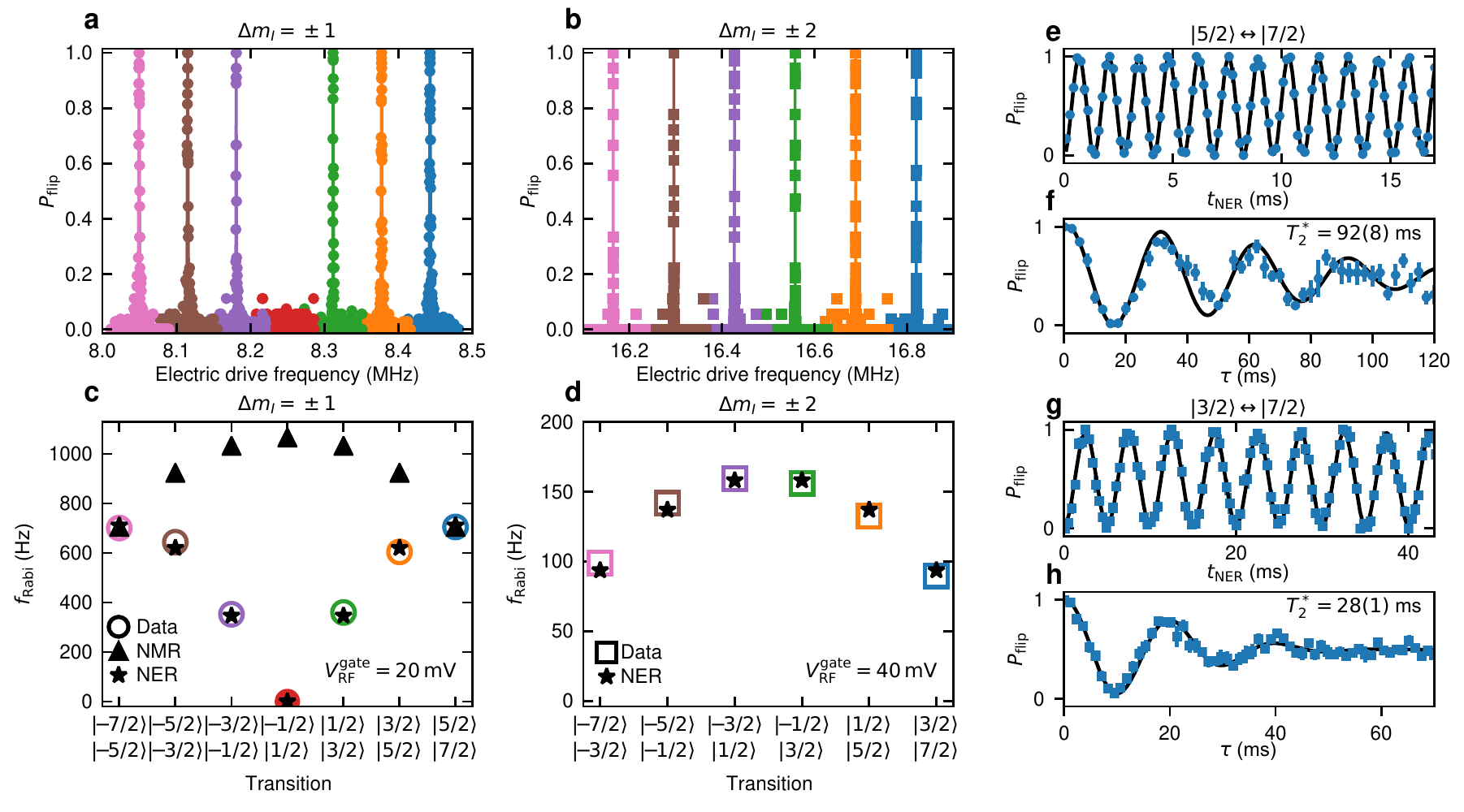}}
    \caption{
    \textbf{Nuclear electric resonance (NER).}
    \textbf{a}, NER spectrum for the $\Delta m_I = \pm 1$  and \textbf{b} $\Delta m_I = \pm 2$ transitions, obtained by applying $V_\mathrm{RF}^\mathrm{gate}$ to a donor gate (see Fig.~1A).
    The $m_I = -1/2 \leftrightarrow +1/2 $ transition (\textbf{a}, red) was not observed, as expected in NER. To acquire the complete $\Delta m_I = \pm 1$, the $m_I = -1/2 \leftrightarrow +3/2$ transition was used to switch between positive and negative $m_I$ values.
    \textbf{c}, Rabi frequencies of the $ \Delta m_I = \pm 1$ and \textbf{d} $\Delta m_I = \pm 2$ transitions, each measured at constant NER drive amplitude.
    Measured values (dots \textbf{c} or squares \textbf{d}) are compared to the theoretical predictions for NER (stars) and NMR (triangles), using the drive amplitude as the single free scaling parameter to match the experimental values. 
    All Rabi frequencies closely follow the NER prediction, including the absence of the $m_I = -1/2 \leftrightarrow +1/2$ transition (red circle in \textbf{c}, missing peak (red dots) in \textbf{a}), and are incompatible with NMR.
    \textbf{e}, Nuclear Rabi oscillations on the $m_I = +5/2 \leftrightarrow +7/2 $ and \textbf{g} $m_I = +3/2 \leftrightarrow + 7/2$  transitions. 
    A sinusoid with no decay fits the data.
    \textbf{f}, Nuclear Ramsey fringes to extract the pure dephasing time $T_2^*$ on a $\Delta m_I = \pm 1$ transition and \textbf{h} a $\Delta m_I = \pm 2$ transition.
    The fits are sinusoids with envelope decaying as $\exp{[-(\tau/T_2^*)^2]}$
    }
    \label{fig:figure2}
\end{figure*}

The application of an RF electric field of amplitude $E_1$ modulates the nuclear quadrupole energies by $\delta Q_{xz}$ and $\delta Q_{yz}$, and induces transitions between nuclear states at a rate $f^{\textrm{Rabi,NER}}_{m_I-1\leftrightarrow m_I} \propto |\delta Q_{xz} \langle m_I-1|\hat{I}_x\hat{I}_z + \hat{I}_z\hat{I}_x|m_I\rangle|$. 
Notably, the transition rate is predicted to be zero for the $m_I=-1/2 \leftrightarrow +1/2$ transition [see Eq.~15 in \cite{supp}]. 
Figure 2A shows the experimental NER spectrum for $\Delta m = \pm 1$ transitions, containing six sharp resonances separated by  $f_Q = 66$~kHz, with the $m_I=-1/2 \leftrightarrow +1/2$ absent. 

The quadrupole interaction is quadratic in the spin operators. 
Therefore, transitions between spin states that differ by $\Delta m_I = \pm 2$ are allowed, to first-order, and occur at a rate $f^{\textrm{Rabi, NER}}_{m_I-2\leftrightarrow m_I} \propto |\delta Q_{xx} \langle m_I-2|\hat{I}_x^2|m_I\rangle|$ [see Eq.~19 in \cite{supp}]. 
Importantly, all $\Delta m_I = \pm 2$ transitions have nonzero rate (Fig. 2B).  
This allowed us to ``jump over'' the NER-forbidden $m_I=-1/2 \leftrightarrow +1/2$ transition, reaching the $m_I=-1/2$ state via the $m_I = -1/2 \leftrightarrow +3/2$ transition. 
From there, the 3 remaining $\Delta m_I = \pm 1$ transitions between states with negative $m_I$ could be accessed.

Figure 2C,D presents the observed transition rates between each pair of states, in excellent agreement with the predicted trends from NER theory. 
For the $\Delta m_I = \pm 1$ transitions, the NMR Rabi frequencies would follow $f^{\textrm{Rabi,NMR}}_{m_I-1\leftrightarrow m_I} = |\gamma_n B_1 \langle m_I-1|\hat{I}_x|m_I\rangle|$, which is notably maximal for $m_I=-1/2 \leftrightarrow +1/2$ transition. The  $\Delta m_I = \pm 2$ NMR transitions are forbidden, to first-order. These results prove decisively that our experiments do not constitute a form of magnetic resonance.

As observed in earlier experiments on $^{31}$P \cite{Saeedi2013,Muhonen2014}, the nuclear spins of ionized donors in $^{28}$Si have exceptional quantum coherence properties. 
We performed a Ramsey experiment (Fig. 2G) on the $m_I = +5/2 \leftrightarrow +7/2$ ($\Delta m_I = \pm 1$) transition to extract the pure dephasing time $T_{2n+}^*(+5/2 \leftrightarrow +7/2) = 92(8)$~ms, which corresponds to an NER broadening (full width at half maximum) $\Gamma_n = \ln{2}/(\pi T_{2n+}^*) = 2.4(2)$ Hz. 
The $m_I = +3/2 \leftrightarrow +7/2$ ($\Delta m_I = \pm 2$) transition has shorter $T_{2n+}^*(+3/2 \leftrightarrow +7/2) = 28(1)$~ms (Fig. 2H). 
Both values, while extremely long in absolute terms, are noticeably shorter than the $T_{2n+}^* \approx 250 - 600$~ms measured on the $^{31}$P nucleus in two other similar devices \cite{Muhonen2014}, fabricated on the same $^{28}$Si wafer. 
Since the $^{31}$P nucleus has zero quadrupole moment, this suggests that the $^{123}$Sb coherence may be affected by electrical noise \cite{Franke2017}, in a way that the $^{31}$P is not. 
Therefore, the $^{123}$Sb nucleus could become a useful tool for spectroscopy of very slow electrical noise. 
Nonetheless, our dephasing time remains 2 orders of magnitude longer than that observed in $^{31}$P when adding a hyperfine-coupled electron, $T_{2n0}^* \approx 430 - 570$~$\mu$s \cite{Muhonen2014}, and 3 orders of magnitude longer than the $T_2^* = 64$~$\mu$s observed in a Tb nucleus in a single-atom magnet \cite{Thiele2014}. This observation highlights the benefit of a pure electrical control mechanism that does not rely upon hyperfine interactions.

\begin{figure*}
    \centering
    \includegraphics[width=.7\linewidth]{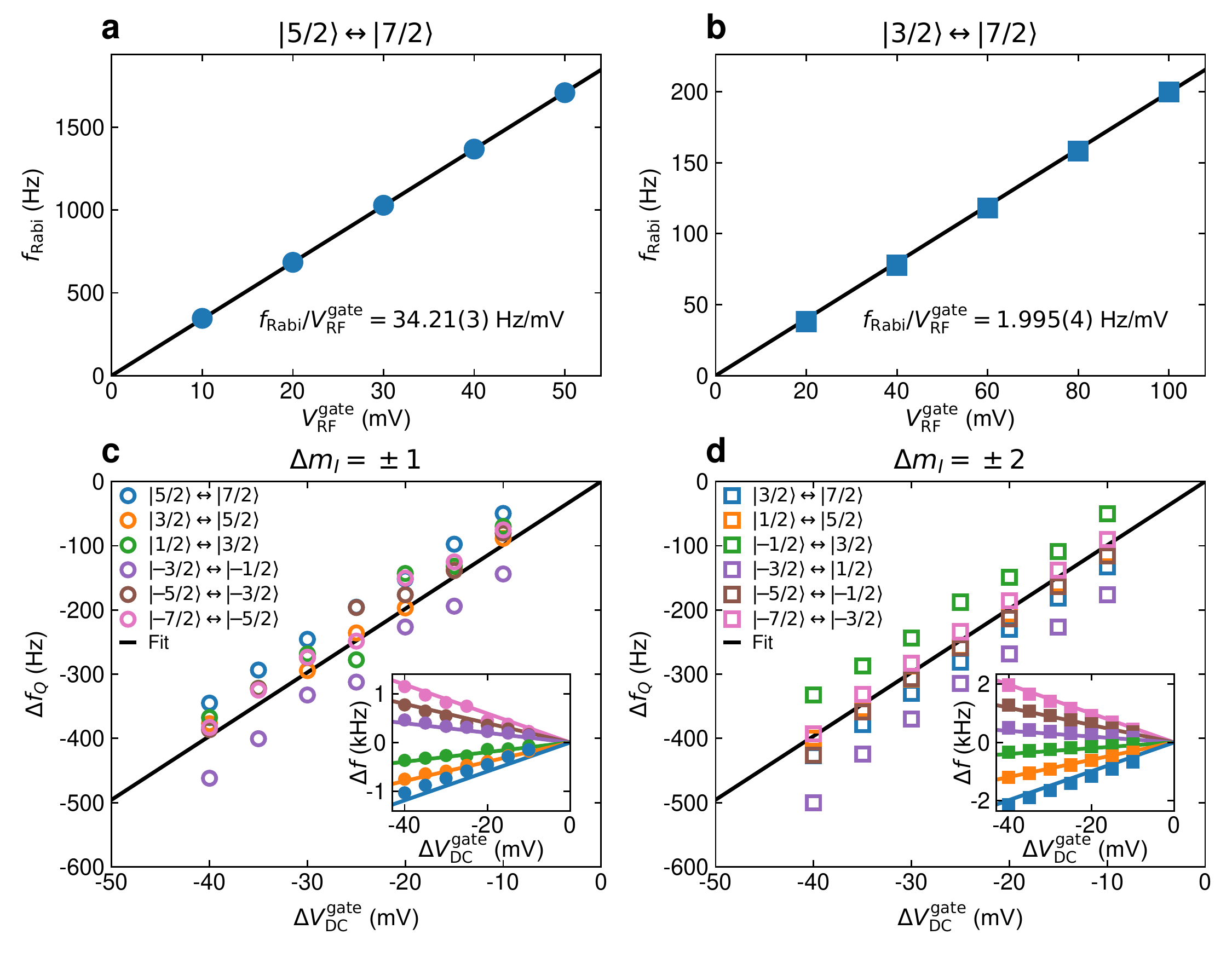}
    \caption{
        \textbf{Linear quadrupole Stark effect (LQSE).}
        \textbf{a}, Rabi frequencies $f_{\rm Rabi}$ for varying electric drive peak amplitude $V_\mathrm{RF}^\mathrm{gate}$, measured on the $\Delta m_I = \pm 1$ transition $\ket{5/2}\leftrightarrow\ket{7/2}$ and \textbf{b}, the $\Delta m_I = \pm 2$ transition $\ket{3/2}\leftrightarrow\ket{7/2}$.
        The linear relation between $V_\mathrm{RF}^\mathrm{gate}$ and $f_{\rm Rabi}$ is consistent with a first-order transition induced by the LQSE.
        \textbf{c, d}, Quadrupole shift $\Delta f_Q$ measured while applying an additional DC voltage $\Delta V_\mathrm{DC}^\mathrm{gate}$ on a donor gate. 
        $\Delta V_\mathrm{DC}^\mathrm{gate}$ causes each transition frequency $f_{m_I -\Delta m_I \leftrightarrow m_I}$ to shift $\Delta f = \Delta m_I(m_I - \Delta m_I / 2) \Delta f_Q$ (inset).
        A combined fit through all \textbf{c} $\Delta m_I = \pm 1$ and \textbf{d} $\Delta m_I = \pm 2$ frequency shifts results in a LQSE coefficient $\Delta f_Q/\Delta V_\mathrm{DC}^\mathrm{gate}=9.9(3) \mathrm{Hz/mV}$.
    }
    \label{fig:Figure3}
\end{figure*}

We measured the Rabi frequencies of the $\Delta m_I = \pm 1$ and $\Delta m_I = \pm 2$ NER transitions as a function of the amplitude of the RF voltage applied to the gate, finding $g_{\rm E,1} = 34.21(3)$~Hz/mV (Fig. 3A) and $g_{\rm E,2} = 1.995(4)$~Hz/mV (Fig. 3B). 
These transition rates show that NER is a weak effect, but due to the long nuclear spin coherence in $^{28}$Si we were able to perform high-fidelity Rabi flops persisting for tens of milliseconds (Fig. 2E,F).

In addition to driving nuclear spin transitions with an RF voltage, we were able to Stark shift the resonance frequencies using an additional DC voltage $\Delta V$ on the gates (Fig. 3C,D). 
All NER frequencies shifted according to $\delta f = \Delta f_Q \Delta V f(m_I,\Delta m_I)$, with $\Delta f_Q = 9.9(3)$~Hz/mV and $f(m_I,\Delta m_I)$ a factor of order unity that represents the matrix element of the electric quadrupole interaction between the initial and final state of each transition \cite{supp}.

The results reported here constitute the first observation of coherent, purely electrical control of a single nuclear spin. 
Achieving this in silicon is, at first sight, remarkable: no effect of electric fields on nuclear spins has ever been observed in a non-polar, non-piezoelectric material in the absence of a hyperfine-coupled electron. 
To gain a microscopic understanding of this phenomenon, we conjectured that our results are a form of LQSE \cite{Dixon1964b}. 
Resonant transitions between nuclear levels induced by electric fields (NER) require that the crystal does not possess point inversion symmetry at the atomic site \cite{Bloembergen1961}, as is indeed the case for silicon.
However, the static shift of the resonance lines (Fig. 3C,D) from the nuclear quadrupole interaction requires breaking the $\mathrm{T}_d$ symmetry of the silicon crystal. 
This can be caused by uniaxial strain ($\epsilon_{zz}$), which lowers the symmetry to $\mathrm{D}_{2d}$, or shear strain ($\epsilon_{xx} - \epsilon_{zz}$), which lowers it to $\mathrm{C}_{2v}$. 
Even in the absence of strain, an electric field can break the $\mathrm{T}_d$ symmetry by polarizing the atomic bonds.

The larger and charged donor atom introduces a local lattice distortion, displacing its four coordinating Si atoms by 0.2~\AA, and polarizes the charge density along the bonds (Fig. 4b,d). 
This, however, does not yet break the $\mathrm{T}_d$ symmetry. 
An EFG is obtained by further introducing strain. 
The $S$-tensor that links EFG to strain has two unique components, $S_{11}$ (uniaxial) and $S_{44}$ (shear). 
We conducted a first-principles, density functional theory (DFT) calculation to extract $S_{11} = 2.4 \times 10^{22}$~V/m$^2$ and $S_{44} = 6.1 \times 10^{22}$~V/m$^2$ \cite{supp}.
With a finite-element numerical model we computed the strain profile in our device, as caused by the different thermal expansions of Si and Al upon cooling to cryogenic temperatures \cite{Thorbeck2015,Pla2018} (Fig. 1B). 
Finally, we triangulated the most likely location of the $^{123}$Sb$^+$ donor by combining the implantation depth profile with a modeling of the relative capacitive coupling between the donor and different pairs of control gates, extracted from the experimental charge stability diagrams \cite{supp}.
Combining these three pieces of information, we arrived at a spatial map of quadrupole splittings $f_Q$ (Fig. 4C), which shows good agreement between the models and the experiment around the predicted location of the donor under study.

The effect of electric fields, both static (LQSE) and dynamic (NER), can be understood as arising from the single unique component of the $R$-tensor, $R_{14}$ \cite{supp}.
By combining a finite-element model of the electric field in the device, the estimated $^{123}$Sb$^+$ donor position, and the experimental values of LQSE and NER Rabi frequencies, we extracted $R_{14} = 1.7 \times 10^{12}$~m$^{-1}$ \cite{supp}.
The strength of this coupling is comparable to prior bulk measurements of LQSE on $^{75}$As in GaAs \cite{gill1963linear}. 
This can be understood by observing that, while the Sb$^+$-Si bond has a weaker ionic character compared to the Ga-As bond, $R_{14}$ scales with atomic number, leading to a similar overall value. 
Since our model agrees with the experiment within a factor of order unity, and no alternative explanation comes within orders of magnitude of the results \cite{supp}, we conclude that we have observed the manifestation of LQSE and NER in a single nuclear spin in silicon.

\begin{figure}
    \centering
    \includegraphics[width=75mm]{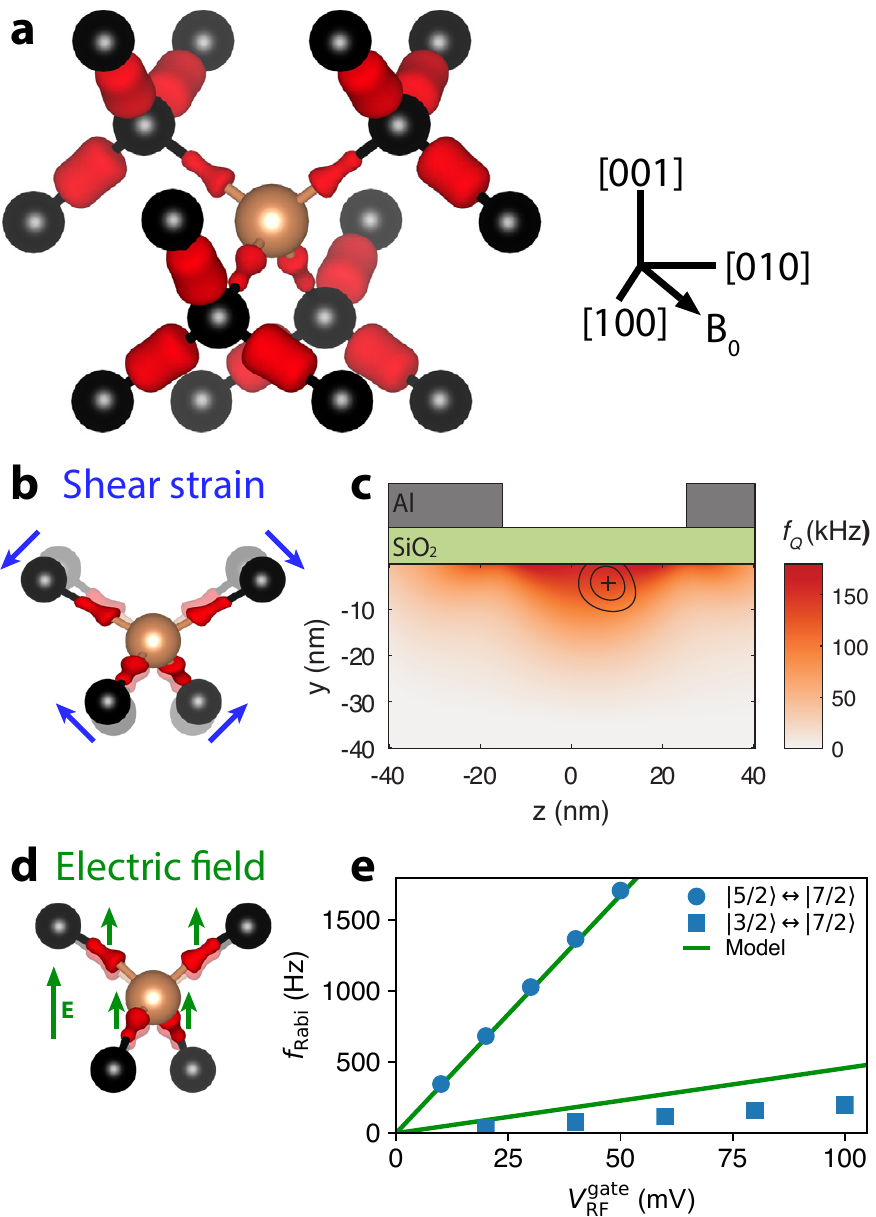}
    \caption{
    \textbf{Microscopic origins of the quadrupole interaction.}
    \textbf{a}, Valence charge density near the Sb$^+$ atom (gold) and its 16 closest Si atoms (black) with a charge density isosurface (red).
    The donor's positive charge causes an asymmetric charge density along the Sb$^+$-Si bond but, in the absence of strain or external electric fields, the \acrshort*{efg} at the $^{123}$Sb site vanishes by symmetry. 
    \textbf{b}, Shear strain displaces the Si atoms and covalent bonds neighboring the $^{123}$Sb nucleus, creating an \acrshort{efg} which results in a quadrupole shift. 
    \textbf{c}, Quadrupole splitting $f_Q$, predicted by combining density functional theory calculations and finite element simulations \cite{supp}. 
    Black contours enclose the 68\% and 95\% confidence regions for the location of the donor, as obtained from capacitance triangulation and donor implantation profile \cite{supp}.
    \textbf{d}, Electric fields applied via gate voltages distort the charge distribution, resulting in both linear frequency shifts (LQSE) and coherent spin transitions (NER). 
    \textbf{e}, Calculation of the NER Rabi frequencies caused by electrical \acrshort*{efg} modulation (green lines), compared to experimental results for a $\Delta m_I = \pm 1$ (dots) and a $\Delta m_I = \pm 2$ (squares) transition. 
    All $f_{\rm Rabi}$ are determined by a single parameter, $R_{14}$, calculated via finite-element modeling and electronic structure theory. No free fitting parameters were used.
    }
    \label{fig:Figure4}
\end{figure}

Our results have significant consequences for the development of nuclear spin based quantum computers, and the design of nanoscale quantum devices. 
The Hilbert space of the $I=7/2$ $^{123}$Sb nucleus has 8 dimensions. 
It can encode the equivalent of 3 quantum bits of information, allowing simple quantum algorithms \cite{Godfrin2017} or quantum error correction codes \cite{Waldherr2014}, all solely using electric fields. 
The donor electron and nuclear spins combined form a ``flip-flop'' qubit \cite{Tosi2017}, controllable by electric dipole spin resonance.
This scheme normally requires a magnetic antenna to reset the nuclear state in the appropriate qubit subspace. 
This need for could be removed completely by using an electrically-drivable high-spin nucleus like $^{123}$Sb. 
A recent result showed that lithographic quantum dots in silicon can be entangled with nuclear spins, and that the nuclear coherence can be preserved while shuttling the electron between different dots \cite{hensen2019silicon}.
Electron spin qubits in silicon can be coherently controlled by electric fields with high speed and high fidelity \cite{Yoneda2018}. 
Adding the ability to electrically control quadrupolar nuclei paves the way to quantum computer architectures that integrate fast electron spin qubits with long-lived nuclear quantum memories, while fully exploiting the controllability and scalability of silicon metal-oxide-semiconductor devices, without the complication of routing RF magnetic fields within the device. 

The experimental validation of a microscopic 
model of the relation between strain and quadrupole splitting, obtained in a functional silicon electronic device, suggests the use of quadrupolar nuclei as single-atom probes of local strain, which has a key role in enhancing the performance of ultra-scaled transistors \cite{Thompson2006}. 
The observation of a large quadrupole splitting $f_Q = 66$~kHz in a high-spin nucleus creates a platform in which to study quantum chaotic dynamics in a single particle \cite{Mourik2018}.
This has further applications in quantum information science, e.g. because of the remarkable analogies between chaotic spin models and digital quantum simulations \cite{Sieberer2018}. 
Although the strain in the present device is static, our work allows us to predict the nuclear Rabi frequencies that would arise from time-dependent strain \cite{supp}. 
A dynamical strain $\approx 5 \times 10^{-8}$ would cause a Rabi frequency of 10 Hz, comparable to both the inhomogeneous nuclear linewidth $\Gamma_n \approx 2.4$~Hz, and to the linewidth $\Gamma_m$ of high-quality silicon mechanical resonators in the MHz range \cite{Ghaffari2013}.
Therefore, it is conceivable to achieve the strong-coupling limit of cavity-quantum electrodynamics between a single nuclear spin and a macroscopic mechanical oscillator, adding a novel spin-mechanical coupling pathway to the toolbox of hybrid quantum systems for quantum information processing and precision sensing \cite{Kurizki2015}.

\bibliographystyle{naturemag}

\textbf{Acknowledgments}
We thank T. Botzem and J.T. Muhonen for useful discussions. 
The research was funded by the Australian Research Council Discovery Projects (Grants No. DP150101863 and No. DP180100969) and the Australian Department of Industry, Innovation and Science (Grant no. AUSMURI00002). 
V.M. acknowledges support from a Niels Stensen Fellowship. 
M.A.I.J. and H.R.F. acknowledge the support of Australian Government Research Training Program Scholarships.
J.J.P. is supported by an Australian Research Council Discovery Early Career Research Award (DE190101397).
A.M. was supported by a Weston Visiting Professorship at the Weizmann Institute of Science during part of the writing of this manuscript. 
We acknowledge support from the Australian National Fabrication Facility (ANFF), and from the laboratory of Prof. R. Elliman at the Australian National University for the ion implantation facilities. 
Sandia National Laboratories is a multi-missions laboratory managed and operated by National Technology and Engineering Solutions of Sandia, LLC, a wholly owned subsidiary of Honeywell International Inc., for DOE's National Nuclear Security Administration under contract DE-NA0003525. 
The views expressed in this manuscript do not necessarily represent the views of the U.S. Department of Energy or the U.S. Government. 
K.M.I. acknowledges support from Grant-in-Aid for Scientific Research by MEXT. 

\textbf{Author contributions} 
S.A. and M.A.I.J. performed the measurements under supervision of V.M., A.L. and A.M., with the assistance of V.S., M.T.M., and H.R.F.; S.A. and M.A.I.J. analyzed the data under supervision of V.M. and A.M., with the assistance of H.R.F., V.S., J.J.P. and A.L.; A.D.B., S.A., V.M., B.J., and A.M. articulated a microscopic theory supported by finite-element modeling by B.J. and electronic structure calculations by A.D.B.; F.E.H. partially fabricated the device under supervision of A.S.D., on isotopically-enriched material supplied by K.M.I., and M.T.M. subsequently fabricated the aluminum gate structures under supervision of V.M. and A.M.; J.C.McC. designed and performed the $^{123}$Sb ion implantation; S.A., V.M., B.J., M.A.I.J., A.D.B., H.R.F. and A.M. wrote the manuscript and supplementary information, with input from all coauthors; A.M. initiated and supervised the research program. 

\textbf{Competing interests:} S.A, V.M. and A.M. have submitted a patent application that describes the use of electrically-controlled high-spin nuclei for quantum information processing (AU2018900665A).

\textbf{Additional information}

\textbf{Extended data} is available for this paper at https://doi.org/xxxx/xxxx.

\textbf{Supplementary materials} is available for this paper at https://doi.org/xxxx/xxxx.

\textbf{Correspondence and requests for materials} should be addressed to A.M.

\newpage

\section*{Methods}

\subsection*{Fabrication}
 
The device was fabricated on a <100> p-type silicon wafer, with a 900 nm thick epitaxial layer of isotopically purified $^{28}$Si on top (concentration of residual $^{29}$Si is 730 ppm). 
Metallic leads for the SET are formed using optical lithography and phosphorus diffusion.
The substrate is subsequently covered with a 250 nm thick field oxide, with a small central window (10 x 20 $\mu$m) containing a high quality, thermally grown layer of SiO$_2$ with a thickness of 8 nm.
Using a combination of standard optical and electron beam lithography techniques, the device is fabricated on this thin oxide window. 
First, a small (90 x 100 nm) window is defined, through which \Sb ions are implanted at an energy of 8 keV and fluence of 2 x 10$^{11}$/cm$^2$, corresponding to an average of 14 donors in the implantation window.
Donors are activated using a rapid thermal anneal at 1000 C. 
Next, in two electron beam lithography steps, the gates forming the SET, the donor gates, and the microwave antenna are created, using thermally evaporated aluminum and lift off, with aluminum oxide as gate dielectric. 
Ohmics to the n-doped SET leads are formed using optical lithography, thermally evaporated aluminium and lift off, followed by a forming gas anneal. 
A detailed step-by-step process flow is given in \cite{supp}. 

\subsection*{Experimental setup}
The sample is cooled to a temperature of ~20 mK in a dilution refrigerator (Bluefors BF-LD400) with superconducting magnet.
During measurements, arbitrary waveform generators (Signadyne M3201A and M3300A) were used to tune the donor electrochemical potential, generate NER pulses, and IQ-modulate the microwave signals generated by a vector microwave source (Keysight E8267D).
The SET current was amplified with a transimpedance amplifier (FEMTO DLPCA-200 in combination with Stanford Instruments SIM911), and subsequently measured with a digitizer (Signadyne M3300A).
Full details of the experimental setup including wiring schematic can be found in \cite{supp}.

\subsection*{Nuclear spin readout}
The nuclear spin state is measured through the electron spin state, by using ESR pulses conditional on the nuclear spin eigenstates. 
Electron spin readout is achieved by spin-to-charge conversion through spin-dependent tunneling onto an SET and subsequent detection of the change in charge occupation of the donor (see \cite{supp} for details). 

An NER pulse has a probability $P_\mathrm{flip}$ of flipping the nuclear spin between two spin states.
Both nuclear spin states are measured after each NER pulse, and are compared to the previous measurement.
This reveals if the NER pulse has flipped the nuclear spin.
This process is repeated $N_\mathrm{repetitions}$ times, and the resulting nuclear flips $N_\mathrm{flips}$ gives an estimate of the flip probability $P_\mathrm{flip} \approx N_\mathrm{flips} / (N_\mathrm{repeititions}-1)$.

\subsection*{Theoretical modelling}
The spin Hamiltonian of the $^{123}$Sb nucleus takes the form:
\begin{equation}
\hat{\mathcal{H}}/h=\gamma_n B_0\hat{I}_z + \sum_{\mathclap{\alpha, \beta \in \{x,y,z\}}} Q_{\alpha\beta} \hat{I}_{\alpha} \hat{I}_{\beta}, \label{eq:H_NMR}
\end{equation}
where $h=6.626\times10^{-34}$~J/Hz is the Planck constant, $\gamma_n = -5.553$~MHz/T is the nuclear gyromagnetic ratio and $B_0 = 1.496$~T. $\hat{I}_{x}$, $\hat{I}_{y}$ and $\hat{I}_{z}$ are the 8-dimensional operators describing the $x,y,z$ projections of the $I=7/2$ spin.
In the presence of an RF electric field of amplitude $E_1$, the $\Delta m_I = \pm 1$ transitions are driven by an additional Hamiltonian term of the form:
\begin{align} 
\hat{\mathcal{H}}_{m_I-1\leftrightarrow m_I}^{\mathrm{NER}}\left(t\right)/h = & \cos\left(2\pi f t\right) \left[\delta Q_{xz}\left(\hat{I}_{x} \hat{I}_{z}+\hat{I}_{z} \hat{I}_{x}\right)\right.\nonumber\\
& \left. +\delta Q_{yz}\left(\hat{I}_{y} \hat{I}_{z}+\hat{I}_{z} \hat{I}_{y}\right)\right],
\end{align}
The $\Delta m_I = \pm 2$ transitions are driven by a term of the form:
\begin{align} 
\hat{\mathcal{H}}_{m_I-2\leftrightarrow m_I}^{\mathrm{NER}}\left(t\right)/h = & \cos\left(2\pi f t\right)\left[ \delta Q_{xx} I_x^2 + \delta Q_{yy}I_y^2 \right. \nonumber\\
& \left. +\delta Q_{xy}\left(I_x I_y + I_y I_x \right)\right].
\end{align}
A detailed derivation of the matrix elements responsible for driving the $\Delta m_I = \pm 1$ and $\Delta m =_I \pm 2$ NER transitions is given in \cite{supp}. 

A finite element model is used to compute the strain and electric fields in the silicon layer near the donor position, using the COMSOL multiphysics software. 
The donor position has been triangulated by comparing simulated gate-to-donor coupling strengths with the experimentally observed strength, combined with the donor implantation profile (see Extended Data Fig.\ref{fig: ED_least_square} and~\cite{supp}).

Kohn-Sham density functional theory (DFT) has been employed to calculate the S-tensor components that describe the impact of strain on the \gls*{efg}.
To this end, 64- and 512-atom supercells were strained using the Projector Augmented-Wave (PAW) formalism \cite{blochl1994projector} with a plane wave basis, as implemented in the Vienna Ab-Initio Simulation Package (VASP) \cite{kresse1996efficient,kresse1996efficiency,kresse1999ultrasoft}. 
The electric field response tensor is estimated by comparing the data points from the DC LQSE (Fig.~\ref{fig:Figure3}C,D) and the $\Delta m_I=\pm 1$ (Fig.~\ref{fig:figure2}C) $\Delta m_I=\pm 2$ (Fig.~\ref{fig:figure2}D) Rabi frequencies with the simulated electric fields at the triangulated donor position.
The final $R_{14}$ was found by minimizing the normalized residuals of the three separate $R_{14}$ estimates.
Full theoretical modelling details can be found in \cite{supp}.

\textbf{Data availability:} All data necessary to evaluate the claims of this paper is provided in the main manuscript or the supplementary information. 
Raw data files can be obtained from the corresponding author upon request.

\onecolumn

\section*{Extended data}
\renewcommand{\figurename}{Extended Data Fig.}
\setcounter{figure}{0} 

\begin{figure}[h]
    \centering
    \includegraphics[width=\linewidth]{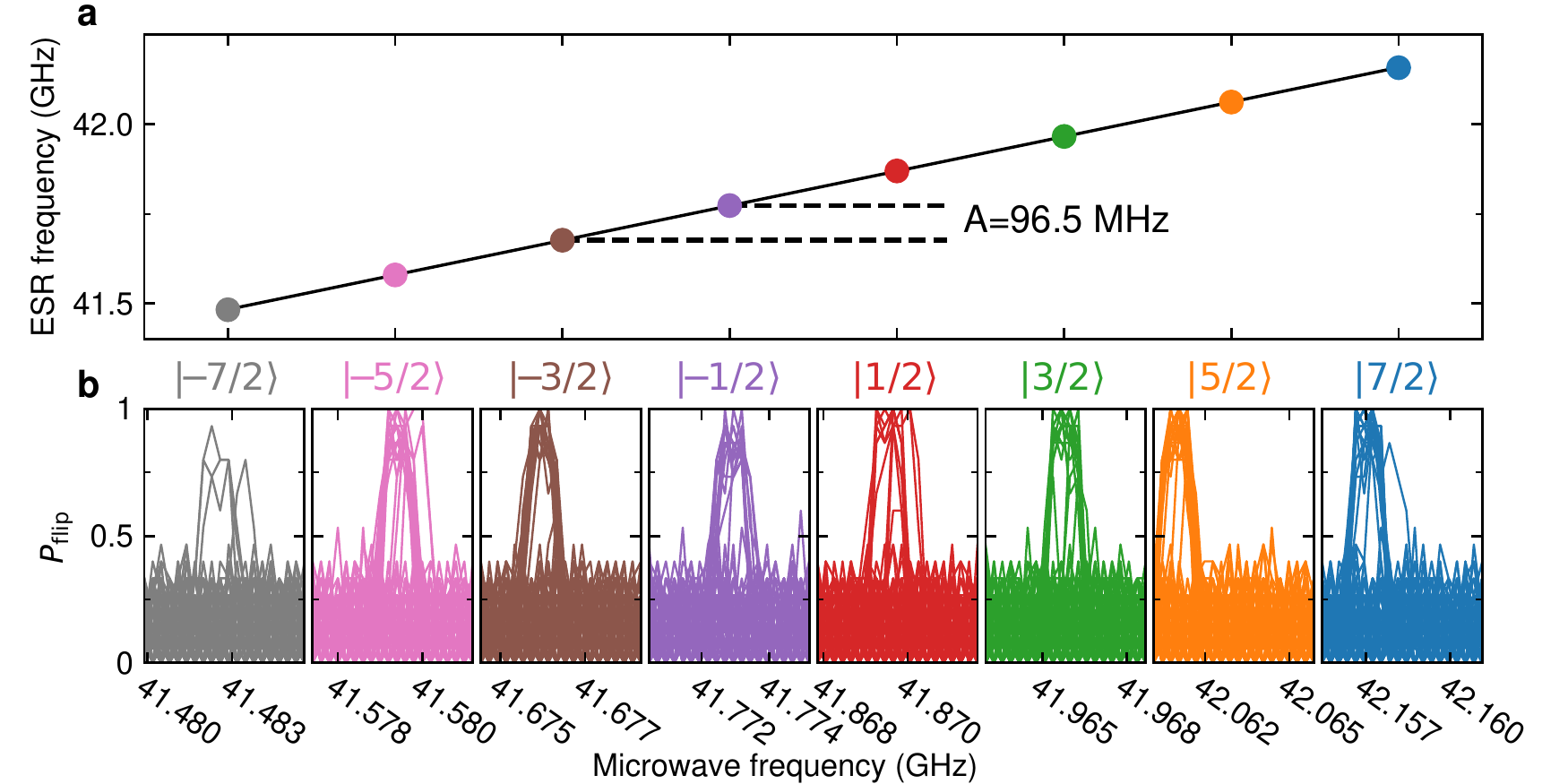}
    \caption{
    \textbf{\acrshort*{esr} spectrum at magnetic field $B_0 = \SI{1.496}{\tesla}$.}
    \textbf{a}, \acrshort*{esr} frequencies for eight nuclear states.
    The average difference between successive \acrshort*{esr} transition frequencies (black line) gives a hyperfine interaction $A = \SI{96.5}{\mega \hertz}$, significantly lower than the bulk value of $\SI{101.52}{\mega\hertz}$.
    One possible cause for this deviation is due to strain, which is known to modify the hyperfine interaction \cite{mansir2018linear}.
    \textbf{b}, \acrshort*{esr} spectral lines.
    For each nuclear state, the nucleus was initialized at the start of each microwave sweep, and adiabatic \acrshort*{esr} pulses with $\SI{1}{\mega \hertz}$ frequency deviation were applied to excite the electron.
    }
    \label{fig:ESR spectrum}
\end{figure}

\begin{figure}[ht]
    \centering
    \includegraphics[width=\linewidth]{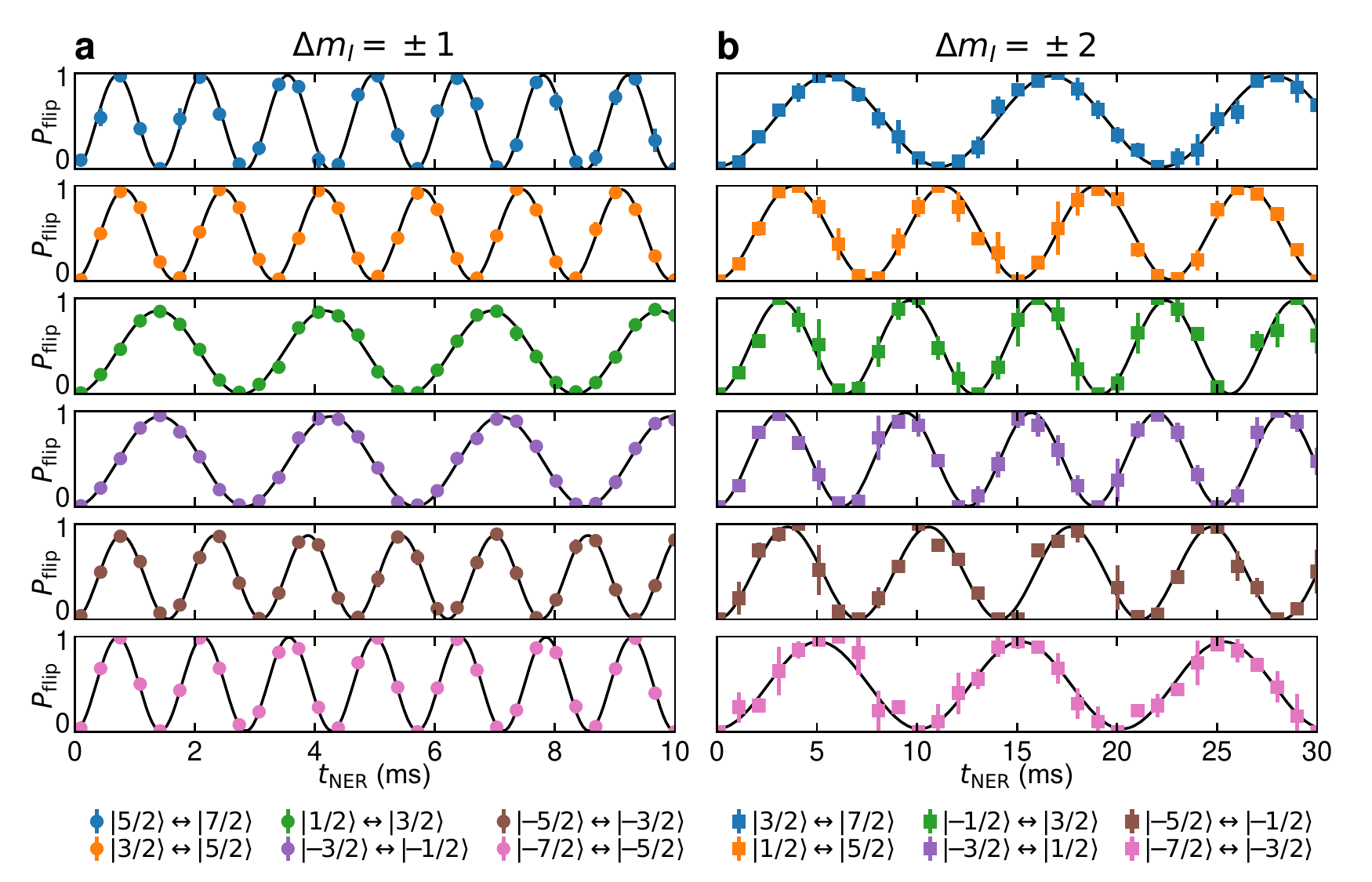}
    \caption{
    \textbf{\acrshort*{ner} Rabi oscillations on resonance.}
    The \acrshort*{ner} pulse durations $t_\mathrm{NER}$ was varied, while the pulse amplitude was fixed at \textbf{a} $V_\mathrm{RF}^\mathrm{gate} = \SI{20}{\milli\volt}$ for $\Delta m_I = \pm 1$ transitions, and \textbf{b} $V_\mathrm{RF}^\mathrm{gate} = \SI{40}{\milli\volt}$ for $\Delta m_I = \pm 2$ transitions.
    Black lines are non-decaying sinusoidal fits to the data, and error bars show 95\% confidence interval.
    }
    \label{fig:NER Rabi oscillations}
\end{figure}

\begin{figure}[ht]
    \centering
    \subfloat{\includegraphics[width=0.49\textwidth]{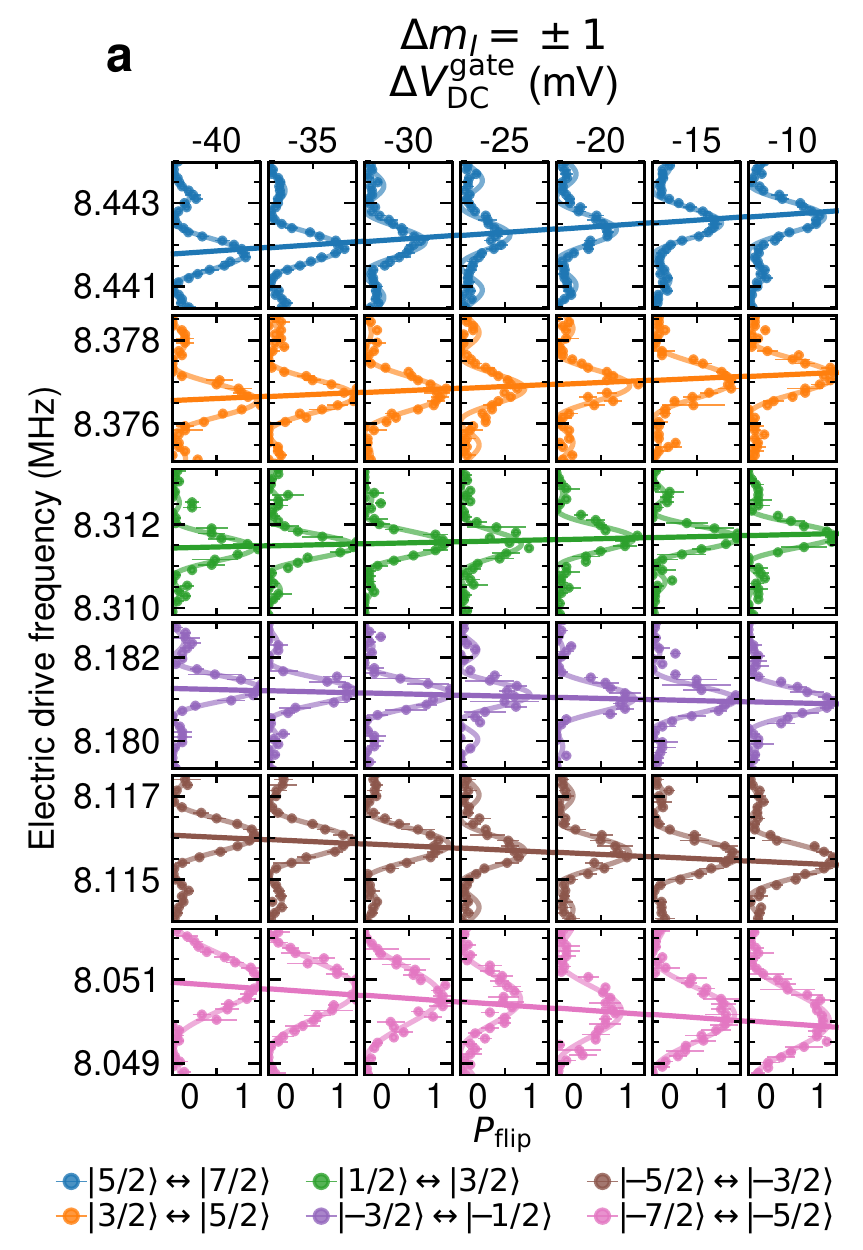}}
    \subfloat{\includegraphics[width=0.49\textwidth]{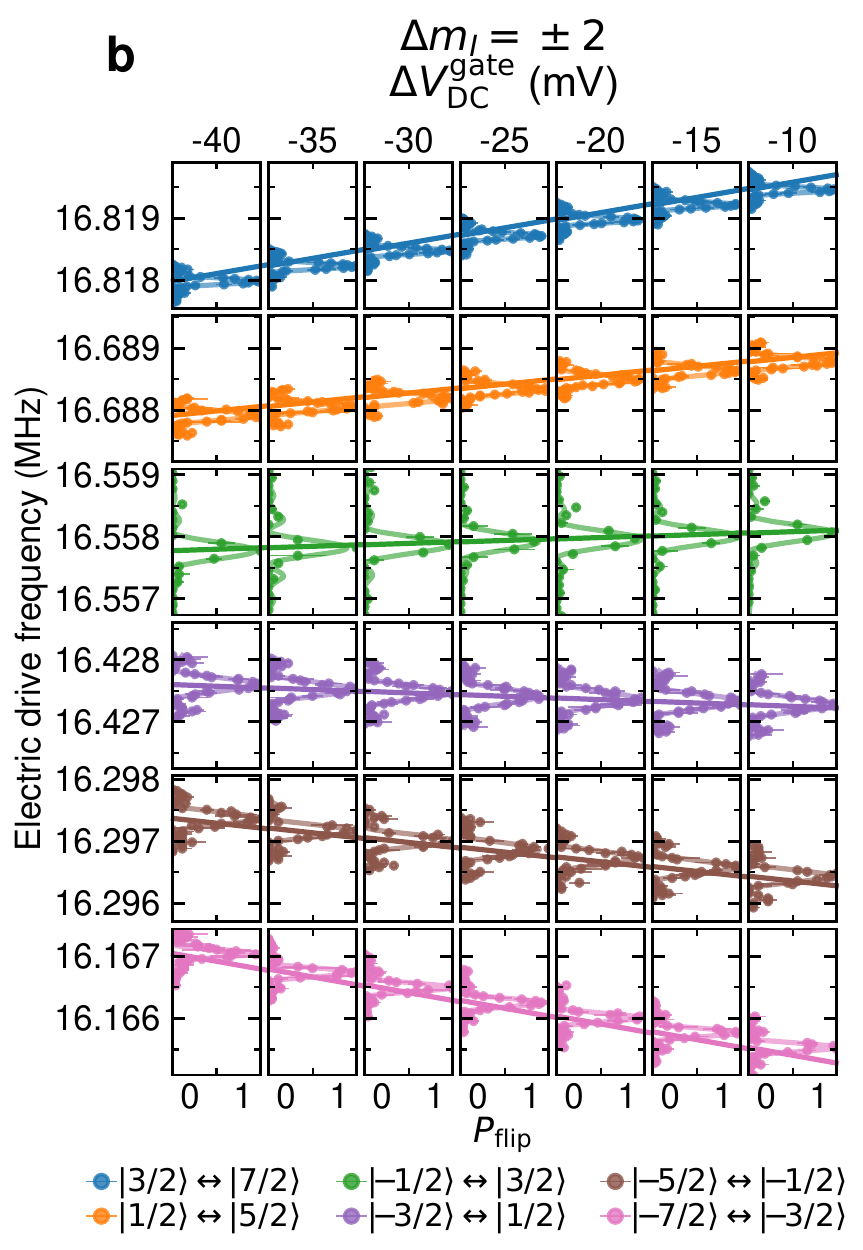}}
  \caption{
    \textbf{\acrshort*{ner} spectral line shifts for varying DC gate voltage.}
  The spectral lines of all \textbf{a} $\Delta m_I = \pm 1$ transitions and \textbf{b} $\Delta m_I = \pm 2$ transitions are measured, while the DC gate voltage bias $\Delta V_\mathrm{DC}^\mathrm{gate}$ during the \acrshort*{ner} pulse is varied (columns).
  Note that this change in $V_\mathrm{DC}^\mathrm{gate}$ happens on top of significant gate voltages, of order 0.5 V, necessary to electrostatically tune the device to enable its operation.
  The varying $\Delta V_\mathrm{DC}^\mathrm{gate}$ modifies the quadrupole interaction via the \acrshort*{lqse} (see \cite{supp}), resulting in shifts of the resonance peaks.
  A single fit to the resonance frequency shifts of all $\Delta m_I = \pm 1$ and $\Delta m_I = \pm 2$ transitions (solid lines) estimates the gate-dependent quadrupole shift as $\Delta f_Q/\Delta V_\mathrm{DC}^\mathrm{gate}=\SI{9.9(3)}{\kilo \hertz \per \volt}$.
  Ordered from top to bottom transition, the drive strengths $V_\mathrm{RF}^\mathrm{gate}$ are $\left[ \SI{20}{\milli \volt}, \SI{20}{\milli \volt}, \SI{25}{\milli \volt}, \SI{25}{\milli \volt}, \SI{20}{\milli \volt}, \SI{25}{\milli \volt}  \right]$ for $\Delta m_I = \pm 1$, and $\left[ \SI{30}{\milli \volt}, \SI{30}{\milli \volt}, \SI{40}{\milli \volt}, \SI{40}{\milli \volt}, \SI{40}{\milli \volt}, \SI{40}{\milli \volt}  \right]$ for $\Delta m_I = \pm 2$.
  Error bars show the 95\% confidence interval.
  }
  \label{fig: DC LQSE spectra}
\end{figure}

\begin{figure}[ht]
  \centering
  \subfloat{\includegraphics[width=0.49\textwidth]{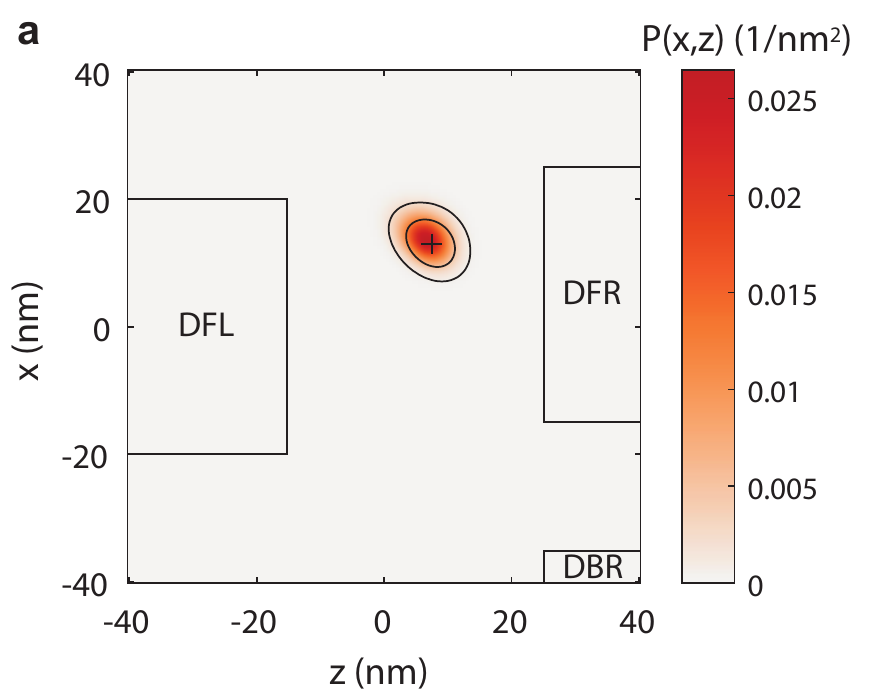}}
  \subfloat{\includegraphics[width=0.49\textwidth]{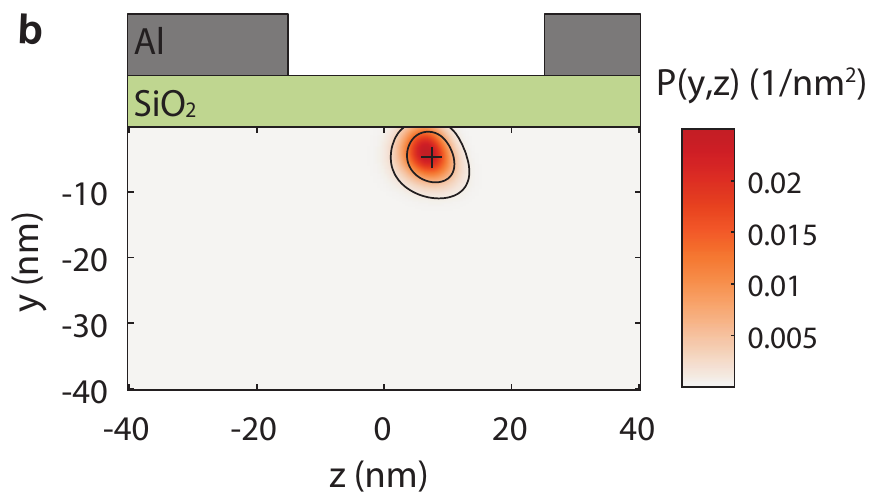}}
  \caption{
    Position triangulation of the \Sb donor. 
  The colormap shows the probability of finding the donor in a certain location. 
  \textbf{a}, \textbf{b}, Probability density function found using a least-squares estimate comparing simulated gate-to-donor coupling strengths with the experimentally observed strengths (see~\cite{supp} for details, including locations of donor gates DFR, DFL, and DBR). 
  To improve on the low resolving power of the triangulation method in the $y$ direction, the triangulation probability density function is multiplied with the donor implantation probability density function~\cite{supp}. 
  This has little effect laterally, but significantly confines the likely depth range of the donor. 
  within the range expected based on the donor implantation parameters. 
  The most likely donor position, indicated by a cross, is at a lateral location $(x,z) = (\SI{13}{\nano \meter},\SI{8}{\nano \meter})$ at a depth of $y = \SI{-5}{\nano \meter}$.
  Probability density functions are normalized over the model volume and are integrated over the out-of-plane axis in both panels, specifically $P(x,z)=\int P(\Vec{r})dy$ and $P(y,z)=\int P(\Vec{r})dx$. 
  The contour lines mark the 68\% and 95\% confidence regions.
  }
  \label{fig: ED_least_square}
\end{figure}

\FloatBarrier
\newpage
\newcommand{\citenum}[1]{\cite{#1}}
\newcommand{\citep}[1]{\cite{#1}}
\pagenumbering{arabic}
\renewcommand*{\thepage}{SM-\arabic{page}}

\setcounter{figure}{0} 
\makeatletter
\renewcommand{\fnum@figure}{\figurename~S\thefigure}
\makeatother
\renewcommand{\figurename}{Fig.}

\begin{titlepage}
	\centering
	{\Large Supplementary Materials: Coherent electrical control of a single high spin nucleus in silicon}
	\vspace{0.5cm}
\end{titlepage}

\tableofcontents

\newpage
\section{Antimony in silicon}\label{sec:antimony information}

Antimony (Sb) is a group-V atom which behaves as a donor when placed as a substitutional dopant in a silicon crystal.
The neutral D$^0$ state of the Sb donor has five valence electrons, four of which form covalent bonds with neighbouring silicon atoms, while the fifth remains loosely-bounded to the donor with a binding energy of 42.7 meV~\citep{ramdas1981spectroscopy}. 
This outer electron can be removed by tuning the electrochemical potential above the Fermi level of a nearby charge reservoir, resulting in an ionized D$^+$ state of the donor.

The Sb isotope used in this experiment, \Sb, has a nuclear spin quantum number $I=7/2$, and therefore has eight ($2I+1$) nuclear eigenstates.
All core electrons of the \Sb donor, as well as the four electrons participating in the covalent bonds with the silicon lattice are paired in singlet states.
Therefore, the spin Hamiltonian of the D$^0$ state can only include the spin of the loosely-bound outer electron and the nuclear spin.
In frequency units, the D$^0$ spin Hamiltonian is
\begin{equation}
    \hat{\mathcal{H}} = \gamma_e B_0 \hat{S}_z + \gamma_n B_0 \hat{I}_z + A \vec{S} \cdot \vec{I} +\hat{\mathcal{H}}_Q,
\end{equation}
where $\gamma_e = \SI{27.97}{\giga \hertz \per \tesla}$ is the electron gyromagnetic ratio,  $\gamma_n = \SI{-5.55}{\mega \hertz \per \tesla}$ is the nuclear gyromagnetic ratio, $B_0$ is the static magnetic field, $A$ is the hyperfine interaction, equal to $\SI{101.52}{\mega \hertz}$ in bulk~\citep{feher1959electron} and assumed to be isotropic, $\vec{I}=\{\hat{I}_x, \hat{I}_y, \hat{I}_z\}$ ($\vec{S}=\{\hat{S}_x, \hat{S}_y, \hat{S}_z\}$) are the spin operators describing the $x,y,z$ projections of the nuclear (electron) spin, and $\hat{\mathcal{H}}_Q$ is the nuclear quadrupole Hamiltonian (Sec.~\ref{sec: quad_ham}).

In the limit where $\gamma_e B_0 \gg A$, as is the case in our experiment ($B_0 = \SI{1.496}{\tesla}$), the hyperfine interaction $A \vec{S} \cdot \vec{I}$ may be approximated as $A \hat{S}_z \hat{I}_z$.
In this case, the nuclear and electron spin eigenstates are separable, and given by the tensor product of the eigenstates of the nuclear $\hat{I}_z$ and electron $\hat{S}_z$ spin operators.
The hyperfine interaction between electron and nucleus introduces a dependence of the \gls*{esr} frequency on the nuclear spin state.
This dependence is exploited to measure the nuclear spin state via the electron spin (Sec.~\ref{sec: experimental_methods}).

If the D$^0$ donor's outer electron is removed, the resulting D$^+$ donor has a considerably simpler Hamiltonian
\begin{equation}
    \hat{\mathcal{H}} = \gamma_n B_0 \hat{I}_z + \hat{\mathcal{H}}_Q.
    \label{eq: H_system_static}
\end{equation}
The nuclear Zeeman interaction separates the eight energy levels, resulting in seven identical nuclear transition frequencies (Fig. 1C).
A nonzero quadrupole interaction separates these nuclear transition frequencies, allowing addressing of individual nuclear transitions (Fig. 1D).
Throughout the remainder of the text, we will refer to the D$^+$ donor Hamiltonian unless explicitly stated otherwise.

\section{Nuclear electric resonance and nuclear quadrupole interaction}
\label{sec: quad_ham}

\subsection{Nuclear quadrupole interaction}

Assuming perfect spherical symmetry of the nuclear charge distribution, its electrical response to its environment is described by the Coulomb potential of a point charge. 
However, deviations from spherical symmetry require correction terms to this simple model.
The lowest-order, non-zero correction term to the point charge model is the nuclear quadrupole interaction, which captures oblate or prolate deviations from spherical symmetry.
This interaction is quadratic in the nuclear spin operators, and scales linearly with the \gls*{efg} of the local electric field\citep{slichter2013principles}. 
 
A general Hamiltonian $\hat{\mathcal{H}}_Q $ of the nuclear quadrupole interaction may be written as
\begin{align} \label{eq: H_Q_static}
    \hat{\mathcal{H}}_Q=Q_{xx} \hat{I}_x^2+Q_{yy}\hat{I}_y^2+Q_{zz}\hat{I}_z^2+Q_{yz}(\hat{I}_y\hat{I}_z+\hat{I}_z\hat{I}_y)+Q_{xz}(\hat{I}_x\hat{I}_z+\hat{I}_z\hat{I}_x)+Q_{xy}(\hat{I}_x\hat{I}_y+\hat{I}_y\hat{I}_x),
\end{align}
with $\hat{I}_\alpha$ as defined in \ref{sec:antimony information}, $Q_{\alpha\beta}$ with $\alpha , \beta \in \{x, y, z\}$, determines the strength of the different terms and is given by
\begin{align} \label{eq: Q_interaction_strength}
    Q_{\alpha\beta}=\frac{e q_n  \mathcal{V}_{\alpha\beta}}{2I(2I-1)h},
\end{align}
where $e$ is the elementary charge, $h$ is Planck's constant, $I$ is the nuclear spin quantum number, and $q_n$ is the quadrupole moment, a constant specific for each nucleus, which parametrizes the degree of spherical asymmetry of the nuclear wavefunction. 
For $^{123}$Sb, $q_n= -0.69$ barn~\citep{haiduke2006nuclear, pyykko2008year}, i.e. \SI{-0.69e-28}{\meter\square}.
The \gls*{efg} tensor components $\mathcal{V}_{\alpha\beta}=\partial^2 V(x,y,z)/\partial\alpha\partial\beta$ are defined as the second partial derivatives of the electric potential $V(x,y,z)$ experienced by the nucleus.
Note that energy is given in units of frequency throughout this work, hence the division in Eq.~\eqref{eq: Q_interaction_strength} by $h$. 
Furthermore, all Sternheimer anti-shielding factors~\cite{sternheimer1963quadrupole} are incorporated into $\mathcal{V}_{\alpha\beta}$. 
Finally, the \gls*{efg} tensor is traceless and symmetric, as $V$ experienced at the nucleus obeys the Laplace equation $\nabla^2 V=0$ and its partial derivatives commute. 
Therefore, $Q_{\alpha \beta} = Q_{\beta\alpha}$.

Two important aspects of $\hat{\mathcal{H}}_Q$ are worth emphasizing here. 
First, this interaction can only be non-zero for nuclei having $I \geq 1$. 
The quadrupole interaction was therefore irrelevant to the extensive body of earlier work on identical device structures containing a \Ph donor with $I=1/2$.
Second, as $q_n$ has a very small value, the \gls*{efg} necessary to generate a significant quadrupole interaction needs to be large. 
Typically, only a microscopic mechanism in a crystal lattice or molecule, such as the distortion of covalent bonds in the vicinity of the nucleus, creates a significant \gls*{efg}. 
An in-depth discussion of the microscopic origin of the \gls*{efg} at the position of the nucleus in our experiment, and its dependence on electric field, may be found in Sec.~\ref{sec: microscopic_theory}. 
For the following, it suffices to know that such an \gls*{efg} is present, and that it can be changed by varying an external electric field.

\subsection{Estimating nuclear quadrupole interaction parameters}
\label{subsec: nuc_perturb}

As all experiments are performed at $B_0 = \SI{1.496}{\tesla}$, the level splitting due to the Zeeman interaction is of order $\SI{8}{\mega \hertz}$. This is more than a factor 100 larger than the experimentally observed quadrupole shift, which is visible as a splitting of $\SI{66}{\kilo \hertz}$ in the spectrum. 
This justifies treating the quadrupole interaction as a perturbation to the Zeeman interaction.

Eigenstates $\ket{m_I}$ of the unperturbed system are labelled with the secondary spin quantum number $m_I$, ranging from $-I$ to $I$ in steps of one.
Applying standard first-order perturbation theory to find the quadrupole interaction correction to the energy, i.e.
\begin{align}
    E_{m_I} = \gamma_n B_0 m_I + \braket{m_I|\hat{\mathcal{H}}_Q|m_I},
\end{align}
results in a transition frequency between neighboring levels given by
\begin{align}
\label{eq: f_transition}
    f_{m_I-1 \leftrightarrow m_I} = -\gamma_n B_0 + \left(m_I-\tfrac{1}{2}\right)\left(Q_{xx}+Q_{yy}-2Q_{zz}\right),
\end{align}
where the transition frequency $f_{m_I-1 \leftrightarrow m_I}$ corresponds to a transition between states $\ket{m_I-1}$ and $\ket{m_I}$.

The quadrupole splitting $f_Q$ introduced in the main text is now defined as the first-order splitting between two subsequent spectral lines
\begin{align} \label{eq: def_Q}
    f_Q = f_{m_I \leftrightarrow m_I+1}-f_{m_I-1 \leftrightarrow m_I} = Q_{xx}+Q_{yy}-2Q_{zz}.
\end{align}

Thus far, the quadrupole interaction Hamiltonian has been considered in the laboratory frame ($z$-axis along $B_0$, $x$-axis in plane with device substrate). 
However, since the \gls*{efg} is a real, traceless, and symmetric tensor, a set of principal axes $x'$, $y'$, $z'$ exists for which the tensor is diagonal. 
The corresponding diagonalized Hamiltonian $\hat{\mathcal{H}}_Q'$ is given by
\begin{align} \label{eq: H_Q_intrinsic}
    \hat{\mathcal{H}}_Q'= \frac{e q_n \mathcal{V}_{z'z'}}{4I(2I-1)h}\left[3\hat{I}_{z'}^2-I^2+\frac{\mathcal{V}_{x'x'}-\mathcal{V}_{y'y'}}{\mathcal{V}_{z'z'}}\left(\hat{I}_{x'}^2-\hat{I}_{y'}^2\right) \right],
\end{align}
with the pricipal axes chosen such that $|\mathcal{V}_{x'x'}|\leq|\mathcal{V}_{y'y'}|\leq|\mathcal{V}_{z'z'}|$. 
It is standard practice to define an asymmetry parameter $\eta = \frac{\mathcal{V}_{x'x'}-\mathcal{V}_{y'y'}}{\mathcal{V}_{z'z'}}$ to characterize the deviation from cylindrical symmetry ($0\leq \eta \leq 1$, $\eta = 0$ corresponding to perfect cylindrical symmetry). 
In summary, the five independent degrees of freedom of the quadrupole Hamiltonian may now be understood as an overall interaction strength $Q$ (the prefactor to $I_{z'}^2$ in Eq.~\eqref{eq: H_Q_intrinsic}), an asymmetry $\eta$, and three angles defining the orientation of the principal axes. 

We stress that the measured quadrupole splitting $f_Q$, as introduced in the main text and defined by Eq.~\eqref{eq: def_Q}, has no one-to-one correspondence to the `intrinsic' quadrupole interaction strength $Q$. 
This is due to a strong dependence of the spectral line splitting on the value of the asymmetry parameter as well as the angle between external magnetic field and the quadrupole's principal $z'$-axis. 
Both of these quantities are unknown in the absence of any a-priori knowledge of the orientation and strength of the \gls*{efg}.
Given this, $f_Q$ may take on any value in the range $-2Q \leq f_Q \leq 2Q$.
Extracting the five independent quadrupole degrees of freedom requires measuring the nuclear spectrum for different orientations of the external magnetic field, which is beyond the scope of the current work.
Consequently, based on the available experimental data alone we cannot comment on the strength, asymmetry or orientation of the quadrupole interaction in the system.
However, the observed splitting $f_Q$ in the spectrum gives a conservative estimate of a quadrupole interaction strength of at least $\frac{3 e q_n \mathcal{V}_{z'z'}}{4I(2I-1)h}=\SI{33}{\kilo \hertz}$.

\subsection{Nuclear electric resonance}
\label{sec:nuclear_electric_resonance}
Resonant electric driving of transitions between nuclear spin states can be achieved by modulation of the quadrupole interaction.
We postulate (see Sec.~\ref{sec: microscopic_theory} for a thorough introduction and discussion) the existence of a modulation of the \gls*{efg} components at the nucleus. 
This modulation has an amplitude $\delta \mathcal{V}_{\alpha \beta}$ (independent of frequency) and is of the form $\cos{\left(2\pi f t\right)}\delta \mathcal{V}_{\alpha \beta}$, where $f$ is the drive frequency.
This \gls*{efg} modulation causes a modulation of the quadrupole interaction (Eq.~\eqref{eq: Q_interaction_strength}), which we shall refer to as $\delta Q_{\alpha \beta}$. 
The time-dependent addition $\hat{\mathcal{H}}(t)$ to the static Hamiltonian $\hat{\mathcal{H}}$ (given by Eq.~\eqref{eq: H_system_static}) is
\begin{align} \label{eq: H_Q_time}
    \hat{\mathcal{H}}(t) = \cos{(2\pi f t)}\delta \hat{\mathcal{H}}_Q = \cos{(2\pi f t)}\mathlarger{\mathlarger{\sum}}_{\mathclap{\alpha, \beta \in \{x,y,z\}}}\delta Q_{\alpha\beta}\hat{I}_\alpha \hat{I}_\beta.
\end{align}

Without prior knowledge of the exact \gls*{efg} modulation at the nucleus, the absolute value of the Rabi frequencies of the different transitions cannot be predicted. 
Crucially, however, a clear and distinctive prediction about the \emph{relationship} between the Rabi frequencies for the different \gls*{ner} transitions is possible. 
For \Sb with $I=7/2$, the spin matrices are given by 
\begin{align}
    \hat{I}_{x} &\quad = \quad \frac{1}{2}
    \begin{bmatrix}
    0 & \sqrt{7} & 0 & 0 & 0 & 0 & 0 & 0 \\
    \sqrt{7} & 0 & \sqrt{12} & 0 & 0 & 0 & 0 & 0 \\
    0 & \sqrt{12} & 0 & \sqrt{15} & 0 & 0 & 0 & 0 \\
    0 & 0 & \sqrt{15} & 0 & \sqrt{16} & 0 & 0 & 0 \\
    0 & 0 & 0 & \sqrt{16} & 0 & \sqrt{15} & 0 & 0 \\
    0 & 0 & 0 & 0 & \sqrt{15} & 0 & \sqrt{12} & 0 \\
    0 & 0 & 0 & 0 & 0 & \sqrt{12} & 0 & \sqrt{7} \\
    0 & 0 & 0 & 0 & 0 & 0 & \sqrt{7} & 0 \\
    \end{bmatrix}\\
    \hat{I}_{y} &\quad = \quad \frac{i}{2} \begin{bmatrix}
    0 & -\sqrt{7} & 0 & 0 & 0 & 0 & 0 & 0 \\
    \sqrt{7} & 0 & -\sqrt{12} & 0 & 0 & 0 & 0 & 0 \\
    0 & \sqrt{12} & 0 & -\sqrt{15} & 0 & 0 & 0 & 0 \\
    0 & 0 &\sqrt{15} & 0 & -\sqrt{16} & 0 & 0 & 0 \\
    0 & 0 & 0 & \sqrt{16} & 0 & -\sqrt{15} & 0 & 0 \\
    0 & 0 & 0 & 0 &\sqrt{15} & 0 & -\sqrt{12} & 0 \\
    0 & 0 & 0 & 0 & 0 & \sqrt{12} & 0 & -\sqrt{7} \\
    0 & 0 & 0 & 0 & 0 & 0 & \sqrt{7} & 0 \\
    \end{bmatrix}\\
    \hat{I}_{z} &\quad = \quad \frac{1}{2} \begin{bmatrix}
    7 & 0 & 0 & 0 & 0 & 0 & 0 & 0 \\
    0 & 5 & 0 & 0 & 0 & 0 & 0 & 0 \\
    0 & 0 & 3 & 0 & 0 & 0 & 0 & 0 \\
    0 & 0 & 0 & 1 & 0 & 0 & 0 & 0 \\
    0 & 0 & 0 & 0 & -1 & 0 & 0 & 0 \\
    0 & 0 & 0 & 0 & 0 & -3 & 0 & 0 \\
    0 & 0 & 0 & 0 & 0 & 0 & -5 & 0 \\
    0 & 0 & 0 & 0 & 0 & 0 & 0 & -7 \\
    \end{bmatrix}.
\end{align}
Transitions between neighboring states $\ket{m_I-1}$ and $\ket{m_I}$ require the first off-diagonal matrix elements to be non-zero. 
Since $\hat{\mathcal{H}}(t)$ contains only products of the spin operators, the \gls*{ner} transitions between neighboring states $\ket{m_I-1}$ and $\ket{m_I}$ are only possible for terms involving the products $\hat{I}_x \hat{I}_z$, $\hat{I}_z \hat{I}_x$, $\hat{I}_y \hat{I}_z$, and $\hat{I}_z \hat{I}_y$.
More formally, when $f$ equals the transition frequency for the transition between $\ket{m_I-1}$ and $\ket{m_I}$ ($\Delta m_I = \pm 1$), the resonant \gls*{ner} Rabi frequency $f^{\textrm{Rabi, NER}}_{m_I-1\leftrightarrow m_I}$ is given by 
\begin{align} \label{eq: frabi_single}
    f^{\textrm{Rabi, NER}}_{m_I-1\leftrightarrow m_I} = |\braket{m_I-1|\delta \hat{\mathcal{H}}_Q|m_I}|=|\delta Q_{xz}\braket{m_I-1|\hat{I}_x \hat{I}_z + \hat{I}_z \hat{I}_x |m_I}+ \delta Q_{yz}\braket{m_I-1|\hat{I}_y \hat{I}_z + \hat{I}_z \hat{I}_y |m_I}|.
\end{align}
The operator $\hat{I}_x \hat{I}_z + \hat{I}_z \hat{I}_x$ is given by 
\begin{align}
    \hat{I}_x \hat{I}_z + \hat{I}_z \hat{I}_x \quad =  \quad  \begin{bmatrix}
    0 & \sqrt{63} & 0 & 0 & 0 & 0 & 0 & 0 \\
    \sqrt{63} & 0 & \sqrt{48} & 0 & 0 & 0 & 0 & 0 \\
    0 & \sqrt{48} & 0 & \sqrt{15} & 0 & 0 & 0 & 0 \\
    0 & 0 & \sqrt{15} & 0 & 0 & 0 & 0 & 0 \\
    0 & 0 & 0 & 0 & 0 & -\sqrt{15} & 0 & 0 \\
    0 & 0 & 0 & 0 & -\sqrt{15} & 0 & -\sqrt{48} & 0 \\
    0 & 0 & 0 & 0 & 0 & -\sqrt{48} & 0 & -\sqrt{63} \\
    0 & 0 & 0 & 0 & 0 & 0 & -\sqrt{63} & 0 \\
    \end{bmatrix},
\end{align}
and the purely imaginary operator $\hat{I}_y \hat{I}_z + \hat{I}_z \hat{I}_y$ has matrix elements with equal modulus. 
The relationship between the $\Delta m_I = \pm 1$ Rabi frequencies is determined by the relative  magnitudes of the first off-diagonal matrix elements. 
In particular, the matrix element corresponding to the $\ket{-1/2} \leftrightarrow \ket{1/2}$ transition is zero and therefore this transition cannot be driven via \gls*{ner}.
Taking the modulus of the matrix elements of Eq.~\eqref{eq: frabi_single} yields the Rabi frequencies $f^{\textrm{Rabi,NER}}_{m_I-1 \leftrightarrow m_I}$ of the $\Delta m_I = 1$ transitions, which are given by
\begin{align} \label{eq: frabi_single_all}
   f^{\textrm{Rabi, NER}}_{m_I-1 \leftrightarrow m_I} = \alpha_{m_I-1\leftrightarrow m_I} \sqrt{\delta Q^2_{xz}+\delta Q^2_{yz}}
\end{align}
with $\alpha_{m_I-1 \leftrightarrow m_I}$ given by
\begin{align}
    \alpha_{m_I-1 \leftrightarrow m_I} = \frac{1}{2}\left| 2m_I-1\right|\sqrt{I(I+1)-m_I(m_I-1)}.
\end{align}
Values of $\alpha_{m_I-1 \leftrightarrow m_I}$ for the different transitions are given in Table~\ref{tab: rabi_constants}.

Transitions between next-nearest-neighboring levels $\ket{m_I-2} \leftrightarrow \ket{m_I}$ ($\Delta m_I = \pm 2$) are allowed in first order for quadratic interactions such as the quadrupole interaction.
Again, inspecting the spin matrices reveals that matrix elements coupling such states must stem from operator products $\hat{I}^2_x$, $\hat{I}^2_y$, $\hat{I}_x \hat{I}_y$, and $\hat{I}_y \hat{I}_x$, as only such products result in non-zero matrix elements on the second off-diagonal. 
Explicitly, the Rabi frequency for such transitions, $f^{\textrm{Rabi, NER}}_{m_I-2 \leftrightarrow m_I}$, is given by
\begin{align}
    f^{\textrm{Rabi, NER}}_{m_I-2\leftrightarrow m_I}=|\delta Q_{xx}\braket{m_I-2| \hat{I}^2_x |m_I}+\delta Q_{yy} \braket{m_I-2| \hat{I}^2_y|m_I}+\delta Q_{xy}\braket{m_I-2|\hat{I}_x \hat{I}_y+ \hat{I}_y \hat{I}_x|m_I}|.
\end{align}
The operator $\hat{I}^2_x$ is given by 
\begin{align}
    \hat{I}^2_x \quad =  \quad \frac{1}{4} \begin{bmatrix}
    7 & 0 & \sqrt{84} & 0 & 0 & 0 & 0 & 0  \\
    0 & 19 & 0 & \sqrt{180} & 0 & 0 & 0 & 0 \\
    \sqrt{84} & 0 & 27 & 0 & \sqrt{240} & 0 & 0 & 0 \\
    0 & \sqrt{180} & 0 & 31 & 0 & \sqrt{240} & 0 & 0 \\
    0 & 0 & \sqrt{240} & 0 & 31 & 0 & \sqrt{180} & 0 \\
    0 & 0 & 0 & \sqrt{240} & 0 & 27 & 0 & \sqrt{84} \\
    0 & 0 & 0 & 0 & \sqrt{180} & 0 & 19 & 0 \\
    0 & 0 & 0 & 0 & 0 & \sqrt{84} & 0 & 7 \\
    \end{bmatrix},
\end{align}
showing the coupling matrix elements on the second off-diagonal. 
The operator $\hat{I}^2_y$ has elements on the second off-diagonal with the same magnitude  but opposite sign, whereas the operator $\hat{I}_x \hat{I}_y + \hat{I}_y \hat{I}_x$ is purely imaginary and its second off-diagonal elements have twice the strength of those of the other operators. 
This leads to a Rabi frequency given by 
\begin{align} \label{eq: frabi_double_all}
    f^{\textrm{Rabi, NER}}_{m_I-2\leftrightarrow m_I} = \beta_{m_I-2\leftrightarrow m_I} \sqrt{(\delta Q_{xx}-\delta Q_{yy})^2+4\delta Q^2_{xy}}
\end{align}
with $\beta_{m_I-2\leftrightarrow m_I}$ given by
\begin{align}
    \beta_{m_I-2\leftrightarrow m_I} = \frac{1}{4}\sqrt{(I-m_I-7)(I-m_I-6)(I-m_I+1)(I-m_I+2)}
\end{align}
Values of $\beta_{m_I-2 \leftrightarrow m_I}$ for the different transitions are given in Table~\ref{tab: rabi_constants}.

The \gls*{ner} Rabi frequencies for the $\Delta m_I = \pm 1$ and $\Delta m_I = \pm 2$ transitions, as given by Eq.~\eqref{eq: frabi_single_all} and Eq.~\eqref{eq: frabi_double_all} respectively, are only rescaled  by the particulars of the \gls*{efg} modulation. 
Importantly, the \emph{relationship }between them presents a trend unique to electric drive via the nuclear quadrupole interaction. 

These predictions for \gls*{ner} should be compared against a similar prediction for \gls*{nmr}.
In the case of \gls*{nmr}, the driving of transitions is caused by an interaction with a time-varying magnetic field of the form $\cos{(2 \pi f t)\gamma_n B_1 \hat{I}_x}$. 
For transitions between neighboring levels, the \gls*{nmr} Rabi frequency is now determined by
\begin{align}
    f^{\textrm{Rabi, NMR}}_{m_I-1\leftrightarrow m_I} = |\braket{m_I-1|\gamma_n B_1 \hat{I}_x|m_I} |,\\
\end{align}
which leads to 
\begin{align} \label{eq: frabi_single_all_NMRl}
    f^{\textrm{Rabi, NMR}}_{m_I-1\leftrightarrow m_I} = |\gamma_n| B_1 \zeta_{m_I-1\leftrightarrow m_I}
\end{align}
with $\zeta_{m_I-1 \leftrightarrow m_I}$ given by
\begin{align}
    \zeta_{m_I-1 \leftrightarrow m_I} = \frac{1}{2}\sqrt{I(I+1)+m_I(m_I-1)}.
\end{align}
In case an additional driving component along $\hat{I}_y$ is present, $B_1$ is replaced by the modulus of the total field, but $\zeta_{m_I-1 \leftrightarrow m_I}$ remains unchanged. 
Values of $\zeta_{m_I-1 \leftrightarrow m_I}$ for the different transitions are given in Table~\ref{tab: rabi_constants}.
Importantly, transitions between next nearest neighboring states ($\Delta m_I = \pm 2$) are to first order forbidden under \gls*{nmr}. 

This experiment resulted in two key observations, as shown in the main text. 
First, the Rabi frequencies of the transitions between neighboring states accurately follow the trend predicted for \gls*{ner} in Eq.~\eqref{eq: frabi_single_all}  (including the absence of the $\ket{-1/2}\leftrightarrow \ket{1/2}$ transition, which is predicted to have zero matrix element), and does not follow the prediction of Eq.~\eqref{eq: frabi_single_all_NMRl} for \gls*{nmr}.
Second, transitions between next-nearest-neighboring states, which are first-order forbidden for \gls*{nmr}, accurately follow the prediction of Eq.~\eqref{eq: frabi_double_all} for \gls*{ner}. 
These combined observations enable the main conclusion of this work: the demonstration of coherent electrical control of the \Sb nuclear spin via \gls*{ner}, without transducing the RF electric field into a magnetic field.

\begin{table}[h]
\renewcommand{\arraystretch}{1.4}
    \centering
        \caption{Rabi frequency coefficients for \acrshort*{ner} and \acrshort*{nmr} transitions. Transitions are labeled by their secondary spin quantum number $m_I$. The coefficient  $\alpha_{m_I-1\leftrightarrow m_I}$ governs the \acrshort*{ner} transition between neighboring levels with $\Delta m_I = \pm 1$, $\beta_{m_I-2\leftrightarrow m_I}$ governs the \acrshort*{ner} transition between next nearest neighboring levels with $\Delta m_I = \pm 2$ (hence the last transition in the table not existing), and $\zeta_{m_I-1\leftrightarrow m_I}$ governs the \acrshort*{nmr} transition between neighboring levels with $\Delta m_I = \pm 1$.
    }
    \begin{tabular}{c|c c c c c c c c}
         $m_I$ &  $7/2$ &  $5/2$ &  $3/2$ &  $1/2$ &  $-1/2$ & $-3/2$ & $-5/2$ \\ \hline
         $\alpha_{m_I-1\leftrightarrow m_I}$ & $\sqrt{63}$ & $\sqrt{48}$ & $\sqrt{15}$ & 0 & $\sqrt{15}$ & $\sqrt{48}$ & $\sqrt{63}$\\
         $\beta_{m_I-2\leftrightarrow m_I}$ & $\sqrt{21}/2$ & $\sqrt{45}/2$ & $\sqrt{60}/2$ & $\sqrt{60}/2$ & $\sqrt{45}/2$ & $\sqrt{21}/2$ & -\\
         $\zeta_{m_I-1\leftrightarrow m_I}$ & $\sqrt{7}/2$ & $\sqrt{12}/2$ & $\sqrt{15}/2$ & $\sqrt{16}/2$ & $\sqrt{15}/2$ & $\sqrt{12}/2$ & $\sqrt{7}/2$
    \end{tabular}
    \label{tab: rabi_constants}
\end{table}

\section{Device fabrication} \label{sec: fab}

\subsection{Fabrication protocol} \label{subsec: fab_prot}

Devices are fabricated using standard microelectronic fabrication techniques. 
The substrate is a p-type <100> silicon (Si) wafer (resistivity 10-\SI{20}{\ohm \centi \meter}), with a \gls*{lpcvd} grown epitaxial layer of isotopically enriched $^{28}\mathrm{Si}$ (concentration of residual $^{29}\mathrm{Si}$ = 730~ppm) of \SI{900}{\nano\meter} thickness on top. 

The main fabrication steps are:
\begin{enumerate}
    \item Definition of negative optical alignment markers via tetramethylammonium hydroxide (TMAH) etch of the silicon, using optical lithography and a wet thermally grown masking oxide. 
    The masking oxide is subsequently removed using buffered \gls*{hf} etching. 
    \item Creation of p-doped regions in between the n-doped metallic leads of step~4, defined using optical lithography and a wet thermally grown masking oxide. 
    This step is designed to suppress spurious leakage currents flowing at the silicon / silicon dioxide (Si/SiO\textsubscript{2}) interface between the n-doped leads. 
    Doping is achieved via thermal diffusion of boron (B). The masking oxide is subsequently removed using buffered \gls*{hf} etching.
    \item Two-step thermal oxidation process to repair defects and drive B into the silicon, resulting in a final wet thermal oxide of \SI{200}{\nano\meter} thickness. 
    \item Creation of n-doped metallic leads, defined by optical lithography using the thermally grown masking oxide of the previous step. Doping is achieved via thermal diffusion of phosphorus (P). Masking oxide is subsequently removed via buffered \gls*{hf} etching. 
    \item Single-step thermal oxidation to drive P into the silicon, resulting in a wet thermal oxide of \SI{200}{\nano\meter} thickness which serves as thick field oxide.
    \item Etching of a central window of $\SI{20}{\micro\meter}\times\SI{40}{\micro\meter}$ in the field oxide using buffered \gls*{hf} etching, and subsequent growth of an \SI{8}{\nano\meter} thick, high quality dry thermal oxide. The p-doped regions and n-doped leads of steps 2 and 4 extend for $\sim \SI{2}{\micro\meter}$ underneath the high quality gate oxide. 
    \item Definition of positive alignment markers via optical lithography. 
    Markers are created by electron beam evaporation of \SI{15/75}{\nano\meter} titanium/platinum (Ti/Pt) and lift-off using warm \gls*{nmp}.
    \item Definition of positive electron beam alignment markers by patterning a \gls*{pmma} mask aligned to the optical alignment markers using \gls*{ebl}. 
    Markers are created by electron beam evaporation of \SI{15/75}{\nano\meter} titanium/platinum (Ti/Pt) and lift-off using warm \gls*{nmp}. 
    \item Definition of a \SI{90}{\nano\meter}$\times$\SI{90}{\nano\meter} implantation window, patterned in a \gls*{pmma} mask using \gls*{ebl}. 
    \item Low-dose implantation of \Sb (see Sec.~\ref{subsec: fab_implant} for details), followed by a rapid thermal anneal for \SI{5}{\second} at \SI{1000}{\celsius} to achieve donor activation and repair implantation damage to the silicon lattice.
    \item Definition of ohmic contacts to the n-doped metallic leads via optical lithography; ohmic contact is achieved by etching through the field oxide using buffered \gls*{hf}, evaporation of \SI{200}{\nano\meter} \gls*{al}, and a \SI{15}{\minute} forming gas (5\% hydrogen in nitrogen) anneal at \SI{400}{\celsius}. 
    \item Definition of all gates necessary to control and read out the \Sb donor, including gates to tune the electrochemical potential of the donor, a \gls*{set} for read-out, and a microwave antenna. This is achieved via two steps of standard \gls*{ebl} using \gls*{pmma} masks, thermal evaporation of \gls*{al} and lift-off using warm \gls*{nmp}. The first layer consists of \SI{20}{\nano\meter} of \gls*{al}, and contains the \gls*{set} barrier gates and two of the donor tuning gates. The second layer consists of \SI{40}{\nano\meter} of \gls*{al}, and contains the top-gate of the \gls*{set}, the two remaining donor tuning gates, and the microwave antenna. The two layers are electrically isolated by the native oxide formed on the first \gls*{al} layer upon exposure to air. 
    \item Forming gas anneal for \SI{15}{\minute} at \SI{400}{\celsius}, to passivate traps in the gate oxide. 
    \item Dicing and packaging, including bonding (see also Sec.~\ref{sec: exp_setup}). 
\end{enumerate}

\subsection{\texorpdfstring{\Sb}{123-Sb} implantation parameters} \label{subsec: fab_implant}
\begin{figure}
    \centering
    \includegraphics{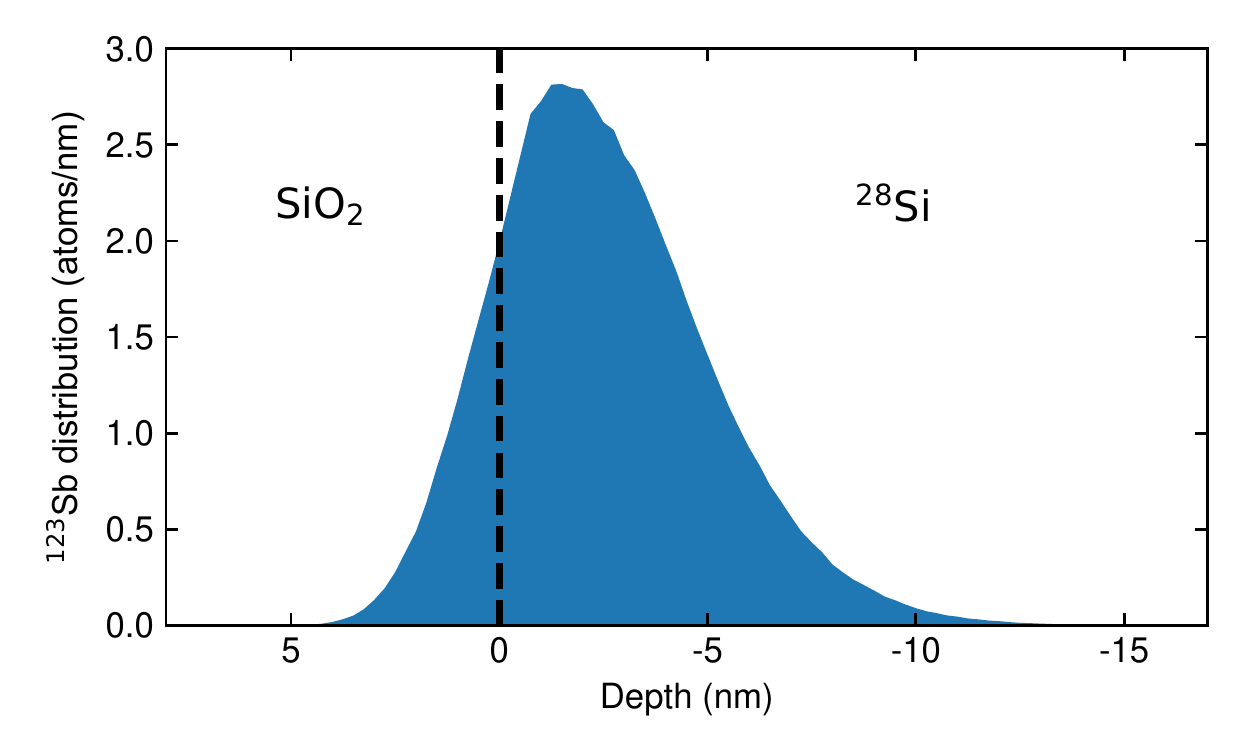}
    \caption{Expected \Sb concentration versus depth, integrated over the implantation window area. 
    Ion implantation parameters are an acceleration voltage of \SI{8}{\kilo\electronvolt} at a fluence of \SI{2e11}{\per\square\centi\meter}. 
    The distribution is simulated with SRIM / TRIM and scaled by the 90~nm~$\times$~90~nm area of the implantation window. 
    The dashed vertical line at \SI{0}{\nano\meter} denotes the SiO\textsubscript{2}/Si interface. 
    The peak of the implantation profile is at  \SI{2}{\nano\meter} below this interface; integrating the distribution over the Si volume underneath the implantation window results in an expected number of 14 \Sb donors per implantation window. 
    }
    \label{fig:implantation distribution}
\end{figure}

Standard modeling using SRIM/TRIM software was used to predict the \Sb dopant profile. 
The implantation energy was chosen such that the peak of the \Sb dopant profile is located \SI{2}{\nano\meter} below the SiO\textsubscript{2}/Si interface. 
Such a shallow implantation was chosen to maximize the donor’s exposure to static strain upon cool-down of the sample (see Sec.~\ref{sec: comsol_model} for details on strain calculations). 
Given a gate oxide thickness of \SI{8}{\nano\meter}, the optimal implantation energy was found to be \SI{8}{\kilo\electronvolt}; the simulated vertical doping profile is shown in Fig.~\ref{fig:implantation distribution}. 
A low dose of \SI{2e11}{\per\square\centi\meter} was implanted. 
This corresponds to an average of 14 donors per implantation window, i.e. each device is expected to have a few donors tunnel-coupled to the single-electron transistor.

\subsection{Microwave antenna} \label{subsec: fab_antenna}

\begin{figure}
    \centering
    \includegraphics{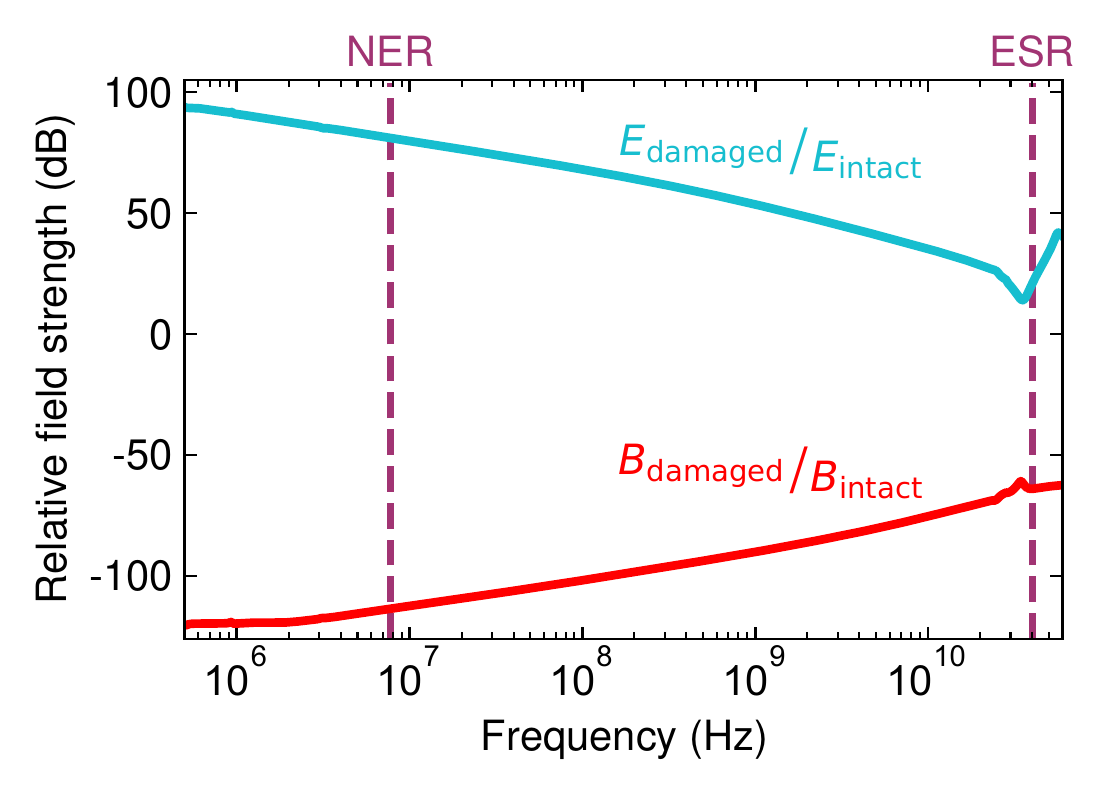}
    \caption{Finite-element simulation of the AC electric and magnetic fields produced by either an intact or a damaged antenna (see Fig. 1A) at the triangulated donor location (Sec.~\ref{sec: donor_triangulation}), using the CST Studio 2018 software package. The ratios $E_{\rm damaged} / E_{\rm intact} \gg 0$~dB and $B_{\rm damaged} / B_{\rm intact} \ll 0$~dB indicate that the damaged antenna produces chiefly an electric field at its tip, rather than a magnetic field, due to the nanoscale open-circuit termination at its tip, in proximity of the donor.
    These effects are more pronounced at the lower frequencies, explaining the driving of nuclear transitions via \gls*{ner}.
    With increasing frequency, the capacitive impedance created by the gap in the damaged antenna becomes small enough to allow some AC current to flow.
    This produces an oscillating magnetic field at microwave frequencies that is sufficient to drive \gls*{esr} transitions, used for the purpose of nuclear state readout.}
    \label{fig:antenna_transmission}
\end{figure}
The antenna that was originally designed to perform \gls*{nmr} and \gls*{esr} is an on-chip coplanar waveguide \gls*{cpw} terminated by a short circuit. 
Due to the short circuit termination, the electric field at the tip of the antenna will be minimized, whilst a time-varying magnetic field is generated by the current flowing through the short.

In an effort to maximize the magnetic field $B_1$ at the donor site, multiple steps were taken to ensure the donor is located as close as possible to the antenna. 
This includes placing the implantation window no further than \SI{150}{\nano\meter} away from the antenna and making the shorting wire as narrow as possible. 
The microwave antenna is made of aluminum (see step~12 of the fabrication procedure outlined in Sec.~\ref{sec: fab}) and was designed to have a length of \SI{500}{\nano\meter} on each side of the \gls*{cpw}, a thickness of \SI{40}{\nano\meter} and a width of \SI{50}{\nano\meter}. 

However, we have noticed across multiple devices that a short of these dimensions is not robust enough to withstand our standard experimental procedures such as connecting and disconnecting the microwave line, or applying high-power RF pulses (of the order of $-10$ to $0$~dBm at the antenna). 
Upon inspection with a \gls*{sem}, the short typically appears `molten' (Fig.~1A), which suggests that the cross-sectional area of the short is too small to sustain high-power microwaves or current spikes.

One can approximate the behaviour of a broken antenna, to first order, by modelling it as a series RC circuit.
The electric field produced is proportional to the voltage at the capacitor, and the magnetic field is proportional to the current flowing through the circuit. 
Thus, the electric field will be at a relatively constant value for all frequencies until it starts rolling off at higher frequencies. 
The capacitor strongly attenuates any low-frequency current and thus causes the magnetic field to increase with frequency. 
At high frequencies, the impedance of the broken part of the antenna decreases to the point where a reasonable amount of current can flow. 
This all points to the antenna acting as an \emph{electric} antenna at low frequencies ($\sim \SI{8}{\mega\hertz}$ needed for \gls*{ner}) and a (albeit poor) magnetic antenna at higher frequencies ($\sim \SI{40}{\giga\hertz}$ needed for \gls*{esr}).

To study the electromagnetic response in some depth, the width and location of the antenna breaks were extracted from the \gls*{sem} (Fig. 1A) image and added to the original antenna design.
We simulated the broken antenna using CST Studio Suite, and extracted the electric and magnetic fields as a function of frequency at the expected donor site (Sec.~\ref{sec: donor_triangulation}).
Furthermore, the original, intact antenna was simulated in CST Studio Suite. 
This allowed the definition of a relative antenna performance, in which the electric and magnetic fields generated by the broken antenna at the donor site are divided by those of the working antenna.

The simulation results show a general trend that the magnetic field is reduced for a broken antenna and the electric field is enhanced, which match the experimental observations.
When analyzing the simulation results, we find that the simulations for a broken antenna suggest that the magnetic field is on the order of \SI{60}{\deci \bel} lower than that of a working antenna at \gls*{esr} frequencies. 
A significantly reduced magnetic field at \gls*{esr} frequencies has indeed been observed, but comparison to previous functional devices in our group suggests that the magnetic field is \SI{\sim30}{\deci\bel} weaker. 
The difference between experimentally observed values and simulation is likely due to the simplifications (exact shape of the discontinuity, grain structure, etc.) that were made in the CST model for ease of simulation.

The antenna simulations in the range of the nuclear resonance frequencies (\SI{\sim 8}{\mega \hertz}) show a highly-attenuated magnetic field (\SI{\sim 100}{\deci \bel} attenuation), which would inhibit any coherent driving of the \gls*{nmr} transitions.
Instead in this frequency range, the structure acts as a good electric antenna with $\sim \SI{80}{\deci\bel}$ higher electric fields, enabling driving via \gls*{ner}.
This explains how we were able to drive transitions via \gls*{ner} instead of \gls*{nmr} with a structure that was designed to be a broad-band magnetic antenna (Fig.~\ref{fig:antenna Rabi}). 
As outlined in the main text, the validity of this interpretation is confirmed by our observation of \gls*{ner} using a nearby electrostatic gate as an electrical antenna.

\section{Experimental setup} \label{sec: exp_setup}
\begin{figure}[ht]
    \centering
    \includegraphics{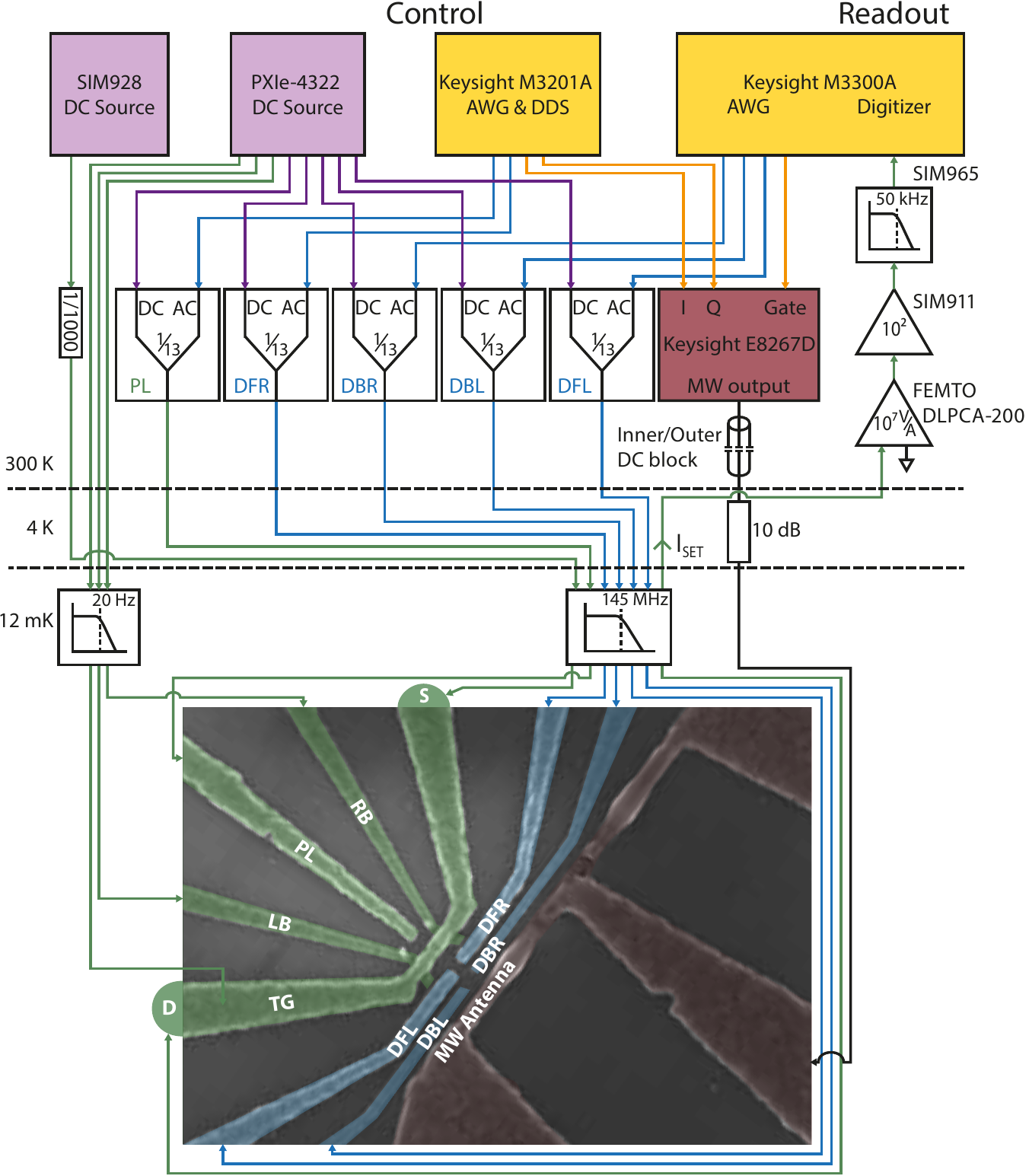}
    \caption{Schematic of the experimental measurement setup. 
    A DC source (PXIe 4322) is used to define gate voltages and tune the \acrshort*{set} via the TG, LB, RB and PL gates. 
    A total of eight \acrshort*{awg} channels (Keysight M3300A and M3201A) are used to apply dynamical electrical stimuli to the donor gates DFL, DFR, DBL and DBR of the device, and IQ modulation inputs to the microwave source. 
    The DC and AC signals are combined in a resistive voltage adder, reducing each input by a factor of 13. 
    A vector microwave source (Keysight E8267D) is used to perform \acrshort*{esr} via the microwave antenna.
    A small voltage bias (SIM928) is applied to the ohmic source contact, and the resulting \acrshort*{set} current is subsequently measured from the drain contact.
    The \acrshort*{set} current passes through a transimpedance amplifier (FEMTO DLPCA-200) with a gain of \SI{1e7}{\volt\per\ampere} and bandwidth of \SI{50}{\kilo\hertz}. 
    This is then further amplified (SIM911) by a factor of $10^2$ and filtered to a \SI{50}{\kilo\hertz} cut-off (SIM965) which is then recorded with a digitizer (Keysight M3300A). 
    }
    \label{fig:experimental_setup}
\end{figure}

Device packaging consists of a custom-made \gls*{pcb} with microwave and low-frequency lines, positioned around a central cutout accommodating the fabricated chip. 
The \gls*{pcb} is mounted to a copper enclosure with SK, i.e. \SI{2.92}{\milli\meter}, (rated to \SI{40}{\giga\hertz}) and MMCX (rated to \SI{6}{\giga\hertz}) connectors.
The device is glued to the enclosure and Al wire bonded to the \gls*{pcb}.
The copper enclosure is mounted on a gold-plated copper cold-finger bolted to the mixing chamber of a Bluefors BF-LD400 cryogen-free dilution refrigerator at a base temperature of \SI{12}{\milli\kelvin}. 
The static magnetic field in the experiment is produced by a superconducting magnet kept in persistent mode. 
The magnet is equipped with a low-drift persistent mode switch that results in a typical magnetic field drift of less than 50 ppb/hour.

Three different types of control lines are present in the experimental apparatus.
A single high-frequency microwave line is used for \gls*{esr} in the $\sim \SI{40}{\giga\hertz}$ regime, and has an inner/outer DC block at room temperature and a \SI{10}{\deci\bel} attenuator at the \SI{4}{\kelvin} stage of the dilution refrigerator. 
The coaxial microwave line has a silver-plated copper-nickel inner conductor, a copper-nickel outer conductor and a PTFE dielectric. 
Six radio-frequency coaxial lines are used for static and dynamic tuning of the donor electrochemical potential, driving via \gls*{ner}, and for the readout signal from the \gls*{set}. 
These lines have a graphite coating on the dielectric to reduce triboelectric noise effects\cite{kalra2016vibration} and are low-pass filtered, with a \SI{145}{\mega\hertz} cut-off.
Three more lines of a Constantan loom, low-pass filtered to a \SI{20}{\hertz} cut-off, are used for the static electrical tuning of the \gls*{set}.

The electronic setup, as depicted in Fig.~\ref{fig:experimental_setup}, includes two instruments as DC sources, a \gls*{srs} SIM928 and \gls*{ni} PXIe-4322.
We use a total of eight \gls*{awg} channels between the Keysight M3201A and M3300A modules, which are bandwidth limited to \SI{200}{\mega\hertz}, and a 100~MSPS digitizer channel of the M3300A is used to record \gls*{set} current traces. 
These modules have on-board \glspl*{fpga} where a variety of custom code is implemented, including an in-house \gls*{dds} system allowing generation and sequencing of sinusoidal and linear chirp pulses.
These are used as the IQ modulation inputs for the Keysight E8267D PSG vector microwave source, which is used to perform \gls*{esr}.

To measure the nuclear flipping probability from \gls*{ner}, two microwave signals with \gls*{esr} frequencies corresponding to the two nuclear states of interest need to be measured (Sec.~\ref{sec: experimental_methods}). Single-sideband modulation, either upper or lower, is used to address the relevant \gls*{esr} transitions.
To avoid simultaneously driving both transitions with leakage of the alternate sideband, the microwave carrier frequency is offset from the center of the two transition frequencies. 
The \gls*{set} current passes through the following chain of amplifiers and filters: FEMTO DLPCA-200 transimpedance amplifier, \gls*{srs} SIM911 BJT amplifier and a \gls*{srs} SIM965 low-pass filter. 
The Python-based QCoDeS data acquisition framework~\cite{qcodes} was used to control the instruments and perform measurements.

\section{Nuclear resonance measurement scheme and analysis} \label{sec: experimental_methods}

\subsection{Electron initialization, control, and readout}

\begin{figure}
    \centering
    \includegraphics{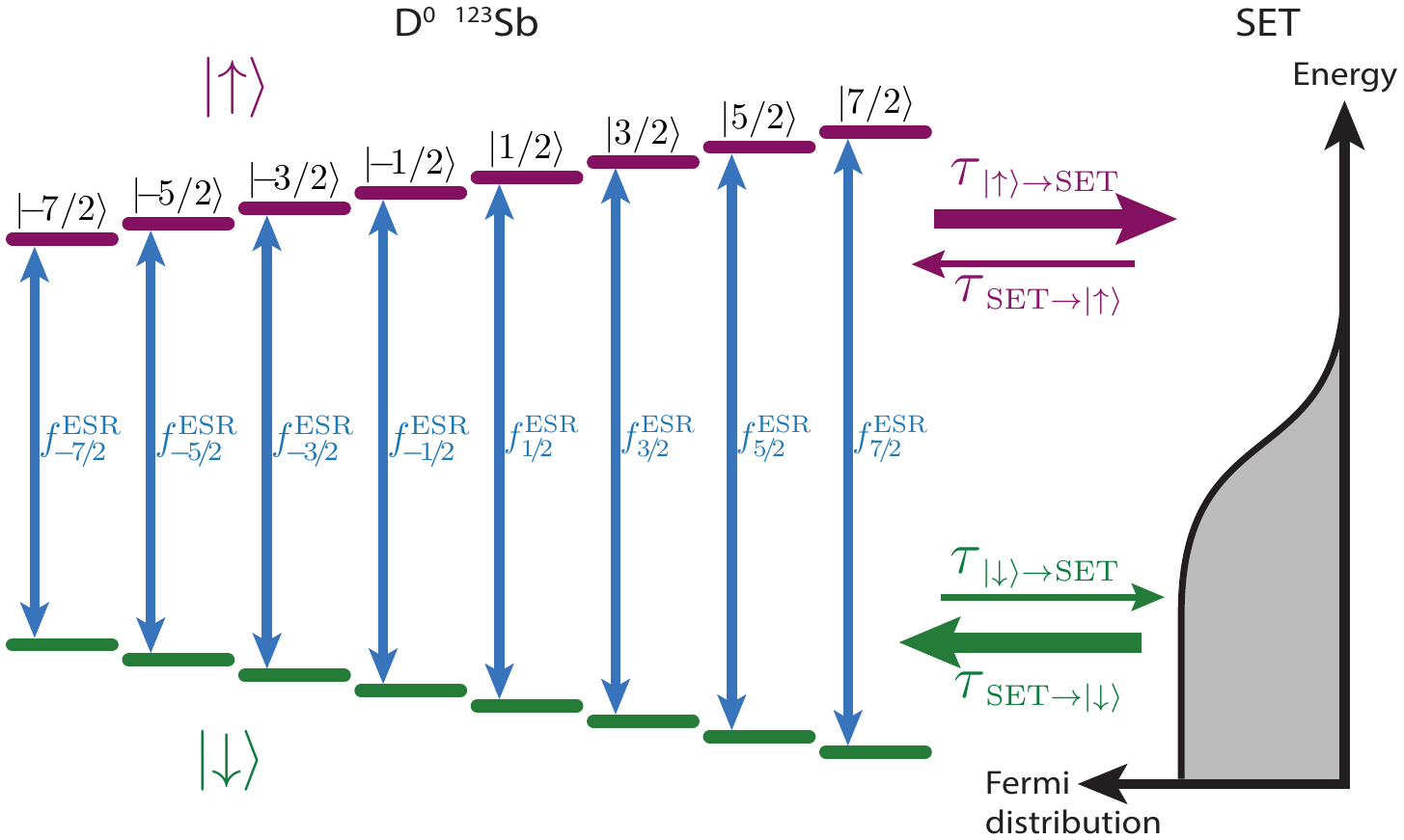}
    \caption{
    Energy level diagram of D$^0$ \Sb.
    This energy level diagram holds for the high magnetic-field limit ($\gamma_e B_0 \gg A$), where the nuclear and electron eigenstates are separable.
    When aligning the donor electrochemical potential with the Fermi energy of the \acrshort*{set}, the number of unoccupied electron states at the \acrshort*{set} (right) is much higher at the energy level of a $\ket{\uparrow}$ electron (blue) than a $\ket{\downarrow}$ electron (red).
    This results in a low (high) tunneling time $\tau_\mathrm{\ket{\uparrow}\rightarrow SET}$ ($\tau_\mathrm{\ket{\downarrow}\rightarrow SET}$) for a $\ket{\uparrow}$ ($\ket{\downarrow}$) electron from the donor to the \acrshort*{set}
    Conversely, the tunneling time from the \acrshort*{set} to the donor $\tau_\mathrm{SET \rightarrow \ket{\uparrow}}$ ($\tau_\mathrm{SET \rightarrow \ket{\downarrow}}$) is low (high) for a $\ket{\uparrow}$ ($\ket{\downarrow}$) electron.
    The \acrshort*{esr} frequencies (green) are different for each of the eight nuclear states, enabling conditional flipping of the electron spin dependent on the nuclear state.
    Energy levels are not to scale, as the electron Zeeman interaction $\gamma_e B_0$ is reduced for illustrative purposes.
    }
    \label{fig:energy level diagram}
\end{figure}

In the experiments outlined in this paper, the D$^0$ donor's outer electron is used as a tool to measure the nuclear spin state \cite{pla2013high}.
In the presence of a static magnetic field $B_0$ along the $z$-axis, the Zeeman interaction causes an energy splitting $\gamma_e B_0$ between a spin-up $\ket{\uparrow}$ and spin-down $\ket{\downarrow}$ electron projected along the $z$-axis.
If the electrochemical potential of a D$^0$ donor is tuned such that a $\ket{\uparrow}$ electron is above the Fermi level, and a $\ket{\downarrow}$ electron below the Fermi level, the characteristic electron tunnel time from donor to \gls*{set} is much shorter for a $\ket{\uparrow}$ electron ($\tau_{\ket{\uparrow}\rightarrow\mathrm{SET}}$) than a $\ket{\downarrow}$ electron ($\tau_{\ket{\downarrow}\rightarrow\mathrm{SET}}$).
Note that $\tau_{\ket{\downarrow}\rightarrow\mathrm{SET}}$ is still finite, primarily due to thermal broadening of the states that are occupied in the \gls*{set}.
Therefore, if we keep these electrochemical potentials fixed for a duration $t$, where $\tau_{\ket{\uparrow}\rightarrow\mathrm{SET}} < t < \tau_{\ket{\downarrow}\rightarrow\mathrm{SET}}$, only a $\ket{\uparrow}$ electron will likely have tunneled onto the \gls*{set}, an effect known as spin-dependent tunneling~\cite{elzerman2004single}.
Capacitive coupling between a donor electron and the \gls*{set} enables the \gls*{set} to be in Coulomb blockade exclusively when the donor is in the D$^0$ charge state.
Consequently, the tunneling of an electron from the donor onto the \gls*{set} can be measured as an onset of \gls*{set} current $I_\mathrm{SET}$ which, combined with spin-dependent tunneling, enables readout of the donor electron spin state.

Initialization of a $\ket{\downarrow}$ electron is performed using the electrochemical configuration described above, this time using the property that the electron tunnel time from \gls*{set} to D$^+$ donor is much shorter for a $\ket{\downarrow}$ electron.
Readout and initialization of the donor's outer electron can therefore be combined by remaining in the spin-dependent tunneling configuration for a fixed duration while measuring the \gls*{set} current.
Within this time, a $\ket{\uparrow}$ electron will likely tunnel onto the \gls*{set}, followed by a $\ket{\downarrow}$ electron tunneling back onto the donor, which is measured as a brief increase in \gls*{set} current.
A $\ket{\downarrow}$ electron, on the other hand, will likely not tunnel out, and no \gls*{set} current measured.
In both cases, the final electron is in the $\ket{\downarrow}$ state.
A more detailed description of readout and initialization of the donor electron is described in Ref.~\citenum{morello2009architecture,Morello2010}.

Microwave-frequency linear chirp pulses are applied to adiabatically flip the electron spin via \gls*{esr}.
During the \gls*{esr} pulse, the electrochemical potential of the donor is lowered well below the Fermi level, to ensure that the electron remains bound to the donor.
The chirp pulse frequency range is chosen such that the \gls*{esr} frequency is close to the center frequency, and the frequency ramp rate (chirpyness) is chosen to be below the adiabatic limit set by the Landau-Zener equation~\cite{laucht2014high}, and above the limit set by the homogeneous linewidth.
The main advantage of adiabatic pulses is that they give a high inversion fidelity without the need for a well-calibrated pulse amplitude, and are insensitive to slow resonance frequency drifts.

\subsection{Manipulation and readout of the nuclear state}

\begin{figure}[ht]
    \centering
    \includegraphics{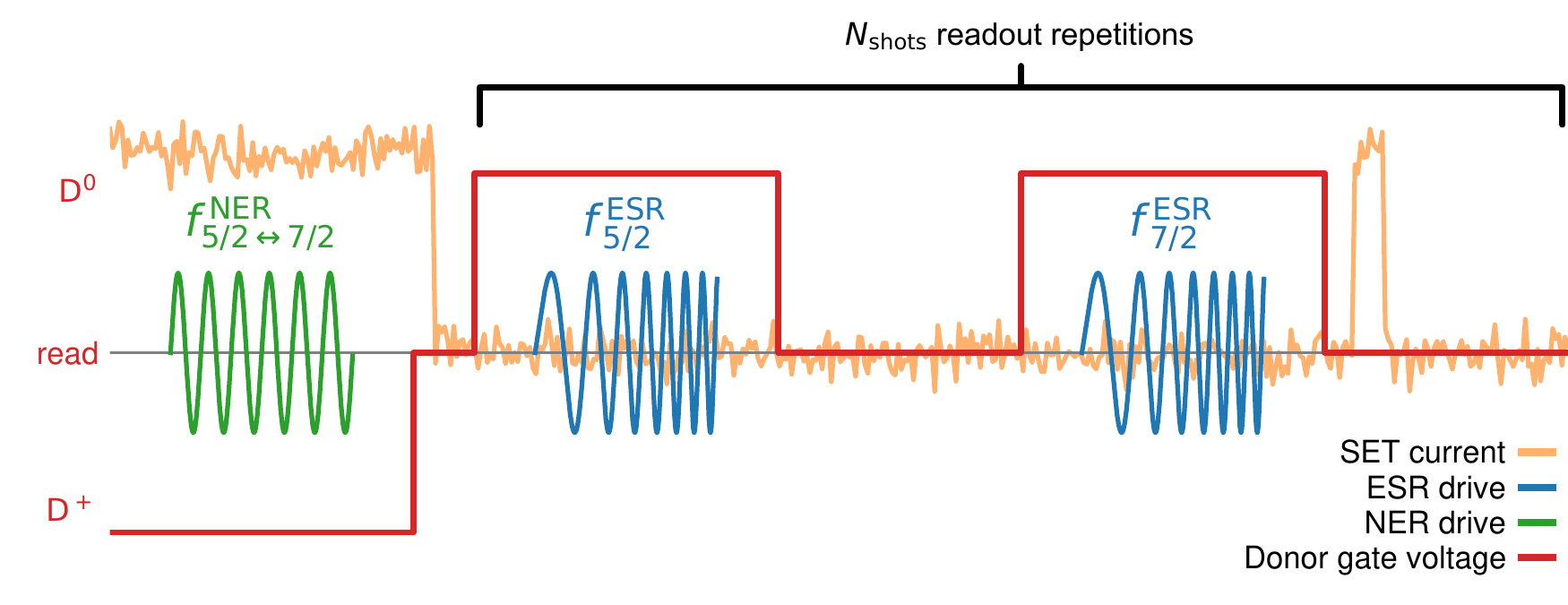}
    \caption{Schematic of \acrshort*{ner} pulse sequence.
    Donor gate voltages (red line) modify the donor electrochemical potential, and thereby whether the donor is ionized (D$^+$) or neutral (D$^0$).
    The \acrshort*{set} current (yellow) is high for a D$^+$ donor, as is the case during the start of the pulse sequence.
    An AC electric pulse (blue) with nuclear transition frequency $f^\mathrm{NER}_{5/2\leftrightarrow 7/2}$ is first applied to the D$^+$ donor to drive the $m_I = 5/2 \leftrightarrow 7/2$ transition.
    The two nuclear states are then measured sequentially by initializing a $\ket{\downarrow}$ electron, and then applying an \acrshort*{esr} chirp pulse with center frequency $f^\mathrm{ESR}_{5/2}$ ($f^\mathrm{ESR}_{7/2}$), which flips the electron to $\ket{\uparrow}$ if the nucleus is in state $\ket{5/2}$ ($\ket{7/2}$).
    The brief increase in \acrshort*{set} current after the second \acrshort*{esr} pulse with frequency $f^\mathrm{ESR}_{7/2}$ indicates that the nuclear spin state is $\ket{7/2}$.
    The readout sequence is repeated $N_\mathrm{shots}$ times to suppress measurement errors and ensure a correct nuclear readout.
    }
    \label{fig:nuclear pulse sequence}
\end{figure}

\begin{figure}[ht]
    \centering
    \includegraphics{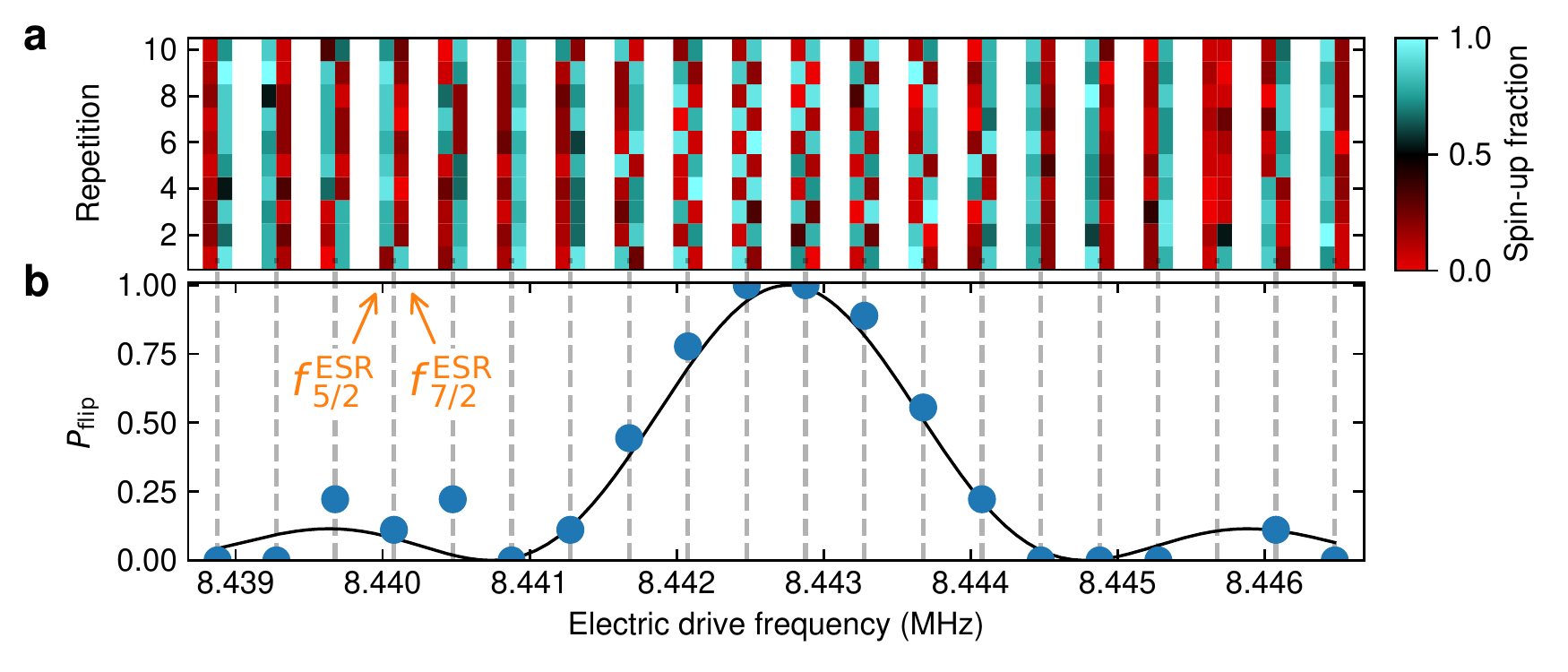}
    \caption{Nuclear resonance spectrum for $\ket{5/2} \leftrightarrow \ket{7/2}$ transition.
        \textbf{a}, Electron spin-up fraction of two \acrshort*{esr} frequencies for varying \acrshort*{ner} frequency, using the pulse sequence described in Fig.~\ref{fig:nuclear pulse sequence}.
        For each \acrshort*{ner} frequency, the left (right) column is the measured electron spin-up fraction for \acrshort*{esr} frequency $f^\mathrm{ESR}_{5/2}$ $\left(f^\mathrm{ESR}_{7/2}\right)$.
        The \acrshort*{esr} frequency with spin-up fraction above a preset threshold (here 0.5, black) indicates the nuclear state.
        The rows in each column show subsequent repetitions of the pulse sequence, from which a nuclear flipping probability $P_\mathrm{flip}$ is extracted. 
        If both electron spin-up fractions are low, this indicates that the nucleus is in one of the other six nuclear states (e.g. electric drive frequency \SI{8.466}{\mega \hertz}).
        In this case, an initialization routine is performed to return the nuclear spin to one of the two measured nuclear spin states (Sec.~\ref{subsec:nuclear spin initialization}).
        \textbf{b}, Nuclear flipping probability $P_\mathrm{flip}$.
        A nuclear flip is counted whenever the nuclear state switches between the two states in subsequent pulse sequence iterations.
        The number of nuclear flips are then divided by the total number of possible flips, resulting in $P_\mathrm{flip}$.
        If the nuclear state cannot be determined for each pulse sequence repetition, the data point is discarded.
        }
    \label{fig:nuclear flip probability}
\end{figure}

The nuclear resonance measurements always focus on two chosen states out of the eight nuclear spin states of \Sb.
We assume the nucleus to be initialized in one of these two relevant spin states.
The general pulse sequence for measurements of the nucleus consists of two stages (Fig.~\ref{fig:nuclear pulse sequence}).
The first stage is nuclear spin manipulation, which we performed on the ionized donor (D$^+$).
A nearby AC gate applies oscillating electric fields that can drive Rabi oscillations between two nuclear spin states via \gls*{ner}.

After completing the nuclear spin manipulation, the final state is measured via the electron.
To this end, a $\ket{\downarrow}$ electron is initialized, followed by an \gls*{esr} chirp pulse that flips the electron only if the nucleus is in the first of the two states.
The electron spin state is then read out by spin-dependent tunneling, where an electron tunneling event indicates that the electron was successfully flipped, and hence that the nucleus is in the nuclear state corresponding to the applied \gls*{esr} frequency.
This process of electron initialization, nuclear-state dependent flipping, and electron readout is then also performed for the \gls*{esr} frequency corresponding to the other nuclear state.
Since the electron readout method does not have a 100\% readout fidelity, the two electron reads are repeated $N_\mathrm{shots} \approx 10$ times and the proportion of electron reads with tunneling event is counted for each nuclear state. 

The nuclear state is successfully measured if exactly one of the two nuclear reads has a fraction of tunneling events above a preset threshold (Fig. ~\ref{fig:nuclear flip probability}A).
A subsequent repetition of the pulse sequence will have a probability of flipping the nuclear spin to the other measured state, in which case the nuclear state with a high fraction of tunneling events will be switched.
By repeating the pulse sequence $N_\mathrm{repetitions}$ times and counting the number of nuclear flips $N_\mathrm{flips}$, the resulting flip probability is $P_\mathrm{flip}=\dfrac{N_\mathrm{flips}}{N_\mathrm{repetitions}-1}$ (Fig. ~\ref{fig:nuclear flip probability}B).
Occasionally, both electron spin-up fractions are either above or below the threshold.
This could be caused by the nuclear state randomly flipping to one of the other six nuclear states, or simply due to the electron readout method not having a 100\% fidelity.
As a result, the nuclear state, and consequently $P_\mathrm{flip}$ could not be determined, and thus the data point is discarded.
To account for effects such as slow drifts, multiple repetitions of the varied parameters were measured, and the final $P_\mathrm{flip}$ is taken as the average over the repeated measurements.

\subsection{Nuclear spin initialization}\label{subsec:nuclear spin initialization}

Throughout a measurement, the nuclear spin is operated in a two-dimensional subspace spanned by the two relevant spin states.
However, two causes were found to result in undesired flipping of the nuclear spin out of this subspace.
The first cause is due to the $\mathrm{D}^0$ and $\mathrm{D}^+$ donor having different spin eigenbases, and therefore a non-zero overlap between distinct nuclear spin eigenstates.
The repeated ionization/neutralization that occurs during readout therefore yields a nonzero probability of the nuclear spin to transition to a different spin state.
This effect has previously been observed in $^{31}$P~(Ref.~\citenum{pla2013high}), but is enhanced for \Sb due to the quadrupole interaction.
The second cause is a flip-flop decay process when the donor is in the neutral state, where a $\ket{m_I, \uparrow}$ state can decay into the $\ket{m_I+1, \downarrow}$ state.
While this decay mechanism conserves the total angular momentum, the nuclear spin $z$-projection has increased.
Consequently, the nucleus is likelier to settle at high $m_I$ values, which is in agreement with our experimental observations.

Two methods were employed to initialize the nuclear spin into a target state $\ket{m_I^\mathrm{target}}$.
The first method is based on using calibrated \gls*{ner} $\pi$-pulses according to the following scheme
\begin{enumerate}
    \item Determine the nuclear state by measuring the electron flipping probability for each of the eight \gls*{esr} frequencies. Stop if the nucleus is in the state $\ket{m_I^\mathrm{target}}$.
    \item Apply a sequence of \gls*{ner} $\pi$-pulses on an ionized donor to flip the nucleus to $\ket{m_I^\mathrm{target}}$. 
    The pulse sequence consists of a minimal number of pulses, which can be a combination of $\Delta m_I = \pm 1$ and $\Delta m_I = \pm 2$ pulses.
    \item Repeat from step 1.
    
\end{enumerate}
While this method can be highly successful, it suffers from two significant drawbacks.
The first drawback is that it is only possible to reach $\ket{m_I^\mathrm{target}}$ if all the $\pi$-pulses connecting the current and target state have been calibrated.
The second drawback is that the nuclear resonance frequencies can slowly drift over time, and as a result the $\pi$-pulses need occasional recalibration.

\begin{figure}
    \centering
    \includegraphics[width=\linewidth]{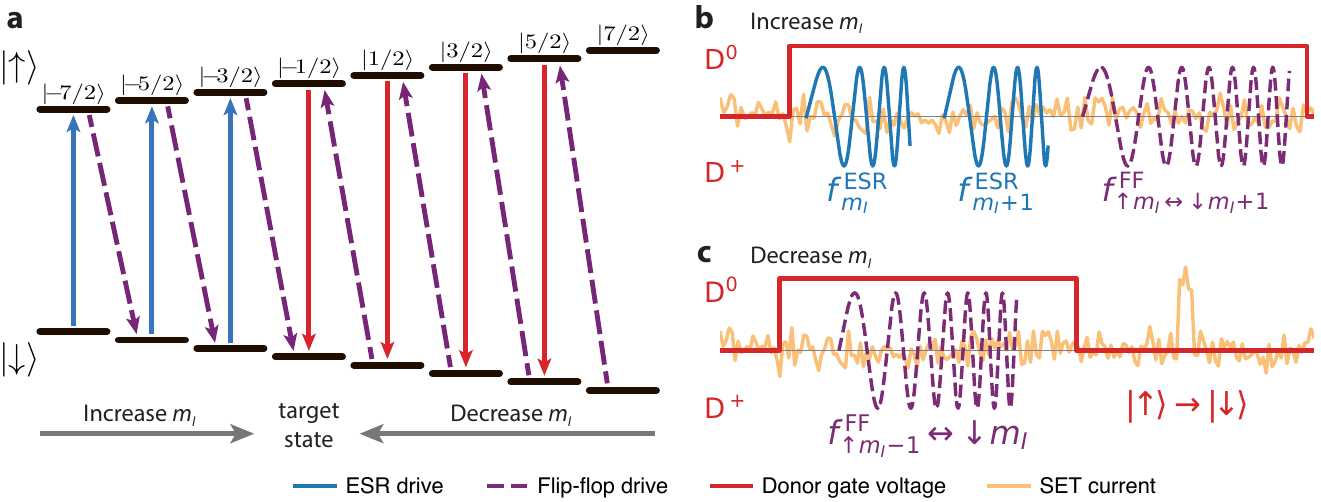}
    \caption{Nuclear spin initialization scheme using flip-flop transitions.
    This scheme pumps the spin into a target state $\ket{m_I^\mathrm{target}}$ without the need for measurements or feedback.
    We illustrate the method with $\ket{-1/2}$ as the target state.
    \textbf{a}, Energy level diagram with nuclear spin initialization transitions.
    Separate pulse sequences are used to increase (left, \textbf{b}) or decrease (right, \textbf{c}) the nuclear spin $m_I$ to $m_I^\mathrm{target} = -1/2$.
    \textbf{b}, Pulse sequence to increase nuclear spin from $m_I$ to $m_I+1$.
    Starting with an initialized $\ket{\downarrow}$ electron, the donor gate voltage (red) is increased to inhibit donor ionization.
    Two ESR pulses are applied to invert the electron to $\ket{\uparrow}$ if the nuclear spin is $m_I$ or $m_I+1$.
    Next a flip-flop chirp pulse with center frequency $f^\mathrm{FF}_{\uparrow m_I \leftrightarrow \downarrow m_I+1}$ is applied.
    If the nucleus has spin $m_I$, this pulse inverts the state to $\ket{\downarrow} \otimes \ket{m_I+1}$, effectively increasing the nuclear spin.
    If the nucleus starts with spin $m_I+1$, the flip-flop pulse is off resonant and the $\ket{\uparrow}$ electron is reinitialized to $\ket{\downarrow}$ in the next iteration of the pulse sequence.
    All other nuclear spin states are unaffected.
    \textbf{c}, Pulse sequence to decrease nuclear spin from $m_I$ to $m_I-1$.
    A flip-flop chirp pulse with center frequency $f^\mathrm{FF}_{\uparrow m_I-1 \leftrightarrow \downarrow m_I}$ is applied to a donor with initialized $\ket{\downarrow}$ electron.
    This pulse adiabatically inverts an initial spin state $\ket{\downarrow} \otimes \ket{m_{I}}$ to $\ket{\uparrow}\otimes \ket{m_I-1}$. 
    After an electron readout phase, the $\ket{\uparrow}$ electron is replaced by a $\ket{\downarrow}$.
    This effectively decreases the nuclear spin projection, while all other nuclear spin states remain unaffected.
    }
    \label{fig:flip-flop initialization}
\end{figure}

Over the course of an experiment, a second nuclear spin initialization scheme was found and utilized that did not suffer from either of these drawbacks.
The scheme relied on spin pumping using the flip-flop transitions, in which both the nuclear and electron spin states flipped (Fig.~\ref{fig:flip-flop initialization}).
Using high-power microwave chirp pulses (henceforth referred to as a flip-flop pulse), a transition between two flip-flop states could be addressed with a fidelity of $\sim 0.5$.
For an initial $\ket{\downarrow}$ electron, application of a flip-flop pulse followed by a donor ionization results in a 50\% chance of decreasing the nuclear spin ($\Delta m_I = -1$), being limited by decoherence.
By repeatedly cycling through all flip-flop transition frequencies $f^{FF}_{\uparrow m_I-1\leftrightarrow \downarrow m_I}$ where the initial nuclear spin state $\ket{m_I}$ is above the target state ($m_I > m_{I,\mathrm{target}}$), the higher nuclear states are effectively pumped down to $\ket{m_I^\mathrm{target}}$ (Fig.~\ref{fig:flip-flop initialization}C).
A similar sequence is used for all nuclear spin states $\ket{m_I}$ below the target state $m_I < m_I^\mathrm{target}$, with the modification of starting with an initial $\ket{\uparrow}$ electron by applying two \gls*{esr} pulses with frequencies $f^\mathrm{ESR}_{m_I}$ and $f^\mathrm{ESR}_{m_I+1}$.
In this case, the flip-flop transition will increase the nuclear spin ($\Delta m_I = +1$), and cycling through the flip-flop transition frequencies effectively pumps all lower nuclear states up to $\ket{m_I^\mathrm{target}}$ (Fig.~\ref{fig:flip-flop initialization}B).

One of the main advantages of this initialization scheme is that it does not require knowledge of the initial nuclear state, and instead cycles through the different flip-flop frequencies to pump the nuclear spin into a target state.
This is an entirely open-loop control sequence to initialize the nuclear spin.
The use of chirp pulses removes the requirement of an accurately-tuned flip-flop frequency, thereby negating the need to periodically recalibrate the flip-flop pulses.
An additional benefit is that the flip-flop transition frequencies can be accurately estimated from the \gls*{esr} frequencies alone, and thus do not require knowledge of nuclear transition frequencies.

\section{Ancillary measurements}

\subsection{Slope in \texorpdfstring{$\Delta m_I = \pm 2$}{Delta m = 2} Rabi frequencies}
\label{subsec:DQT f_Rabi slope}
\begin{figure}[ht]
    \centering
    \includegraphics{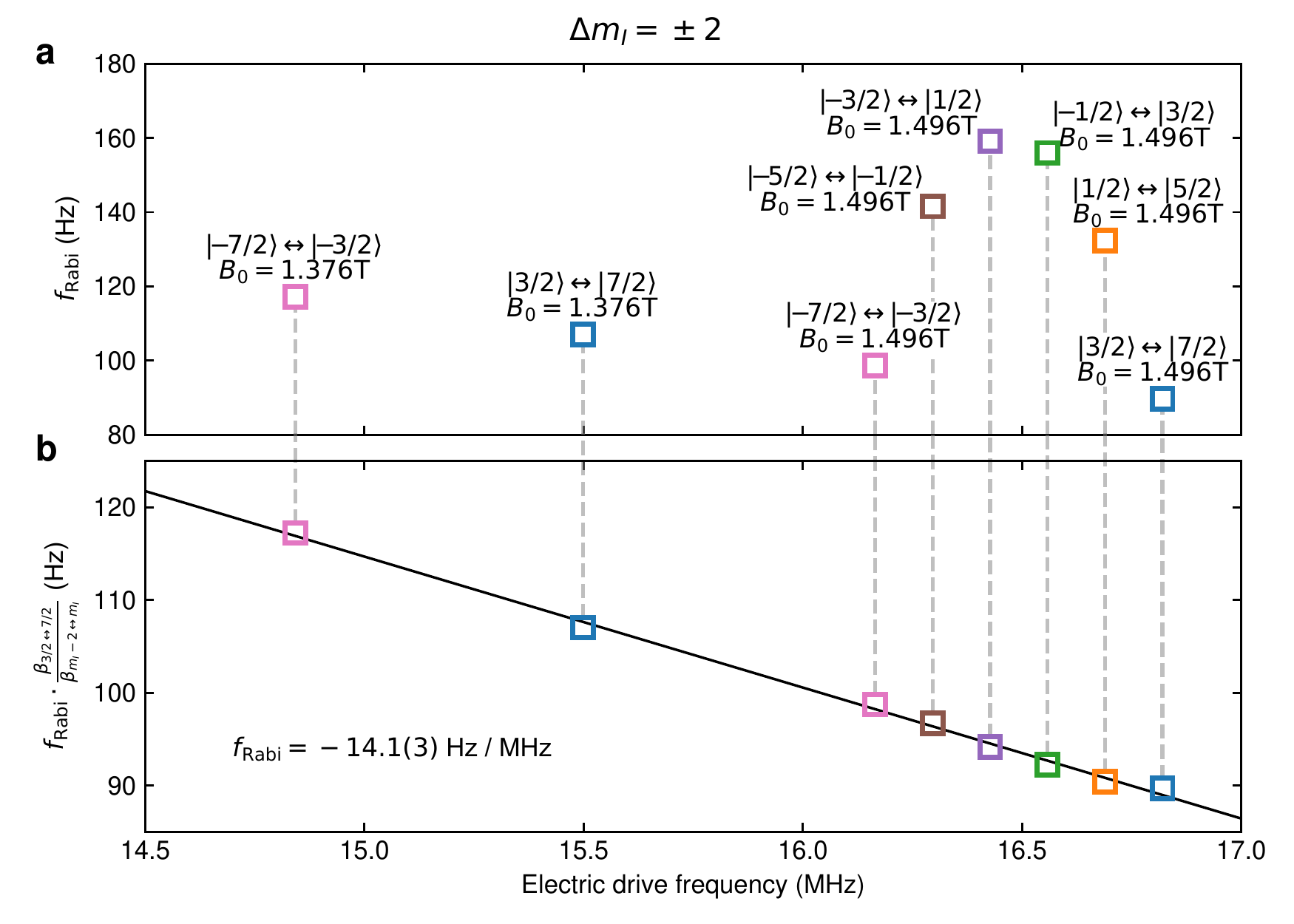}
    \caption{
    Rabi frequencies for varying resonance frequencies.
    \textbf{a}, Measured Rabi frequencies for $\Delta m_I = \pm 2$ transitions.
    The electric drive amplitude applied to the gate line input (i.e. before cable transmission losses) is kept constant for all measurements.
    While the Rabi frequencies are supposed to be symmetric about the center, the measured values at $B_0 = \SI{1.496}{\tesla}$ exhibit a small decrease with increasing electric drive frequency.
    The outer two transitions have been remeasured at a lower magnetic field \SI{1.376}{\tesla} and hence at a lower drive frequency. 
    Both Rabi frequencies are higher than their counterparts at higher magnetic field. 
    \textbf{b}, Rescaled Rabi frequencies accounting for the Rabi frequency coefficient $\beta_{m_I \leftrightarrow m_I-2}$ (Table.~\ref{tab: rabi_constants}).
    The Rabi frequency of each transition $m_{I-2} \leftrightarrow m_I$ is scaled down by $\beta_{m_I-2 \leftrightarrow m_I} / \beta_{3/2\leftrightarrow 7/2}$.
    All scaled Rabi frequencies accurately follow a straight line with slope $-14.1(3) \mathrm{Hz/MHz}$, indicating a dependence of the electric drive amplitude on the electric drive frequency, which affects the Rabi frequency.
    This effect is not observed for $\Delta m_I = \pm 1$ transitions. 
    Since the electric drive amplitude applied to the gate line is kept constant, this effect is likely an experimental artifact caused by a non-uniform transmission profile of the gate line in the frequency regime of the $\Delta m_I = \pm 2$ transitions.
    }
    \label{fig:DQT Rabi slope}
\end{figure}
Both the $\Delta m_I = \pm 1$ and $\Delta m_I = \pm 2$ \gls*{ner} Rabi frequencies should be symmetric about the center transition (Table.~\ref{tab: rabi_constants}). 
While this symmetry has been largely observed (Fig.~2, C and D), the $\Delta m_I = \pm 2$ Rabi frequencies exhibit a small negative slope with increasing frequency.
This slight asymmetry either has a physical cause not captured by the model Hamiltonian of the ionized nucleus, or is caused by an experimental artifact external to the donor, such as a frequency-dependent attenuation of the gate line in this frequency range.

To discriminate between these two possible causes, we reduced the static magnetic field $B_0$ from $\SI{1.496}{\tesla}$ to $\SI{1.376}{\tesla}$ (Decreasing transition frequencies by $20f_Q$), and remeasured the Rabi frequencies of the outer transitions $\ket{3/2}\leftrightarrow \ket{7/2}$ and $\ket{-7/2}\leftrightarrow \ket{-3/2}$ (Fig.~\ref{fig:DQT Rabi slope}A) keeping the driving amplitude constant.
The two measured Rabi frequencies are both higher than their counterparts at higher magnetic field, thereby excluding that the effect is associated with specific nuclear transitions.
Furthermore, by scaling the Rabi frequencies by their coefficients $\beta_{m_I-2 \leftrightarrow m_I}$ (Table.~\ref{tab: rabi_constants}), all Rabi frequencies follow a straight line with a gradient of $\SI{14.1(3)}{\hertz \per \mega \hertz}$ (Fig.~\ref{subsec:DQT f_Rabi slope}B).
This is strongly suggestive of a dependence of the observed Rabi frequencies on the electric drive frequency, but not due to an underlying microscopic cause. 
The likely cause is a  frequency dependence of the gate line attenuation, as room-temperature measurements of identical transmission lines have shown variations up to $\SI{1}{\deci \bel \per \mega \hertz}$ in the few-MHz regime. 
Although the electric drive amplitude is kept constant at the gate line input, this frequency-dependent attenuation causes the drive amplitude to vary at the donor site.
This interpretation is further confirmed by the lack of a slope in the Rabi frequencies when driving with the antenna (Sec.~\ref{subsec: Antenna Rabi oscillations}), as the antenna transmission line (designed to operate up to \SI{40}{\giga \hertz}) does not have any significant frequency dependence in the few-MHz regime.

\subsection{Rabi oscillations driven via the microwave antenna}
\label{subsec: Antenna Rabi oscillations}

\begin{figure}
    \centering
    \includegraphics{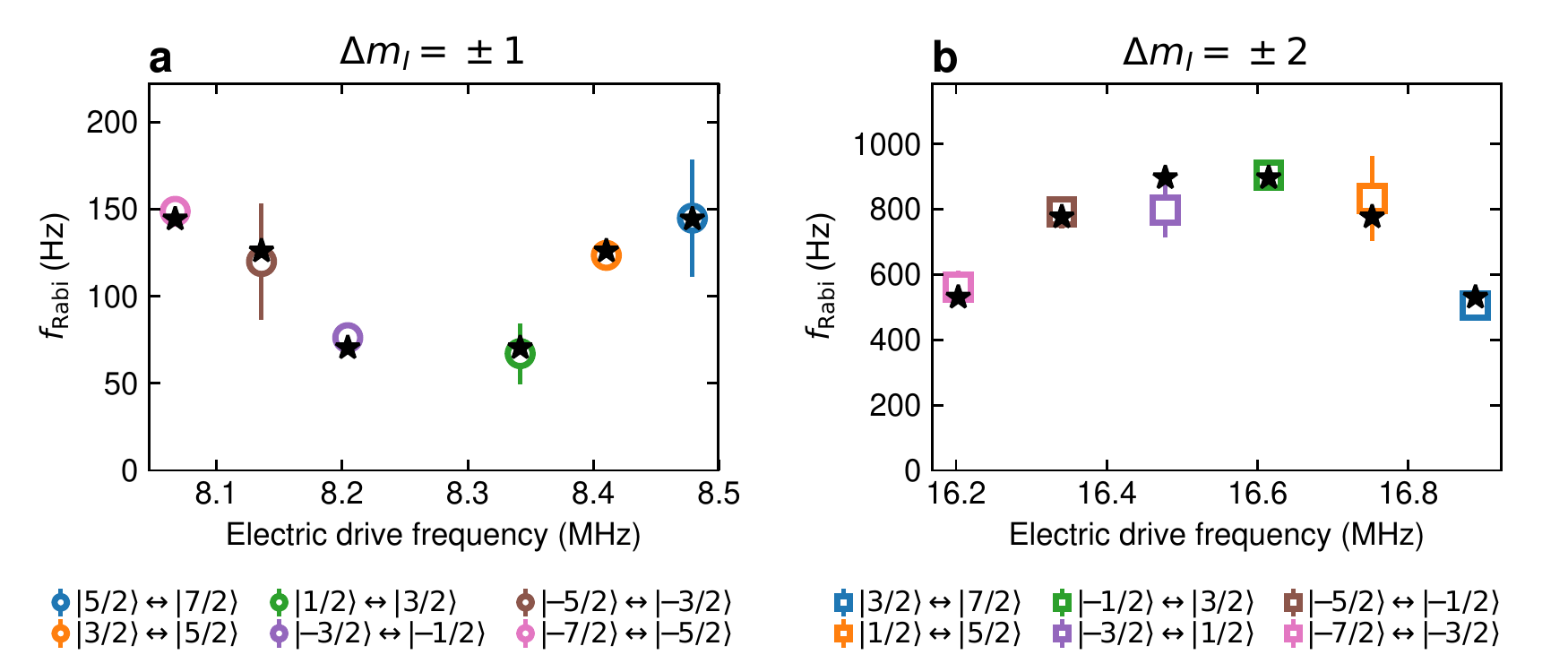}
    \caption{
    \acrshort*{ner} Rabi frequencies while driven via the damaged antenna.
    The Rabi frequencies are the weighted average of those extracted by fitting Rabi's formula to three different \acrshort*{ner} spectrum data sets.
    For both the $\Delta m_I = \pm 1$ and $\Delta m_I = \pm 2$ transitions, the relationship between the Rabi frequencies match well with those expected for \acrshort*{ner} (stars).
    Error bars show the 95\% confidence interval.
    }
    \label{fig:antenna Rabi}
\end{figure}

The \gls*{ner} spectrum has also been measured while driving with the damaged microwave antenna, from which the Rabi frequencies have been extracted for all $\Delta m_I = \pm 1$ (Fig.~\ref{fig:antenna Rabi}A) and $\Delta m_I = \pm 2$ (Fig.~\ref{fig:antenna Rabi}B) transitions.
In both cases, the relationship between the Rabi frequencies match well with those expected for \gls*{ner} (Table~\ref{tab: rabi_constants}).
Additionally, the transition $\ket{-1/2} \leftrightarrow \ket{1/2}$ could not be driven, further confirming that the driving mechanism from the antenna in the $\sim$MHz frequency range is electric instead of magnetic.

The Rabi frequencies of the $\Delta m_I = \pm 2$ transitions are significantly higher than those of the $\Delta m_I = \pm 1$ transitions.
This is opposite to the results obtained while driving with an electric gate, where the $\Delta m_I = \pm 1$ Rabi frequencies are much higher.
This could be caused by the differing AC electric field orientation at the donor site when driving with an antenna versus with a donor gate, which will affect the ratio of $\Delta m_I = \pm 1$ to $\Delta m_I = \pm 2$ Rabi frequencies.
Additionally, the $\Delta m_I = \pm 2$ antenna-driven Rabi frequencies do not exhibit a linear dependence on drive frequency, which is in contrast with the gate-driven Rabi frequencies (Sec.~\ref{subsec:DQT f_Rabi slope}).
This agrees with the estimate of a uniform transmission of electric signals (see Sec.~\ref{subsec: fab_antenna} in this frequency range, and further strengthens the above interpretation of non-uniform gate line transmission.

\subsection{Close-up of nuclear spectrum} \label{subsec:spectrum closeup}
\begin{figure}[ht]
    \centering
    \includegraphics{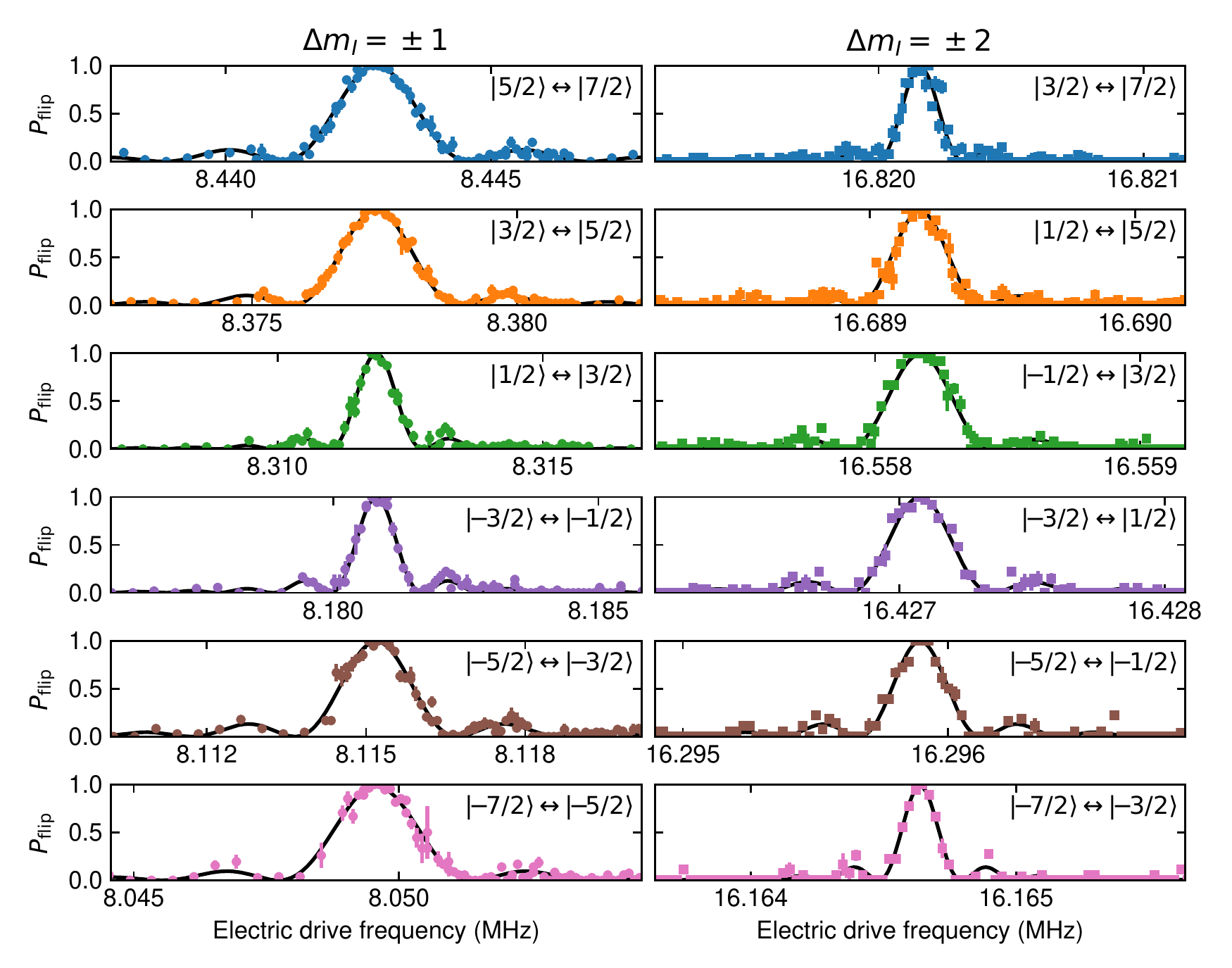}
    \caption{Closeup of \acrshort*{ner} resonance peaks.
    Pulse durations were calibrated to fully invert the populations of a transition when driven on resonance, and drive amplitudes are $V_\mathrm{RF}^\mathrm{gate} = \SI{30}{\milli\volt}$ for $\Delta m_I = \pm 1$ transitions (left column), $V_\mathrm{RF}^\mathrm{gate} = \SI{40}{\milli\volt}$ for $\Delta m_I = \pm 2$ (right column).
    Each of the resonance peaks were fitted using Rabi's formula~\eqref{eq:Rabi formula} with fixed pulse durations (black lines).
    The spectral lines are power broadened by an amount inversely proportional to their Rabi frequencies and hence to their Rabi frequency coefficients (Table~\ref{tab: rabi_constants}).
    Error bars show 95\% confidence interval.
    }
    \label{fig:spectrum closeup}
\end{figure}

The nuclear spectrum in Figs.~2A~and~2B is presented over a wide frequency range, and does not highlight the shape of the spectral lines near the resonance frequencies.
A close-up of each of the measured \gls*{ner} spectral lines is therefore shown in Fig.~\ref{fig:spectrum closeup}.
Each resonance is power-broadened, and its shape is fitted to Rabi's formula, given by
\begin{equation}
    P_\mathrm{flip} = \frac{\gamma^2}{\Omega^2} \sin^2{\Omega t},
    \label{eq:Rabi formula}
\end{equation}
where $\gamma = \pi f_\mathrm{Rabi}$ is the drive strength, $\Omega = \sqrt{\gamma^2 + 2\pi(f - f_{m_I -\Delta m_I \leftrightarrow m_I})/4}$, and $t$ is the \gls*{ner} pulse duration.

\clearpage
\section{Microscopic origin of the quadrupole interaction and nuclear electric resonance} \label{sec: microscopic_theory}

The quadrupole interaction and \gls*{ner} in this device are related to the presence of a strain- and electric field-dependent \gls*{efg} at the site of the \Sb donor in silicon.
In this section, we use finite-element models to estimate the strain and electric fields that are present in our device and to triangulate the position of the donor under study (Sec.~\ref{sec: comsol_model}).
We then develop a microscopic model that relates the strain and electric fields to the \gls*{efg} at the site of the donor, which directly relates to the quadrupole interaction strength. (Sec.~\ref{subsec: Microscopic_EFG}).

By combining the strain profile from our finite element model with \gls*{dft} calculations for the strain-\gls*{efg} coupling, we estimate the portion of the quadrupole interaction that is due to strains arising from the thermal contraction of metallic gates in the device (Sec.~\ref{subsec: strain_quadrupole}). 
Then, combining the electric field profile from our finite element model with the measured \gls*{ner} shifts and Rabi frequencies, we estimate the portion of the quadrupole interaction that is due to external applied electric fields.
We compare the inferred electric field \gls*{efg} coupling to historical measurements of \gls*{lqse} in bulk materials (Sec.~\ref{subsec: electric_quadrupole}).
Concluding this section, we use our finite element model to quantitatively rule out other conceivable physical mechanisms that could lead to  \gls*{ner} (Sec.~\ref{subsec: unlikely_causes}).

\subsection{Finite elements simulations of the device} \label{sec: finite_element}

\subsubsection{The model set up} \label{sec: comsol_model}

\begin{figure}[ht]
\centering
\includegraphics[width=0.7143\textwidth]{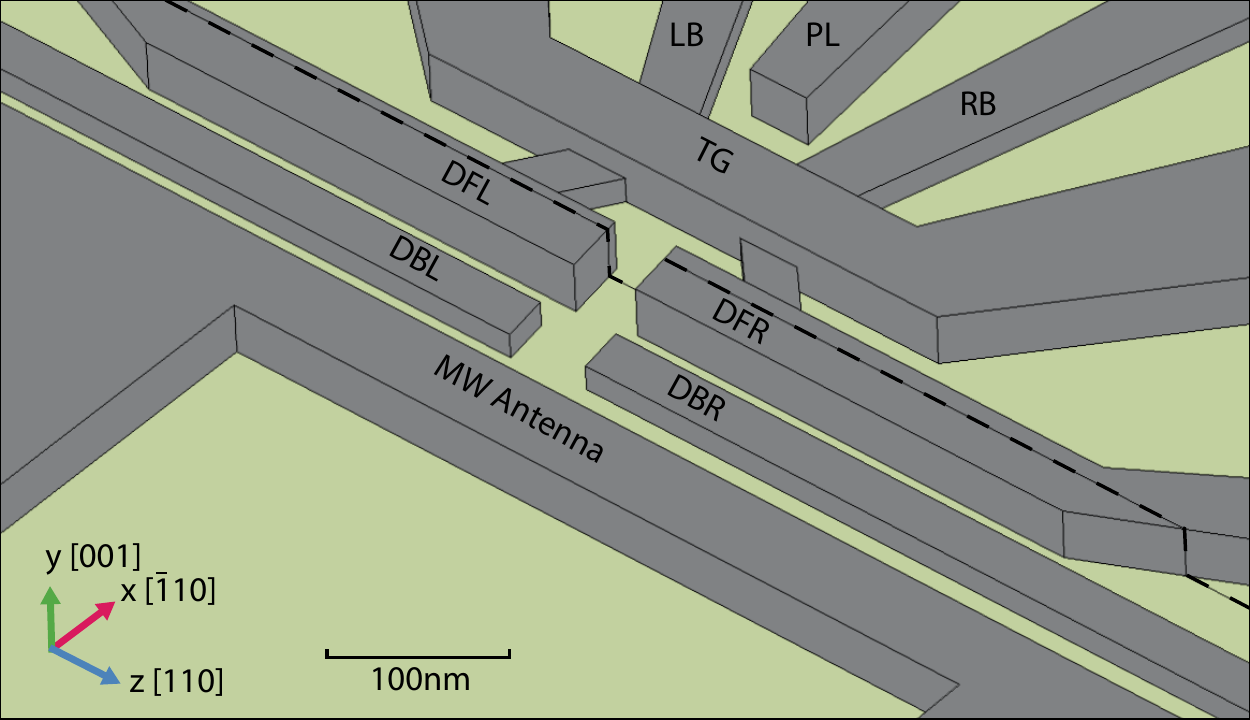}
\caption{Model geometry as defined in COMSOL.
The gate layout is adapted to match the imaged device in Fig.~\ref{fig:experimental_setup}.
All gates are labeled for identification.
The coordinate system on the bottom left indicates the axis convention used in this work as well as the silicon crystal orientation. 
The [001] crystal direction is along the $y$ axis normal to the device surface and the static magnetic field $B_0$ is applied along the [110], $z$ axis.
The origin of this model is located at the Si/SiO$_2$ interface (not visible here) in the center of the implantation region between DFL and DFR.
The dashed line marks the $y-z$ cut-plane at $x=\SI{13}{\nano\meter}$ that cuts through the most probable donor position (Sec.~\ref{sec: donor_triangulation}).
This plane is chosen to show simulations results throughout this section.}
\label{fig: COMSOLmodel}
\end{figure}

The finite element solver COMSOL was used for simulating electrostatics and strain due to the different thermal contraction of Si and Al upon cooling down the device.
To achieve this, the electrostatics package (part of the AC/DC package), and the solid mechanics package (part of the structural mechanics package) were used, respectively. 
The model consists of a $\SI{2}{\micro\meter} \times\SI{2}{\micro\meter} \times \SI{2}{\micro\meter}$ silicon substrate that is mechanically fixed and electrically grounded at the bottom.
An $\SI{8}{\nano\meter}$ thick silicon dioxide layer is defined on top of the silicon, and the aluminum gates are defined on top of this layer. 
Following the fabrication procedure described in Section~\ref{sec: fab}, the aluminum is covered by $\SI{2}{\nano\meter}$ of aluminum oxide, and consists of 2 separate layers with a thickness of \SI{20}{\nano \meter} for the bottom layer and \SI{40}{\nano \meter} for the top layer.
The gate layout is matched to that of the actual device upon its imaging and is shown in Fig.~\ref{fig: COMSOLmodel}.
In the electrostatic simulations, the gate-induced two-dimensional electron gas forming the SET leads and island is modelled by a 1 nm thick metallic layer at the SiO$_2$/Si interface, with the lateral dimensions of the top gate TG, and gaps of the width of the barrier gates LB and RB. 
The two SET leads and island are kept at a potential of 0 V. 
A tetrahedral mesh was used, adjusting the properties to ensure elements do not have dimensions larger than $\SI{2}{\nano\meter}$ in the implantation region of the donor.

The structural deformation during the cooldown is modeled in two simulation steps. 
It is assumed that the device is strain free at $\SI{850}{\celsius}$, i.e. the growth temperature of the silicon oxide. 
At this  stage, no aluminum gates are present.
In the first step, the Si/SiO$_2$ chip is cooled down from 850 to $\SI{400}{\celsius}$. 
Then, the Al gates are added to the structure and, at the temperature of $\SI{400}{\celsius}$, the Al/Al$_x$O$_y$ gates are annealed in forming gas. 
We assume that the $\SI{400}{\celsius}$ forming gas anneal strain-relaxes the aluminum structures at this temperature. 
In the second step, including the initial strain of the silicon/silicon oxide, the deformation during the cooldown from $\SI{400}{\celsius}$ to $\SI{200}{\milli\kelvin}$ is modelled.
At $\SI{200}{\milli\kelvin}$ the thermal deformation has saturated and further cooldown has a negliable effect.

Silicon is an anisotropic material. 
When setting up the model, special care needs to be taken in correctly aligning the orientation of the aluminum gate layout with the silicon crystal axes.
Our sample is fabricated on top of a (100) wafer, where [001] is the crystal direction normal to the wafer surface.
The wafer was diced along the in-plane crystal directions [110] and [$\bar1$10] (the natural cleaving directions) and the device subsequently aligned to these directions. 
Consequently, the [110] and [$\bar1$10] directions correspond to the $z$ and $x$ axis in the COMSOL model, respectively, and this is the coordinate convention used throughout the manuscript.
The stiffness matrix for silicon is commonly defined with respect to the [100], [010] and [001] crystal axes.
We define a new coordinate system with respect to the crystal axis $x''$ along [010], $y''$ along [001], and $z''$ along [100].
To match our chosen lab frame ($x$, $y$, $z$) with the standard crystal orientation ($x''$, $y''$, $z''$), either the model has to be rotated by $\pi$/4 about $y$, or the stiffness tensor has to be rotated.
We chose the first option, as it is more convenient to rotate electric field components and strain matrices than the stiffness tensor.
In what follows, all spatial maps of quantities will be shown in the lab frame.

The aluminum oxide is assumed to be isotropic. All thermal expansion coefficients used in the strain simulations can be found in Table~\ref{tab: sim_parameter}. 
All other material properties are taken from the COMSOL material library.

\begin{table}
    \centering
    \setlength{\extrarowheight}{0.1cm}
    \caption{Simulation parameters used in the COMSOL device model.
    Parameters taken directly from the COMSOL material library are not listed here. 
    The top section of the table lists the different thermal expansion coefficients.
    Thermal deformation in the two cooldown steps is captured by the average coefficient of thermal expansion in this temperature range for all materials present. 
    The bottom section lists all applied gate voltages, see Fig.~\ref{fig: COMSOLmodel} for labeling of the gates.}
    \begin{tabular}{l l l}
        \hline\hline
        Parameter[Source]&	Symbol&	Value\\
        \hline
        Thermal expansion $\SI{850}{\celsius}$ to $\SI{400}{\celsius}$ & &\\
        Silicon\citep{white1997thermophysical}& $\bar\alpha$ & $\SI{4.198d-6}{\per\kelvin}$\\
        Silicon oxide\citep{hahn1972thermal}& $\bar\alpha$ &	$\SI{0.4d-6}{\per\kelvin}$\\
        $\alpha$-Quartz\citep{rosenholtz1941linear}& $\bar\alpha_{\bot}$ & $\SI{23.62d-6}{\per\kelvin}$\\
         & $\bar\alpha_{\parallel}$ & $\SI{13.33d-6}{\per\kelvin}$\\
        Thermal expansion $\SI{400}{\celsius}$ to $\SI{200}{\milli\kelvin}$ & &\\
        Silicon\citep{swenson1983recommended}& $\bar\alpha$ & $\SI{2.28d-6}{\per\kelvin}$\\
        Silicon oxide\citep{hahn1972thermal}& $\bar\alpha$ &	$\SI{0.446d-6}{\per\kelvin}$\\
        $\alpha$-Quartz\citep{rosenholtz1941linear,barron1982thermal}& $\bar\alpha_{\bot}$ &$\SI{14.186d-6}{\per\kelvin}$\\
         & $\bar\alpha_{\parallel}$ & $\SI{7.824d-6}{\per\kelvin}$\\
        Aluminum\citep{nix1941thermal,corruccini1961thermal}& $\bar\alpha$ & $\SI{21.43d-6}{\per\kelvin}$\\
        Aluminum oxide\citep{white1997thermophysical}& $\bar\alpha$ & $\SI{4.98d-6}{\per\kelvin}$\\
        \hline
        Applied gate voltages & & \\
        Top gate & $V_{\rm TG}$ & $\SI{1.881}{\volt}$\\
        Right barrier & $V_{\rm RB}$ & $\SI{0.3421}{\volt}$\\
        Left barrier & $V_{\rm LB}$ & $\SI{0.4293}{\volt}$\\
        Donor gate front left& $V_{\rm DFL}$ & $\SI{0.4823}{\volt}$\\
        Donor gate front right& $V_{\rm DFR}$ & $\SI{0.3915}{\volt}$\\
        Donor gate back left& $V_{\rm DBL}$ & $\SI{0.4177}{\volt}$\\
        Donor gate back right& $V_{\rm DBR}$ & $\SI{0.4059}{\volt}$\\
        Plunger gate& $V_{\rm PL}$ & $\SI{-0.0032}{\volt}$\\
        Source & $V_{\rm SRC}$ & $\SI{0.0002}{\volt}$
    \end{tabular}
    \label{tab: sim_parameter}
\end{table}

\subsubsection{Donor triangulation} \label{sec: donor_triangulation}

\begin{figure}[ht]
    \includegraphics{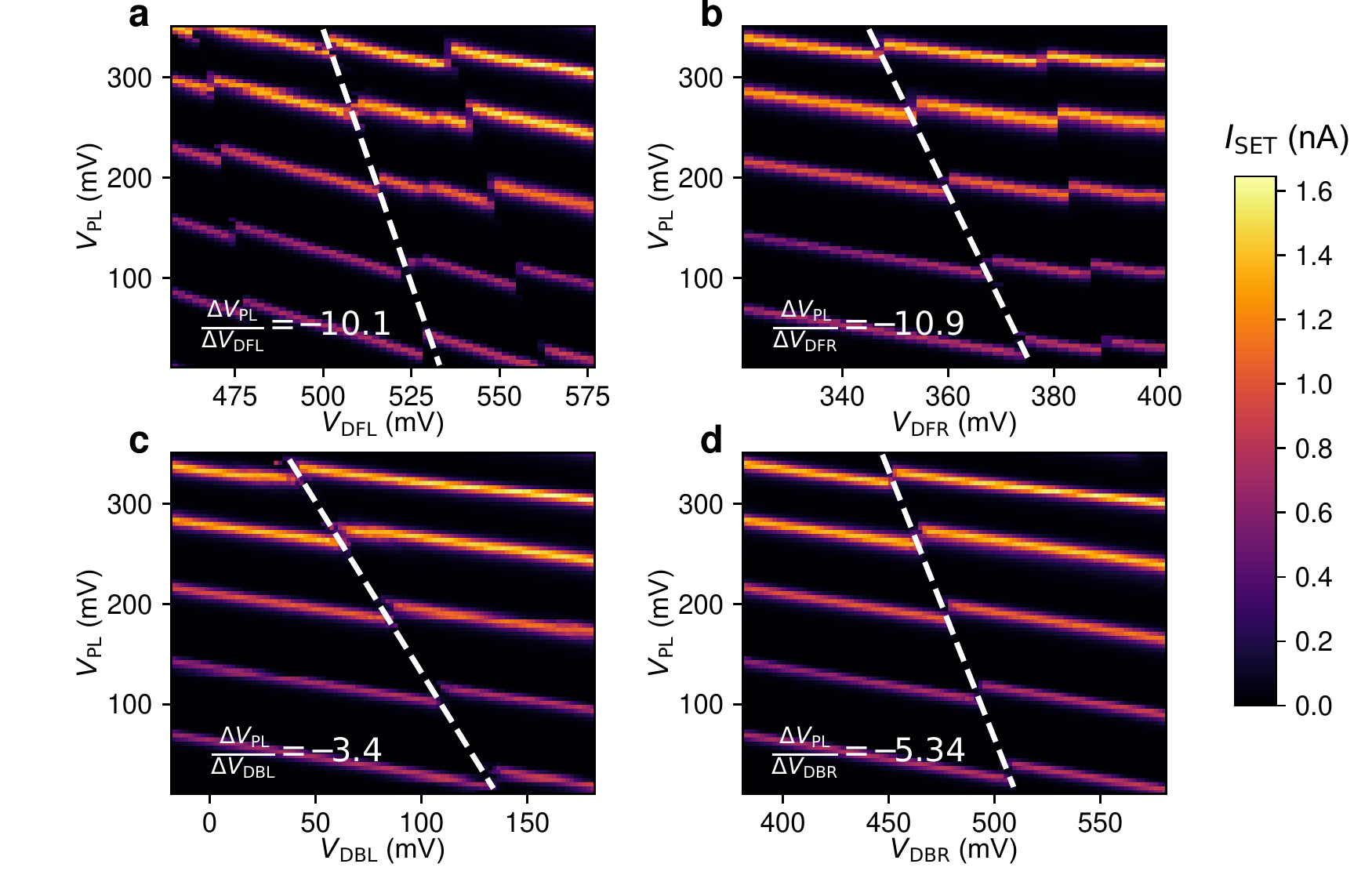}
    \caption{\acrshort*{set} current as a function of different gate voltages. 
    The \acrshort*{set} source is voltage biased at \SI{200}{\micro\volt}. 
    In all panels, the plunger gate voltage is varied on the vertical axis, while the gate voltage of DFL (\textbf{a}), DFR (\textbf{b}), DBL (\textbf{c}), or DBR (\textbf{d}) is varied on the horizontal axis. 
    Bright lines of high current correspond to a Coulomb peak, and the dark regions in between to Coulomb blockade.
    The discrete jump in the Coulomb peaks is caused by the ionization of the donor, with the donor ionized on the low gate voltage side of the jump and neutral on the high gate voltage side of the jump. 
    The slope of the transition, indicated by the white dashed lines, is an iso-potential line of the donor. 
    This slope corresponds to the ratio of the change in electrostatic potential at the donor due to the respective gate voltages.}
    \label{fig: dc_slopes}
\end{figure}

The donor's electrochemical potential, which sets the ionized-to-neutral charge transition, is influenced differently by different gates, according to their capacitive coupling to the donor charge. 
The allows us to triangulate the location of the donor by comparing the experimental ratios of different gate capacitances with electrostatic COMSOL simulations. 

Fig.~\ref{fig: dc_slopes} shows the SET current as a function of the voltages on PL and one of the donor gates, with a different donor gate in each panel. 
A discrete shift in the SET Coulomb peaks indicates a charge transfer, e.g. the ionization of the measured donor marked by the white dashed lines. 
For each gate voltage configuration along these lines, the electrostatic potential at the donor position $V(\Vec{r}_0; V_{\rm PL}, V_{\rm DFL}, ...)$ is constant. 
Changing the gate voltages in Fig.~\ref{fig: dc_slopes}A by $\Delta V_{\rm PL}$ and $\Delta V_{\rm DFL}$ along the dashed line, we may to first order write
\begin{align}
    &\frac{dV(\Vec{r}_0; V_{\rm PL}, V_{\rm DFL}, ...)}{dV_{\rm PL}}\cdot \Delta V_{\rm PL} + \frac{dV(\Vec{r}_0; V_{\rm PL}, V_{\rm DFL}, ...)}{dV_{\rm DFL}}\cdot \Delta V_{\rm DFL} = 0\\
    & -\frac{\Delta V_{\rm PL}}{\Delta V_{\rm DFL}}=\frac{dV(\Vec{r}_0; V_{\rm PL}, V_{\rm DFL}, ...)}{dV_{\rm DFL}} \bigg/ \frac{dV(\Vec{r}_0; V_{\rm PL}, V_{\rm DFL}, ...)}{dV_{\rm PL}}.
    \label{eq: CTSslopes}
\end{align}
We identify the left hand side as the experimentally measured slope
\begin{align}
    s_{\rm DBL} = -\Delta V_{\rm PL} / \Delta V_{\rm DBL} = \tan (\theta),
\end{align}
which we can equivalently define for the three other panels.
To estimate the precision of extracting these slopes, we have introduced the angle $\theta$. 
With an uncertainty $\sigma_\theta=\SI{1}{\degree}$, Gaussian error propagation leads to an estimated standard deviation $\sigma_{\rm DFL}$ given by
\begin{align}
    \sigma_{\rm DFL} = (s_{\rm DFL}^2+1)\cdot 2\pi\cdot \SI{1}{\degree}/\SI{360}{\degree}.
\end{align}

We now triangulate the donor by finding the position $\Vec{r}$ at which an electrostatic simulation of the right hand side in Eq.~\eqref{eq: CTSslopes} best matches the experimentally measured slopes. We use COMSOL to compute the electrostatic potential landscape across the model for the set of gate voltages given in Table~\ref{tab: sim_parameter}.
Subsequently, the relevant gate voltages are varied by \SI{10}{\milli\volt}, and the electrostatic potential landscape across the model is computed again for variation of each gate voltage.
This results in a spatially-varying simulated slope, which, at a given position $\Vec{r}$, is defined as
\begin{align}
    s^{\rm sim}_{\rm DFL}(\Vec{r}) = \frac{V(\Vec{r};V_{\rm PL},V_{\rm DFL}+\SI{10}{\milli \volt},...)-V(\Vec{r};V_{\rm PL},V_{\rm DFL},...)}{V(\Vec{r};V_{\rm PL}+\SI{10}{\milli \volt},V_{\rm DFL},...)-V(\Vec{r};V_{\rm PL},V_{\rm DFL},...)}. 
\end{align}
Here again the DFL gate is used as an example to illustrate the procedure, and the plunger gate is used as the common gate.

\begin{figure}[ht]
  \centering
  \subfloat{\includegraphics[width=0.49\textwidth]{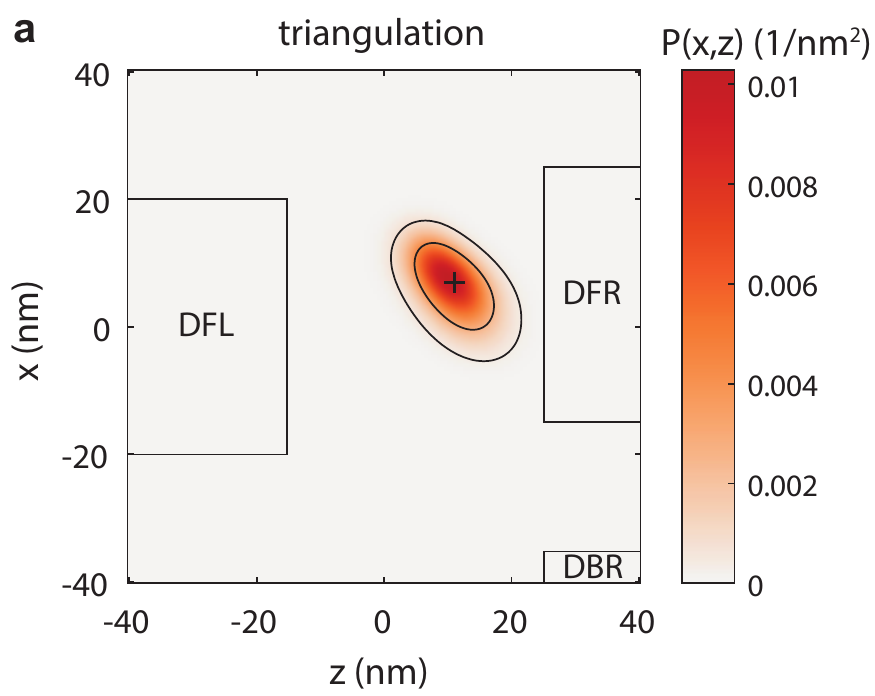}}
  \subfloat{\includegraphics[width=0.49\textwidth]{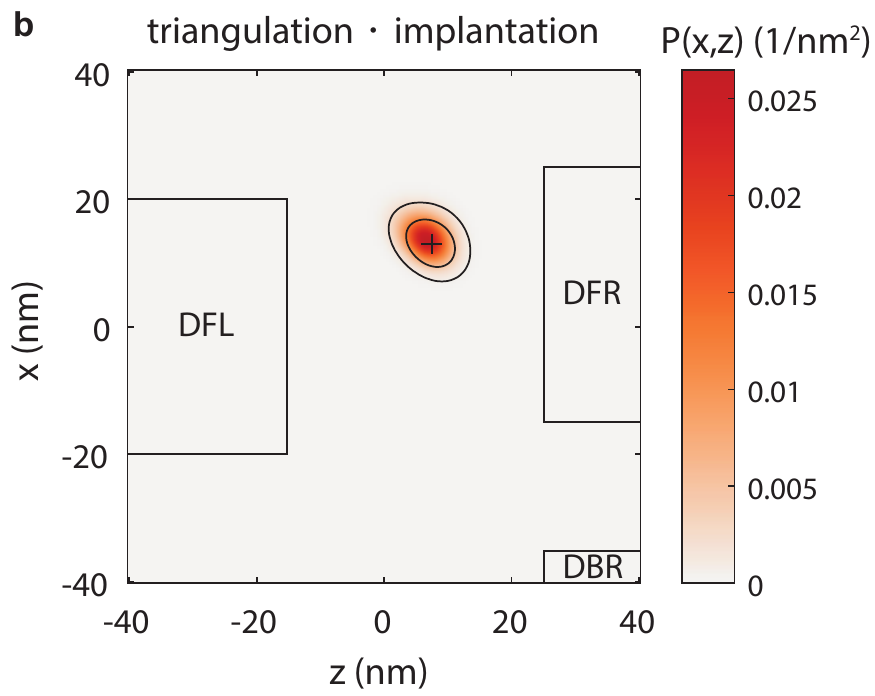}}\\
  \subfloat{\includegraphics[width=0.49\textwidth]{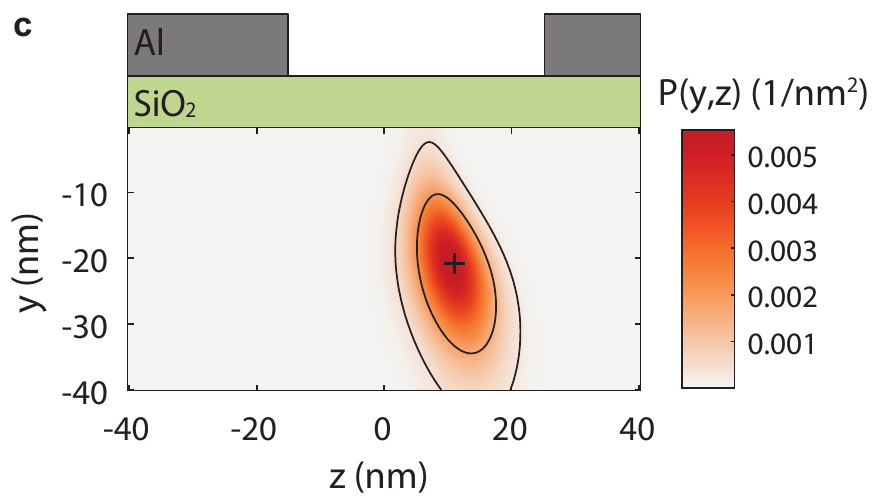}}
  \subfloat{\includegraphics[width=0.49\textwidth]{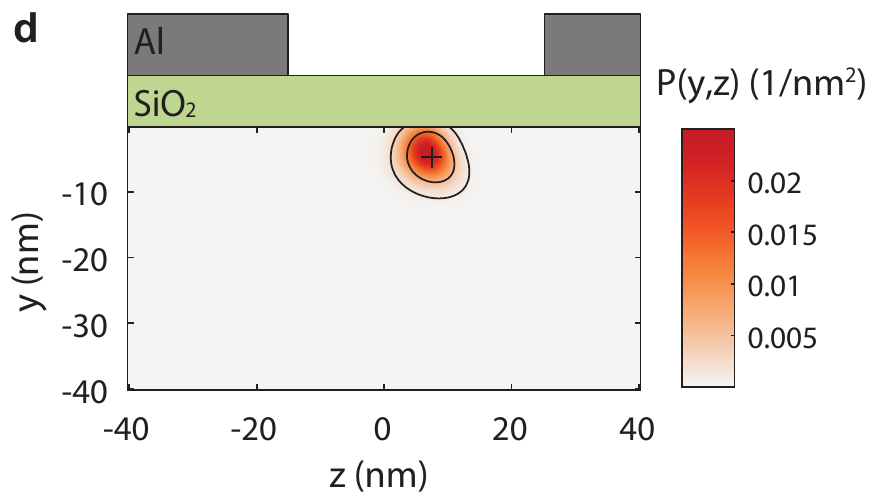}}
  \caption{Position triangulation of the \Sb donor. 
  The colormap shows the probability of finding the donor in a certain location. 
  \textbf{a}, \textbf{c}, Probability density function found using a least-squares estimate comparing simulated gate-to-donor coupling strengths with the experimentally observed strengths (see text for details). 
  A lateral topview (\textbf{a}) and a transverse cross-section (\textbf{c}) are shown with a cross indicating the most likely position. 
  \textbf{b}, \textbf{d}, Probability density including the implantation profile. 
  To improve on the low resolving power of the triangulation method in the $y$ direction, the triangulation probability density function is multiplied with the donor implantation probability density function. 
  This has little effect laterally, but significantly confines the likely depth range of the donor. 
  within the range expected based on the donor implantation parameters (Fig.~\ref{fig:implantation distribution}). 
  A lateral topview \textbf{b} shows $(x,z) = (\SI{13}{\nano \meter},\SI{8}{\nano \meter})$ as the most likely lateral location. 
  A transverse cross-section \textbf{d} indicates a depth of $y = \SI{-5}{\nano \meter}$ as the most likely depth of the donor.
  Probability density functions are normalized over the model volume and are integrated over the out-of-plane axis in both panels, specifically $P(x,z)=\int P(\Vec{r})dy$ and $P(y,z)=\int P(\Vec{r})dx$. 
  The contour lines mark the 68\% and 95\% confidence regions.
  }
  \label{fig: least_square}
\end{figure}

To compare the simulated slopes to the measured ones at each point within the model space, we use a least-squares estimate\cite{sivia2006data} that is defined by
\begin{align}
    P_{\mathrm{triangulation}}(\Vec{r}) = \mathcal{N}\cdot\exp\left[-\frac{1}{2}\sum_{g = \rm gates}\left(\frac{s^{\rm sim}_{g}(\Vec{r})-s_{g}}{\sigma_{g}}\right)^2\right],
\end{align}
with $P_{\mathrm{triangulation}}(\Vec{r})$ the probability density, which we normalize over the given volume via $\mathcal{N}$, and the summation runs over the DFL, DFR, DBL and DBR gates. 

We have verified that the thickness and exact geometry of the two-dimensional electron gas has negligble influence on the triangulation probability density function. 
Furthermore, we have implemented a Thomas-Fermi approximation to the density of the two-dimensional electron gas in one specific geometry and found that this only negligibly alters the triangulation probability density function, and we therefore choose the less computationally expensive, metallic approximation.

The resulting triangulation probability density function is shown in Fig.~\ref{fig: least_square}(A, C). 
The planar layout of the gate structures leads to a low sensitivity of the capacitance triangulation method in the out-of-plane (vertical) $y$ direction, as reflected by a large uncertainty along this axis, visible in Fig.~\ref{fig: least_square}(C). 
However, a significant section of this region constitutes highly unlikely positions of the donor, as revealed by the donor implantation depth profile in Fig.~\ref{fig:implantation distribution}.
A more accurate probability distribution of the donor's position is therefore given by
\begin{align}
     P(\Vec{r}) = \mathcal{N} P_{\mathrm{triangulation}}(\Vec{r})P_{\mathrm{implantation}}(\Vec{r})
\end{align}
with $P_{\mathrm{implantation}}(\Vec{r})$ the dopant probability distribution and $\mathcal{N}$ a normalization factor. 
$P_{\mathrm{implantation}}(\Vec{r})$ is the distribution shown in Fig.~\ref{fig:implantation distribution}, a modified Gaussian distribution homogenous in the lateral dimensions $x$ and $z$ and with mean at -2.5 nm, standard deviation of 2.6 nm and slight skewness (0.54) and kurtosis (3.59).  
This distribution is cut off at the Si/SiO$_2$ interface, and we have verified that donor diffusion due to the rapid thermal anneal of \SI{5}{\second} at \SI{1000}{\celsius} has a negligible effect. 

Further methods to narrow  $P(\Vec{r})$, e.g. by modeling the donor-SET tunnel coupling\cite{mohiyaddin2013}, were not adopted here.

Fig.~\ref{fig: least_square}(B, D) shows color maps of this final probability density function for the donor's location.
The best estimate (maximum of $P(\Vec{r})$) is found at a lateral position of $(x,z) = (\SI{13}{\nano \meter}, \SI{8}{\nano \meter})$ and a depth of \SI{-5}{\nano \meter} below the Si/SiO$_2$ interface.  

\subsection{Microscopic origin of the electric field gradient} \label{subsec: Microscopic_EFG}

As discussed in Section \ref{sec: quad_ham}, the quadrupole interaction is generated by an \gls*{efg} at the site of the \Sb nucleus. 
This \gls*{efg} is due to inhomogeneities in both the external applied electric field and the internal electric field arising from the charged electrons and nuclei comprising the host crystal.
Due to the relatively large physical length scales of the gates that supply the external field, the former contribution can be ignored, as quantitatively supported by a finite element model in Sec.~\ref{subsubsec: direct_gate_effect}. 
Here, we focus entirely on a microscopic model for the distortion of the electronic charge distribution around the \Sb nucleus.
We propose that both strain and externally applied electric fields distort the chemical environment coordinating the \Sb nucleus, and thus give rise to a non-zero \gls*{efg}, providing the physical mechanism that allows us to drive \gls*{ner} with an AC electric field. 
The coupling of the externally applied electric field to the \gls*{efg} also leads to a manifestation of the \gls*{lqse}, in which the quadrupole splitting can be tuned with a DC electric field (see main text for details).

We first consider a generic expression that relates the total charge density of the electrons and nuclei, $\rho(\Vec{r})$, to the \gls*{efg} tensor, $\mathcal{V}_{\alpha\beta}$. Taking the \Sb nucleus as the origin of coordinates,
\begin{align}
    \mathcal{V}_{\alpha\beta}= \frac{1}{4\pi\varepsilon_0}
    \int \left(
    \frac{3 r_{\alpha}r_{\beta}-\Delta_{\alpha\beta}|\Vec{r}|^2}{|\Vec{r}|^5}
    \right) \rho(\Vec{r}) d^3\Vec{r},
\end{align}
where $\Vec{r} = (r_x, r_y , r_z)$ is the position vector, $\Delta_{\alpha\beta}$ is the Kronecker delta, and $\alpha , \beta \in \{x, y, z\}$.
The integration kernel coefficient to $\rho(\Vec{r})$ will remain invariant under the application of strain or an electric field, the effects of which will enter through their coupling to the charge density.
Expanding the charge density as a first-order functional Taylor series about its form at zero strain and zero electric field we arrive at
\begin{align}
    \mathcal{V}_{\alpha\beta}= \frac{1}{4\pi\varepsilon_0}
    \int \left(
    \frac{3 r_{\alpha}r_{\beta}-\Delta_{\alpha\beta}|\Vec{r}|^2}{|\Vec{r}|^5}
    \right)
    \left(\rho(\Vec{r})\rvert_{\epsilon_{\gamma\delta}=0,E_\gamma=0}+
    \sum \limits_{\gamma\delta} \frac{\delta \rho(\Vec{r})}{\delta \epsilon_{\gamma\delta}}\bigg\rvert_{\epsilon_{\gamma\delta}=0,E_\gamma=0} \epsilon_{\gamma\delta}+
    \sum \limits_{\gamma} \frac{\delta \rho(\Vec{r})}{\delta E_\gamma}\bigg\rvert_{\epsilon_{\gamma\delta}=0,E_\gamma=0} E_\gamma
    \right)
    d^3\Vec{r}. 
\end{align}
Here $\epsilon_{\gamma\delta}$ is the strain tensor and $E_\gamma$ is the external applied electric field, where $\gamma, \delta \in \{x, y, z\}$ index their respective components. 
This suggests the simplified functional form,
\begin{align}
    \mathcal{V}_{\alpha\beta}=
    \mathcal{V}^{0}_{\alpha\beta}+
    \sum \limits_{\gamma\delta} S_{\alpha\beta\gamma\delta}\epsilon_{\gamma\delta}+
    \sum \limits_{\gamma} R_{\alpha\beta\gamma}E_\gamma. \label{eq:full_efg_eqn}
\end{align}
$\mathcal{V}^0_{\alpha\beta}$ is the \gls*{efg} arising due to the unperturbed charge density.
$S_{\alpha\beta\gamma\delta}$ is the fourth-rank gradient-elastic tensor that describes the effect of strain on the \gls*{efg} at the \Sb nucleus.
$R_{\alpha\beta\gamma}$ is the third-rank electric field response tensor, capturing the relationship between the electric field and the \gls*{efg} at the \Sb nucleus.
The symmetry properties of the tensors that appear in Eq.~\eqref{eq:full_efg_eqn} suggest a more compact representation using Voigt's notation.
$\mathcal{V}_{\alpha\beta}$ is a symmetric tensor with 6 independent components that we represent with integer indices, i.e., $\mathcal{V}_{1}=\mathcal{V}_{xx}, \mathcal{V}_{2}=\mathcal{V}_{yy}, \mathcal{V}_{3}=\mathcal{V}_{zz}, \mathcal{V}_{4}=\mathcal{V}_{yz}=\mathcal{V}_{zy}, \mathcal{V}_{5}=\mathcal{V}_{xz}=\mathcal{V}_{zx},$ and $\mathcal{V}_{6}=\mathcal{V}_{xy}=\mathcal{V}_{yx}$.
Similarly, because $S$ and $R$ are symmetric they can be represented by 36 and 18 independent components, respectively, rather than 81 and 27.
The tetrahedral T$_d$ symmetry of the \Sb donor site permits an even more dramatic reduction in the number of independent components of $S$ and $R$ that will be discussed in Sec.~\ref{subsubsec: gradient-elastic tensor} and Sec.~\ref{subsubsec:electric_field_response_tensor}.

We only consider the three leading order contributions to the \gls*{efg} in detail.
It is worth noting that the higher-order mixed terms may be of interest in future experiments.
Specifically, shear strains may break symmetries that lead to a permanent dipole moment that can enhance the electric field's coupling to the \gls*{efg}, relative to uniaxial strains.
However, this would require measurements of samples under different or variable strains to be substantiated.

We first consider whether the strain and electric field independent contribution, $\mathcal{V}^0_{\alpha\beta}$, vanishes by symmetry.
This can be determined by identifying whether the representation of the integrand, a product of $\rho(\Vec{r})$ and the \gls*{efg} kernel, includes the totally symmetric irreducible representation of the point group associated with the symmetry of the substitutional site. 
In the absence of strain or an external applied electric field, this point group is T$_{d}$ and it is evident that the totally symmetric irreducible representation does not appear, leading this contribution to vanish by symmetry.
Our measurements indicate a non-zero \gls*{efg}, so the symmetry of the \Sb site must be effectively lowered by strain and/or an external applied electric field.

Both uniaxial and shear strains will change the positions of the silicon atoms that coordinate the \Sb donor such that the T$_{d}$ symmetry of the substitutional site is broken.
Uniaxial strain will reduce the site's symmetry to D$_{2d}$, giving rise to a linear coupling between the strain and the diagonal components of the \gls*{efg} tensor.
Shear strain will reduce the site's symmetry to C$_{2v}$, giving rise to a linear coupling between the strain and the off-diagonal components of the \gls*{efg} tensor.
These couplings correspond to the two non-trivial components of the $S$ tensor for T$_d$ symmetry, $S_{11}$ and $S_{44}$, respectively.
\gls*{dft} calculations of these couplings are given in Sec.~\ref{subsubsec:gradient_elastic_tensor_calculations}. 
These calculations include the direct evaluation of the \gls*{efg} as a function of strain and corroborate a linear relationship that extrapolates to zero in the absence of strain.

Even in the absence of strain, an external applied electric field can create a non-vanishing \gls*{efg}. 
Because the silicon crystal is not piezoelectric, the effect of this electric field will not be to change the positions of the coordinating silicon atoms. 
We substantiate in Sec.~\ref{subsubsec:unlikely_source_of_ner_piezo} that an indirect modulation of the strain through a piezoelectric coupling in the oxide is not a significant effect.
Rather, the primary effect of an electric field will be to mix the electronic ground state of the ionized \Sb with excited states of opposite parity, proportional to the transition dipole moment between them.
This polarization leads to a symmetry breaking in $\rho(\Vec{r})$ that gives rise to a linear coupling between the electric field and the \gls*{efg} through the $R$ tensor.
From symmetry considerations, the only non-trivial component will be $R_{14}$, which gives rise to off-diagonal contributions to the \gls*{efg} tensor.
This direct coupling between the electric field and the \gls*{efg} is the physical mechanism that allows us to drive \gls*{ner} with an external applied AC electric.
That this coupling is present and of the correct order of magnitude to rationalize our observations is further substantiated by the observed shift of the nuclear spectral lines in response to a DC electric field, which results from the same mechanism.
In the \gls*{ner} literature, the resulting shift in transition frequencies is known as the \gls*{lqse}.
The presence of the \gls*{lqse} relies on a product of matrix elements with opposite parities under inversion, and thus it is forbidden by symmetry for nuclei that inhabit sites with inversion symmetry.
Because the \Sb donor does not occupy such a site, it is plausible in principle that such an electric field induced shift can explain our observations. 
We provide an estimate of this coupling from our measurements of the DC electric field induced shifts in Sec.~\ref{subsubsec:electric_field_response_COMSOL} and a comparison to measurements of the \gls*{lqse} in bulk crystals in Sec.~\ref{subsubsec:comparison_to_bulk_lqse}.

\subsection{Strain induced quadrupole splitting} \label{subsec: strain_quadrupole}

\subsubsection{The gradient-elastic tensor} \label{subsubsec: gradient-elastic tensor}

For T$_d$ tetrahedral symmetry the gradient-elastic tensor $S$ has two different nonzero components $S_{11}$ and $S_{44}$. 
Here,  Voigt's notation in the conventional crystallographic, non-rotated basis, and engineering strains, are used. 
In this frame (given by $x''$, $y''$, $z''$, see Sec.~\ref{sec: comsol_model}), the \gls*{efg} tensor is given by
\begin{align}
    \begin{pmatrix}
        \mathcal{V}_{x''x''}\\
        \mathcal{V}_{y''y''}\\
        \mathcal{V}_{z''z''}\\
        \mathcal{V}_{y''z''}\\
        \mathcal{V}_{x''z''}\\
        \mathcal{V}_{x''y''}
    \end{pmatrix}
    =
    \begin{pmatrix}
        S_{11} & -S_{11}/2 & -S_{11}/2 &0&0&0\\
        -S_{11}/2 & S_{11} & -S_{11}/2 &0&0&0\\
        -S_{11}/2 & -S_{11}/2 & S_{11} &0&0&0\\
        0&0&0& S_{44} & 0 & 0\\
        0&0&0& 0 & S_{44} & 0\\
        0&0&0& 0 & 0 & S_{44}
    \end{pmatrix}
    \cdot
    \begin{pmatrix}
        \epsilon_{x''x''}\\
        \epsilon_{y''y''}\\
        \epsilon_{z''z''}\\
        \gamma_{y''z''}\\
        \gamma_{x''z''}\\
        \gamma_{x''y''}
    \end{pmatrix}.
    \label{eq: Stensor}
\end{align}
Note that the relationship between shear components of the  engineering and infinitesimal strain is $\gamma_{y''z''}=2\epsilon_{y''z''}$.
\gls*{dft} calculations have been performed (see Sec.~\ref{subsubsec:gradient_elastic_tensor_calculations}) to evaluate $S_{11}$ and $S_{44}$.
In the experiment, the magnetic field is applied along the [110] crystal direction, e.g. in the frame of the gradient-elastic tensor above this corresponds to $\hat{\mathcal{H}} = \gamma_n B_0 (\hat I_{x''} + \hat I_{z''})/\sqrt{2}$.
Computing the first order perturbative correction, as in Sec.~\ref{subsec: nuc_perturb}, yields the quadrupole splitting
\begin{align}
    f_Q = \frac{e q_n}{2I(2I-1)h}\, \frac{3}{4} \left[S_{11}(2\epsilon_{y''y''}-\epsilon_{x''x''}-\epsilon_{z''z''})-4S_{44}\gamma_{x''z''}\right].
\end{align}
Rotating the strain tensor by $\pi/4$ about $y$ to align the magnetic field with the $z$ axis results in the laboratory frame expression for the spectral quadrupole splitting given by
\begin{align}
        f_Q = \frac{e q_n}{2I(2I-1)h}\, \frac{3}{4} \left[S_{11}(2\epsilon_{yy}-\epsilon_{xx}-\epsilon_{zz})-4S_{44}(\epsilon_{xx}-\epsilon_{zz})\right].
        \label{eq: Q_strain}
\end{align}

\subsubsection{Gradient-elastic tensor calculations} \label{subsubsec:gradient_elastic_tensor_calculations}
We use Kohn-Sham density functional theory (DFT) to develop an atomistic understanding of the impact that strain has on the \gls*{efg} at the ionized \Sb donor.
We perform a series of supercell calculations to determine the manner in which the silicon atoms coordinating the \Sb site relax as a function of strain.
These calculations also provide us with first principles values for the $S$ tensor that avoid the need for empirically derived Sternheimer factors.

For all supercell calculations we use the Projector Augmented-Wave (PAW) formalism\citep{blochl1994projector} with a plane wave basis, as implemented in the Vienna Ab-Initio Simulation Package (VASP) \cite{kresse1996efficient,kresse1996efficiency,kresse1999ultrasoft}. 
Within the PAW formalism the all-electron Kohn-Sham orbitals, and their associated density, can be accessed via an explicit linear transformation on smoother pseudo orbitals that can be efficiently represented in a plane wave basis. 
As such, we can accurately compute properties that depend sensitively on the charge density and local potential near nuclei, while also making use of the frozen core approximation to reduce the number of orbitals that must be explicitly included in a given calculation.
One such property is the \gls*{efg}, which has been shown to be accurately represented within the PAW formalism when compared to both experiments and calculations using the linear augmented plane wave (LAPW) formalism \cite{petrilli1998electric}.

Because we are studying an ionized Sb donor, we do not need to represent its hydrogenic bound state in our calculations and we can thus work with smaller supercells than are required for neutral shallow defects \cite{zhang2013shallow}. 
That we do not need to represent states near the conduction band edge also mitigates concerns about the impact of the band gap problem on our results. 
We thus find the semilocal Perdew-Burke-Ernzerhof (PBE) construction of the generalized gradient approximation to the exchange-correlation functional \cite{perdew1996generalized} to be adequate.
Our calculations are carried out using a plane wave cut-off of \SI{500}{\electronvolt} for the orbitals and \SI{1000}{\electronvolt} for the augmentation charges, with a 3$\times$3$\times$3 Gamma-centered sampling of the first Brillouin zone for 64 atom supercells, and a Gamma-only sampling for 512 atom supercells.
Even though we do not need to represent the neutral donor wave function, finite size effects still impact our results because our supercell calculations are non-cubic and have a net positive charge.
It is well-known that total energy and forces in such calculations are slow to converge with system size. 
Dipole corrections that would otherwise account for finite size errors are only implemented for cubic supercells in the software package that we are using.
The values of the \gls*{efg} computed for the 64 and 512 atom supercells at a given strain are within 10\% of one another and that this level of accuracy suffices to demonstrate a linear trend across different strains.
In the worst case, the difference between the values of the slope obtained regressing on a linear model with a zero $y$-intercept and a non-zero $y$-intercept is 0.1\%.

We self-consistently relax the Si crystal geometry around the ionized \Sb donor for both uniaxial strain on [100] and shear strain at 5 different values (-10$^{-2}$, -3$\times 10^{-3}$, -10$^{-3}$, -3$\times 10^{-4}$, and -10$^{-4}$). 
Our starting point for each calculation is a pristine Si crystal with the strain applied relative to the experimental lattice constant (\SI{5.431}{\angstrom}) and a single substitutional \Sb donor.
To account for the +1 ionization state of the Sb donor, we include 4 electrons per atom in our supercells, and a homogeneous background charge is assumed to maintain the overall neutrality necessary to obtain a finite total energy.
We relax the atomic positions at a fixed supercell volume until the interatomic forces are all less than \SI{1}{\milli\electronvolt\per\angstrom}, and find that in all cases the primary impact of the ionized \Sb is to displace the 4 silicon atoms that coordinate it by $\sim$ \SI{0.2}{\angstrom}.
The next nearest neighbors are displaced by $\sim$ \SI{0.05}{\angstrom}, and beyond that all displacements are below $\sim$ \SI{0.025}{\angstrom}. 
This is consistent with a simple picture in which the ionized Sb occupies a larger volume than neutral Si by merit of its filled M and N shells.
For each strain type and amplitude we compute the \gls*{efg} and perform a linear regression on the computed values to extract estimates for $S_{11}$ and $S_{44}$.

For the case of uniaxial strain, we evaluate $S_{11}$ for both a model in which the Poisson ratio is 0 (i.e., unphysical ``cork-like'' Si) and another in which it takes on its experimental bulk value of 0.28. 
The resulting values of $S_{11}$ are \SI{1.9e22}{\volt\per\square\meter} and \SI{2.4e22}{\volt\per\square\meter}, respectively, noting that this distinction preserves the order of magnitude of the coupling.
For shear strain, we compute $S_{44}$ to be \SI{6.1e22}{\volt\per\square\meter}.
These values are comparable to those reported for As in Si by Franke {\it et al.}, which are respectively \SI{1.5e22}{\volt\per\square\meter} and \SI{6.0e22}{\volt\per\square\meter} \cite{franke2015interaction}.
That the ratio of the shear component to the uniaxial component is larger for As than Sb is to be expected because Sb occupies a larger volume than As and will thus have an \gls*{efg} that couples more strongly to strains that change volume.

Our calculations were carried out using 4 electron PAW potentials for Si and 15 electron PAW potentials for Sb. 
To assess the primary chemical contribution to the \gls*{efg}, we repeated the calculations with a 5 electron PAW potential for Sb and found that the \gls*{efg} decreased by 5$\%$ for both models of uniaxial strain and 2.5$\%$ for shear strain.
This small change confirms that the \gls*{efg} is generated primarily by the contribution to the local potential due to the charge density in the sp$^3$-like Sb-Si bonding orbitals, and not a strain-induced distortion of the on-site contribution to the local potential due to the Sb donor's d electrons.

The relaxed supercell geometries expose microscopic details of the symmetry breaking at the \Sb site. 
In the absence of strain this site has the symmetry of the point group T$_d$, which is apparent from the tetrahedral coordination of the four nearest silicon atoms.
Uniaxial strain reduces the symmetry of this site to D$_{2d}$, whereas shear strain reduces it to C$_{2v}$. 
We note that this detail may become important when considering contributions to the \gls*{efg} beyond first order to which both the strain and electric field are coupled, and will be discussed briefly in Sec.~\ref{subsubsec:comparison_to_bulk_lqse}.

\subsubsection{Calculation of quadrupole splitting due to strain}

As $S_{44}/S_{11}\approx 2.5$, Eq~\eqref{eq: Q_strain} shows that the leading contribution to the quadrupole splitting will be the shear strain along the magnetic field aligned with the [110] axis.
We use COMSOL to compute the thermal deformation and show the shear strain $\epsilon_{xx}-\epsilon_{zz}$ in the main text (Fig.~1B).
We use Eq.~\eqref{eq: Stensor} to relate the computed strains to \gls*{efg} and numerically compute the eigenenergies of Eq.~\ref{eq: H_system_static} and show the estimated quadrupole splitting $f_Q$ in the main text (Fig.~4C).  
The variation in quadrupole splitting shows a good qualitative correspondence with the variation of the shear strain, reflecting the larger contribution of $S_{44}$ to the quadrupole splitting.
Values found for the strain-induced quadrupole splitting range between 90 and \SI{160}{\kilo \hertz} throughout the area where the donor is estimated to be. 
This result represents an outstanding quantitative agreement with the experimentally observed value of $f_Q = \SI{66}{\kilo \hertz}$, in light of the complex combination of strain modeling, DFT calculations and donor triangulation required in our modeling effort.

\FloatBarrier
\subsection{Electric field induced quadrupole splitting and NER} \label{subsec: electric_quadrupole}

\subsubsection{The electric field response tensor} \label{subsubsec:electric_field_response_tensor}

The effect of an applied electric field on the \gls*{efg} can be described by a third rank tensor.
In T$_d$ symmetry we only have one unique non-vanishing element $R_{14}=R_{25}=R_{36}$.
Here, the first index refers to the electric field component at the donor site in the crystallographic frame (i.e. $x''=1, y''=2, z''=3$) and the second index refers to the resulting \gls*{efg} component in Voigt's notation.
In the frame given by the crystal axes ($x''$, $y''$ and $z''$) the tensor acts as follows:
\begin{align}
    \begin{pmatrix}
        \mathcal{V}_{x''x''}\\
        \mathcal{V}_{y''y''}\\
        \mathcal{V}_{z''z''}\\
        \mathcal{V}_{y''z''}\\
        \mathcal{V}_{x''z''}\\
        \mathcal{V}_{x''y''}
    \end{pmatrix}
    =
    \begin{pmatrix}
        0&0&0\\
        0&0&0\\
        0&0&0\\
        R_{14} & 0 & 0\\
        0 & R_{14} & 0\\
        0 & 0 & R_{14}
    \end{pmatrix}
    \cdot
    \begin{pmatrix}
        E_{x''}\\
        E_{y''}\\
        E_{z''}\\
    \end{pmatrix}.
    \label{eq: Rtensor}
\end{align}
We note that the above assumption of T$_d$ symmetry is not entirely obvious, as strain breaks this symmetry. 
However, to first order, the effects of electric field and strain are independent; the development of a more elaborate theory will be the topic of future research. 

The coupling between an external applied electric field and the \gls*{efg} at the \Sb nucleus is the physical origin of both \gls*{ner} and the \gls*{lqse}.
Both AC and DC fields will act to modulate the off-diagonal components of the \gls*{efg} tensor relative to the crystallographic axes of the silicon host.
It is important to note that this is not the same coordinate system as the laboratory frame, which is defined relative to the $B_0$ field. 
Thus, off-diagonal modulation in the crystal frame can lead to modulation of both the off-diagonal (\gls*{ner}) and diagonal components (\gls*{lqse}) in the laboratory frame.
Future experiments could employ a vector magnet to explore these details.
For example, one could imagine aligning the external applied magnetic field such that the laboratory and crystal coordinate frames coincide to suppress \gls*{lqse}.

\subsubsection{Electric field response tensor estimate} \label{subsubsec:electric_field_response_COMSOL}

As for the strain in Sec.~\ref{subsubsec: gradient-elastic tensor}, we can derive an analytic expression for the shift of the quadrupole splitting with electric field, assuming a strong magnetic field applied along the [110] crystal direction. 
We rotate the lab frame expression for the quadrupole splitting (Eq.~\eqref{eq: def_Q}) into the crystal frame
\begin{align}
    f_Q=\frac{e q_n}{2I(2I-1)h}\,(\mathcal{V}_{xx}+\mathcal{V}_{yy}-2\mathcal{V}_{zz})=
    \frac{e q_n}{2I(2I-1)h}\,(-\frac{1}{2}\mathcal{V}_{x''x''}+\mathcal{V}_{y''y''}-\frac{1}{2}\mathcal{V}_{z''z''}-3\mathcal{V}_{x''z''}).
\end{align}
As discussed and shown above (Eq.~\eqref{eq: Rtensor}), an electric field only gives rise to off-diagonal \gls*{efg} tensor components in the crystal frame. 
Thus, only $\Delta\mathcal{V}_{x''z''}=R_{14}\, \Delta E_{y''}$ contributes to the spectral line shift as a function of electric field, 
\begin{align}
    \Delta f_Q = \frac{e q_n}{2I(2I-1)h}\,  (-3)R_{14}\, \Delta E_{y''} 
\end{align}
and since the transformation between crystal and lab frames is a rotation about $y$, the same analytic form is found in the lab frame: 
\begin{align}
    \Delta f_Q = \frac{e q_n}{2I(2I-1)h}\,  (-3) R_{14}\, \Delta E_{y}.
    \label{eq:QfromRE}
\end{align}

Similarly, we can derive analytic expressions for the expected Rabi frequencies of the $\Delta m_I = \pm 1$ and $\Delta m_I = \pm 2$ transitions under \gls*{ner}. 
Applying a sinusoidal pulse to an electric gate with amplitude $\delta V$ will cause an electric field modulation at the donor position, $\delta\Vec{E}$.
This in turn gives rise to the modulation of the \gls*{efg}, $\delta\mathcal{V}_{\alpha\beta}$, postulated in Sec.~\ref{sec:nuclear_electric_resonance}.
There, we derived the lab frame representation of the \gls*{ner} $\Delta m_I=\pm 1$ transition Rabi frequencies in Eq.~\eqref{eq: frabi_single_all}, which, upon rotation into the crystal frame, becomes
\begin{align} 
    \ f^{\textrm{Rabi, NER}}_{m_I-1 \leftrightarrow m_I} &= \left| \alpha_{m_I-1\leftrightarrow m_I}\, \frac{e q_n}{2I(2I-1)h}\, \sqrt{\delta\mathcal{V}_{xz}^2+\delta\mathcal{V}_{yz}^2} \right| \\
    &= \left|\alpha_{m_I-1\leftrightarrow m_I}\, \frac{e q_n}{2I(2I-1)h}\, \sqrt{\frac{1}{4}(\delta\mathcal{V}_{x''x''}-\delta\mathcal{V}_{z''z''})^2+\frac{1}{2}(\delta\mathcal{V}_{x''y''}+\delta\mathcal{V}_{y''z''})^2}\right|
\end{align}
Again, only off-diagonal \gls*{efg} tensor components contribute in the crystal frame, and we find 
\begin{align}
    f^{\textrm{Rabi, NER}}_{m_I-1 \leftrightarrow m_I} = \left| \alpha_{m_I-1\leftrightarrow m_I}\, \frac{e q_n}{2I(2I-1)h}\,  R_{14}\,  \frac{1}{\sqrt{2}}(\delta E_{x''}+\delta E_{z''})\right|,
\end{align}
and for the lab frame
\begin{align} \label{eq: frabi_single_R}
    f^{\textrm{Rabi, NER}}_{m_I-1 \leftrightarrow m_I} = \left| \alpha_{m_I-1\leftrightarrow m_I}\, \frac{e q_n}{2I(2I-1)h}\,  R_{14}\,  \delta E_{z}\right|.
\end{align} 
Note that only the electric field component along the applied magnetic field contributes. 
Repeating the procedure for the $\Delta m_I = \pm 2$ transition Rabi frequencies using Eq.~\eqref{eq: frabi_double_all}, we find
\begin{align}
    f^{\textrm{Rabi, NER}}_{m_I-2\leftrightarrow m_I} &= \left|\beta_{m_I-2\leftrightarrow m_I}\, \frac{e q_n}{2I(2I-1)h}\sqrt{(\delta\mathcal{V}_{xx}-\delta\mathcal{V}_{yy})^2+4\delta\mathcal{V}_{xy}^2}\right|\\
    &=\left| \beta_{m_I-2\leftrightarrow m_I}\, \frac{e q_n}{2I(2I-1)h}\sqrt{(\frac{1}{2}\delta\mathcal{V}_{x''x''}+\frac{1}{2}\delta\mathcal{V}_{z''z''}-\delta\mathcal{V}_{x''z''}-\delta\mathcal{V}_{y''y''})^2+2(\delta\mathcal{V}_{x''y''}-\delta\mathcal{V}_{y''z''})^2}\right|\\
    &=\left| \beta_{m_I-2\leftrightarrow m_I}\, \frac{e q_n}{2I(2I-1)h}\,R_{14}\,\sqrt{(\delta E_{y''})^2+2(\delta E_{x''}-\delta E_{z''})^2}\right|
\end{align}
for the crystal frame, and for the lab frame
\begin{align} \label{eq: frabi_double_R}
    f^{\textrm{Rabi, NER}}_{m_I-2\leftrightarrow m_I} = \left|\beta_{m_I-2\leftrightarrow m_I}\, \frac{e q_n}{2I(2I-1)h}\,R_{14}\,\sqrt{(\delta E_{y})^2+4(\delta E_{x})^2}\right|.
\end{align}

\begin{figure}[ht]
  \centering
  \subfloat{\includegraphics[width=0.45\textwidth]{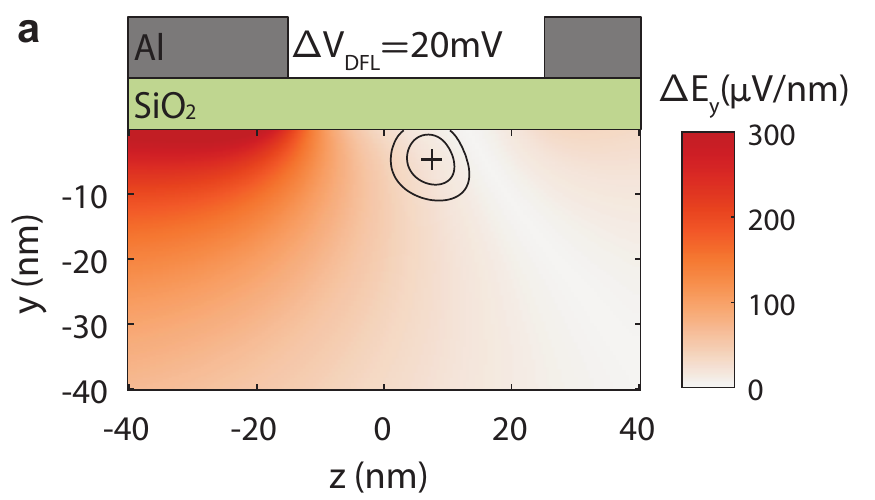}}
  \subfloat{\includegraphics[width=0.45\textwidth]{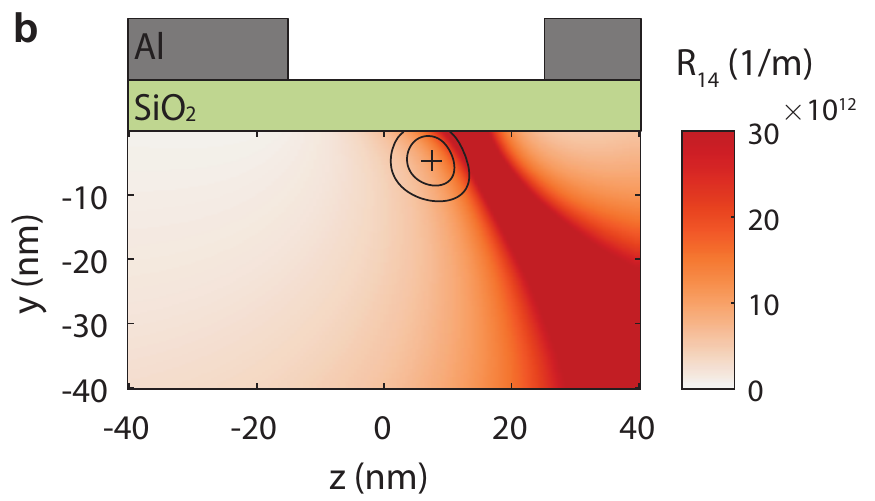}}\\
  \subfloat{\includegraphics[width=0.45\textwidth]{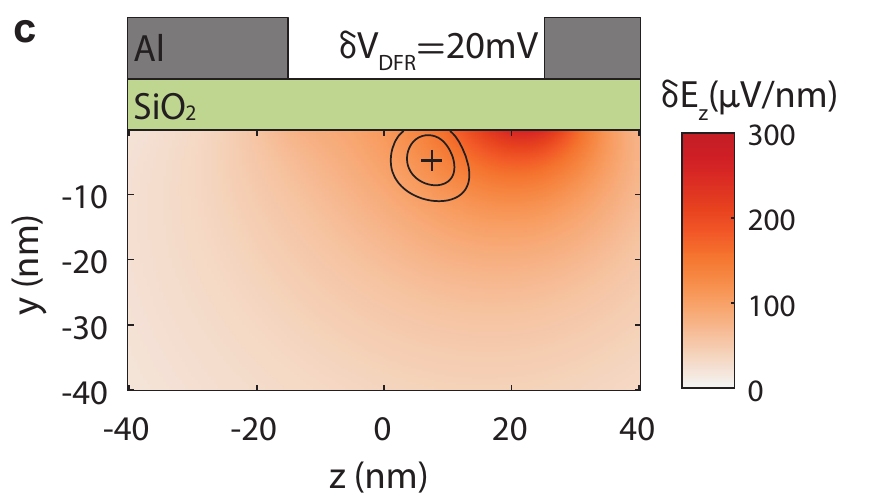}}
  \subfloat{\includegraphics[width=0.45\textwidth]{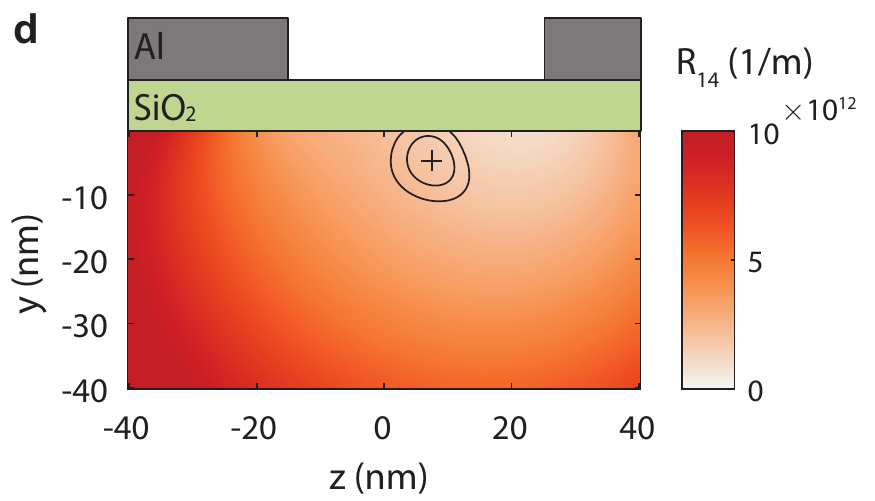}}\\
  \subfloat{\includegraphics[width=0.45\textwidth]{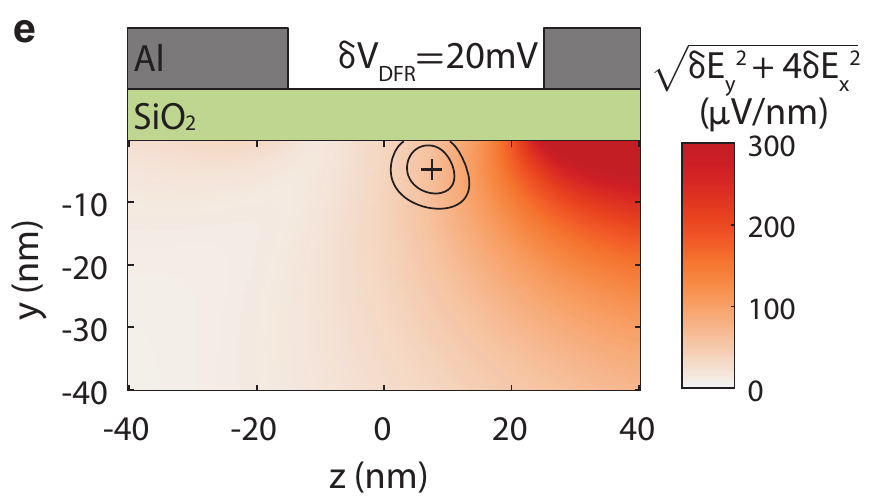}}
  \subfloat{\includegraphics[width=0.45\textwidth]{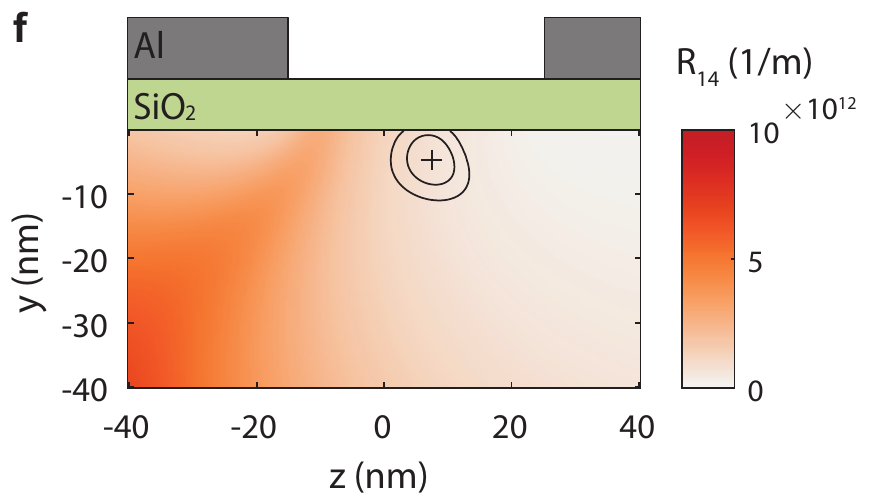}}
  \caption{Effect of an electric field on the ionized \Sb nucleus. 
  \textbf{a}, Spatial dependence of the out-of-plane electric field difference $\Delta E_y$ upon applying \SI{20}{\milli\volt} to the DFL gate (left).
  This is the gate varied to observe the spectral line shift of $\Delta f_Q=\SI{9.9}{\kilo\hertz}$ per applied volt (see main text). 
  \textbf{c, e}, Spatial dependence of the varying electric field components $\delta E_z$ (\textbf{c}) and $\sqrt{ \delta E_y^2+4\delta E_x^2 }$ (\textbf{e}) upon varying the DFR gate (right) by \SI{20}{\milli \volt}. 
  This is the gate varied in the experiment to achieve \acrshort*{ner}, and these electric field components are responsible for the $\Delta m_I = \pm 1$ and $\Delta m_I = \pm 2$ transitions, respectively. 
   \textbf{b, c, d},Spatial dependence of $R_{14}$. 
   Each value is found by solving Eq.~\eqref{eq:QfromRE}, Eq.~\eqref{eq: frabi_single_R} or Eq.~\eqref{eq: frabi_double_R} for $R_{14}$. 
   Here the experimentally observed values for $\Delta f_Q = \SI{199}{\hertz}$ per applied $\SI{20}{\milli\volt}$ on gate DFL, and $f^\mathrm{Rabi, NER}_{5/2\leftrightarrow 7/2}=\SI{684}{\hertz}$ and $f^\mathrm{Rabi, NER}_{3/2\leftrightarrow7/2}=\SI{38.5}{\hertz}$ for $\SI{20}{\milli\volt}$ drive amplitude on gate DFR are used.  
   At the most likely donor position we find \textbf{b} $R_{14}=\SI{10e12}{\per\meter}$, \textbf{d} $R_{14}=\SI{1.7e12}{\per\meter}$ and \textbf{f} $R_{14}=\SI{0.7e12}{\per\meter}$. 
   In all panels, COMSOL is used to compute the change in electric field upon applying the indicated gate potential; the $y-z$ cut-plane is shown along $x=\SI{13}{\nano\meter}$ as indicated in Fig.~\ref{fig: COMSOLmodel}; the cross indicates the most likely donor position and contour lines correspond to the 68\% and 95\% confidence intervals for the donor position based on the triangulation procedure described in Sec.~\ref{sec: donor_triangulation}.
   }
  \label{fig: RTensor}
\end{figure}

We can exploit these linear relationships to estimate the value of $R_{14}$.
We use COMSOL to model the out-of-plane DC electric field shift $\Delta E_y$ for increasing the gate voltage $V_{\rm DFL}$ by \SI{20}{\milli \volt}, as shown across the device in Fig.~\ref{fig: RTensor}A.
This is the relevant electric field component for the LQSE as a function of gate voltage, as shown in Eq.~\eqref{eq:QfromRE}.
Furthermore, the electric drive is modelled by applying the peak voltage of a sinusoidal pulse to the DFR gate, $\delta V_{\rm DFR}=\SI{20}{\milli\volt}$, as used in the experiment.  
The relevant electric field components determining the Rabi frequency as given by Eq.~\eqref{eq: frabi_single_R} and Eq.~\eqref{eq: frabi_double_R} are shown across the device in Fig.~\ref{fig: RTensor}C for the $\Delta m_I = \pm 1$ transitions and in Fig.~\ref{fig: RTensor}E for the $\Delta m_I = \pm 2$ transitions. 

For the DC spectral line shift, the experiment found that $\Delta f_Q=\SI{9.9}{\kilo\hertz}$ per applied volt on gate DFL (main text). 
For the electric drive, the experiment found that $f^\mathrm{Rabi, NER}_{5/2\leftrightarrow 7/2}=\SI{684}{\hertz }$ and $f^\mathrm{Rabi, NER}_{3/2\leftrightarrow 7/2}=\SI{38.5}{\hertz }$ for \SI{20}{\milli\volt} drive amplitude. 
Using these three experimental values and the simulated electric fields throughout the device, $R_{14}$ is the only unknown constant.
Fig.~\ref{fig: RTensor}(B, D, F)  shows the $R_{14}$ values needed to explain the measured DC spectral line shift (B), the Rabi frequencies of the $\Delta m_I = \pm 1$ \gls*{ner} transitions (D) and the Rabi frequencies of the $\Delta m_I = \pm 2$ \gls*{ner} transitions (F). 
At the most likely donor position (black marker), as determined by combining triangulation with the ion implantation profile, we find values of $R_{14}=\SI{10e12}{\per \meter}$ (B), $R_{14}=\SI{1.7e12}{\per \meter}$ (D) and $R_{14}=\SI{0.7e12}{\per \meter}$ (F). 
The order of magnitude spread of these values should be viewed as a remarkable achievement, given the complexity of the models and the absence of any free parameters in them. 

We combine these results and determine a unique value of $R_{14}$ at the triangulated donor position by minimizing the sum of the square of the normalized residuals
\begin{align}
    \chi^2(\vec{r}) = \sum_i\left(\frac{M_i-R_{14}(\vec{r})\cdot S_i(\vec{r})}{\sigma_{M_i}}\right)^2.
    \label{eq: Err}
\end{align}
Here, $M_i$ is the observed measurement value, i.e. the left hand side in Eq.~\eqref{eq:QfromRE}, Eq.~\eqref{eq: frabi_single_R} or Eq.~\eqref{eq: frabi_double_R}. 
$S_i(\vec{r})$ is the product of all known constants on the right hand side of Eq.~\eqref{eq:QfromRE}, Eq.~\eqref{eq: frabi_single_R} or Eq.~\eqref{eq: frabi_double_R} and the electric field component. 
$R_{14}(\vec{r})$ is the targeted minimization value.
$\sigma_{M_i}$ is the uncertainty of the measurement value $M_i$.
Assigning an equal measurement error to each observation, we find at the most likely donor position $R_{14}=\SI{1.7e12}{\per \meter}$, and this is the value used in Fig.~4E. 

In Sec.~\ref{subsec: strain_quadrupole} we explained the observed quadrupole splitting $f_Q$ in the spectrum as caused by static strain. 
However, a small part of this should be due to the LQSE, as static electric fields are present, caused by the gate voltages required for device operation. 
Using the static electric field from our COMSOL model and the value of $R_{14}$ we found, we estimate an electric contribution to $f_Q$ of \SI{1.7}{\kilo \hertz}, i.e. $\sim 2.5 \%$ of the experimentally observed $f_Q$. 
This suggest that indeed the largest fraction of $f_Q$ is caused by strain. 

In what follows we aim to justify the value of $R_{14}$ on the basis of a microscopic theory.

\subsubsection{Comparison to LQSE measurements and empirical microscopic theory for bulk crystals} \label{subsubsec:comparison_to_bulk_lqse}

We have conjectured that the DC electric field induced shift of the \gls*{ner} spectral lines is a manifestation of the \gls*{lqse}, predicted in 1961 by Bloembergen: ``\textit{If a nucleus or paramagnetic ion with spin greater than one-half occupies a site in a crystal lattice without inversion symmetry, an external applied electric field will cause, in general, a change in the quadrupole coupling constant or crystal-line field splitting proportional to E}''~\cite{bloembergen1961science}. 
The effect was first invoked to explain contemporary observations of electric field induced shifts in the zero-field quadrupolar spectra of ionic crystals containing halogen nuclei ($^{81}$Br~\cite{kushida1961shift} and $^{35}$Cl~\cite{armstrong1961linear1,armstrong1961linear2}). 
Subsequently, it was measured for $^{27}$Al in Al$_2$O$_3$~\cite{dixon1964linear} and $^{69}$Ga, $^{71}$Ga, and $^{75}$As in GaAs~\cite{gill1963linear,brun1963electrically}. 
The physical mechanism by which the external applied electric field couples to the nuclear quadrupolar interaction is through a distortion of the chemical environment of a given nucleus, which gives rise to a change in the \gls*{efg} at the nucleus.
Typical measured proportionalities between the external applied electric field and the resulting \gls*{efg} are on the order of \SI{1e12}{\per\meter}.

The manifestation of \gls*{lqse} in our experiment is distinguished from previous measurements not only by being observed in a single nucleus, but one in which the nuclear quadrupole moment is between 2 ($^{75}$As) and 8 ($^{35}$Cl) times larger than any previous measurements and in a host material that is non-polar in bulk.
We stress that the necessary condition for \gls*{lqse} of broken point inversion symmetry is naturally fulfilled by silicon; the $T_d$ symmetry of Si lacks point inversion symmetry at the lattice sites. 
Comparisons to the GaAs measurements are the most relevant because the individual nuclei experience the same tetrahedral coordination as our \Sb donor.
In the preceding section (Sec.~\ref{subsubsec:electric_field_response_COMSOL}) we estimated a value of $R_{14}=\SI{1.7e12}{\per \meter}$ for the \Sb substitutional donor in silicon under study here.
This is comparable to the values reported for $^{75}$As in bulk GaAs~\cite{gill1963linear,brun1963electrically}, respectively \SI{1.55e12}{\per\meter} and \SI{2.0e12}{\per\meter}.

The microscopic theory used to rationalize the bulk GaAs measurements of Gill and Bloembergen\cite{gill1963linear} decomposes the electric field response tensor into ionic and covalent contributions, $R_{14}=R_{14}^\mathrm{ion}+R_{14}^\mathrm{cov}$.
The ionic contribution accounts for the electric field gradient due to an electric field induced distortion of the crystal lattice (i.e. piezoelectricity), whereas the covalent contribution accounts for the electric field gradient due to an electric field induced distortion of the valence orbitals.
In the case of $^{75}$As, their theory predicts that roughly 2/3 of the measured value of $R_{14}$ is covalent, while the remaining 1/3 is ionic, i.e., $R_{14}^\mathrm{cov}/R_{14} \approx 2/3$ and $R_{14}^\mathrm{ion}/R_{14} \approx 1/3$.
Because the silicon host crystal is non-polar, we anticipate that our inferred value of $R_{14}$ is almost entirely covalent, $R_{14} \approx R_{14}^\mathrm{cov}$.
Thus, in comparing the strength of the LQSE in our experiment to bulk GaAs, we compare to Gill and Bloembergen's covalent value of \SI{1.1e12}{\per\meter}.
The nucleus-specific constant of proportionality between the \gls*{lqse} splitting and the external applied electric field is 
\begin{equation}
    K^\mathrm{cov} = \frac{q_n R^\mathrm{cov}_{14}}{2I(2I-1)}.
\end{equation}
Accounting for the difference in nuclear quadrupole moment, nuclear spin, and $R_{14}$, we find $K^\mathrm{cov}$ values of \SI{5.8e-18}{m} for the covalent contribution to $^{75}$As in GaAs and \SI{2.8e-18}{m} for \Sb in silicon.
That these effects have the same order of magnitude suggests that it is reasonable to rationalize the DC bias induced shift in the \gls*{ner} spectral lines with the \gls*{lqse}.
In Sec.~\ref{subsubsec:electric_field_response_COMSOL} it is also demonstrated that this same picture can rationalize the observed $\Delta m_I = \pm 1$ and $\Delta m_I = \pm 2$ Rabi frequencies.

It is also informative to directly compare the values of $R^\mathrm{cov}_{14}$ in the context of the atomistic model Gill and Bloembergen provide for $R^\mathrm{cov}_{14}$ in GaAs. 
Applying second-order perturbation theory to derive a linear coupling between the external applied electric field and the \gls*{lqse} shift, they arrive at the empirical expression
\begin{equation}
R^\mathrm{cov}_{14} = \frac{\sqrt{3}}{2} \frac{\lambda}{1+\lambda^2} \frac{e q_{at}}{4\pi\epsilon_0} \frac{\langle \psi_{c}| e(r_x+r_y+r_z)| \psi_{v}\rangle}{W_c-W_v}, \label{eq:empirical_r_tensor}
\end{equation}
where $\lambda \in \left[0,1\right]$ is related to the ionicity of the Ga-As bond, $e q_{at}/(4\pi\epsilon_0)$ is the \gls*{efg} at the $^{75}$As nucleus due to an electron in a 4p orbital in an atomic model (i.e., not a solid), $|\psi_{c}\rangle$ and $W_c$ are the electronic orbital and orbital energy associated with a bonding orbital near the conduction band edge, and $|\psi_v\rangle$ and $W_v$ are the corresponding quantities for an antibonding orbital near the valence band edge.
For $^{75}$As in GaAs, $\lambda=0.4$ corresponding to $\pm 2$ effective charges on the nuclei, $q_{at}=\SI{40.8e30}{\per\cubic\meter}$, a transition dipole moment of \SI{1.3}{\electronvolt\angstrom}, and an orbital gap of \SI{2}{\electronvolt} are found to be in good agreement with the measured value of $R_{14}$ when combined with a model for $R_{14}^\mathrm{ion}$.
Because the symmetry of the \Sb site is the same, an empirical model for the $R_{14}\approx R_{14}^\mathrm{cov}$ measured in our experiment would take a form similar to Eq.~\eqref{eq:empirical_r_tensor}.
Our inferred value of $R^\mathrm{cov}_{14}$ is only $\sim 1.5$ times larger than the one that was measured for $^{75}$As.
That these values are so close in value can then be rationalized in terms of these empirical parameters, namely the ionicity of the Si-Sb$^+$ bond, the \gls*{efg} at the \Sb nucleus due to electrons in p orbitals, and the electronic structure of the bonding and antibonding orbitals at the defect site.

While we expect that the Si-Sb$^+$ bonding and antibonding orbitals will have some ionic character, we do not expect that they will have a more ionic character than the relevant Ga-As orbitals.
This expectation is verified by preliminary quantum chemistry calculations on cationic hydrogen-terminated Sb-doped crystalline Si clusters using the NWChem software package \cite{valiev2010nwchem}.
Studying finite clusters allows us to circumvent the technical difficulties associated with applying the modern theory of polarization in charged supercells.
These calculations indicate that the relevant transition dipole moment is on the order of \SI{0.1}{\electronvolt\angstrom}, noting that the comparable transition dipole matrix element used by Gill and Bloembergen is an order of magnitude larger.
Because our inferred value of $R^{cov}_{14}$ is the same order of magnitude as for GaAs, this suggests that $q_{at}$ needs to be commensurately larger to compensate. 
We expect that $q_{at}$ will, in fact, be larger because it will be the \gls*{efg} at the \Sb nucleus due to an electron in a 5p orbital, instead of a 4p orbital. 
The Sternheimer effect can also be invoked for even further increases, particularly due to the positive ionization state of the \Sb and the observation that this effect grows with atomic number.

It is worth noting again that the sensitivity of our inference of $R_{14}$ to details in the COMSOL model suggests that we should interpret it as an order of magnitude estimate. 
Given the dependencies on empirical quantities described above, the microscopic physics seems to corroborate that this estimate is quite reasonable. 
Further experiments and the development of a first principles theory that does not rely on empirical expressions of the form in Eq.~\ref{eq:empirical_r_tensor} will provide even more clarity.
Such a theory is a topic of ongoing work.
While there are reasonably mature first principles methods for computing the electric field gradient with and without strain (see Sec.~\ref{subsubsec:gradient_elastic_tensor_calculations}), the direct evaluation of the \gls*{lqse} parameters has not yet been demonstrated.
Such a theory would circumvent the need for a description in terms of empirical parameters like bond ionicities, atomic \gls*{efg}s, and Sternheimer factors.
Another item of interest for a more detailed theory is the possibility of a second-order coupling of the \gls*{efg} to both the strain and electric field, due to the development of a permanent electric dipole moment under shear strains as alluded to in Sec.~\ref{subsec: Microscopic_EFG}.

\FloatBarrier
\subsection{Alternative (unlikely) sources of NER} \label{subsec: unlikely_causes}

To further corroborate the claim that our data is most convincingly explained by an electric-field induced modulation of the nuclear quadrupole tensor caused by a local distortion of the atomic bonds, we quantitatively analyze two possible alternative explanations for the origin of a dynamical EFG.
Both alternatives explanations are found to be in discrepancy with the observed NER rate by many orders of magnitude.

\begin{figure}[ht]
\centering
    \includegraphics[width=0.5\textwidth]{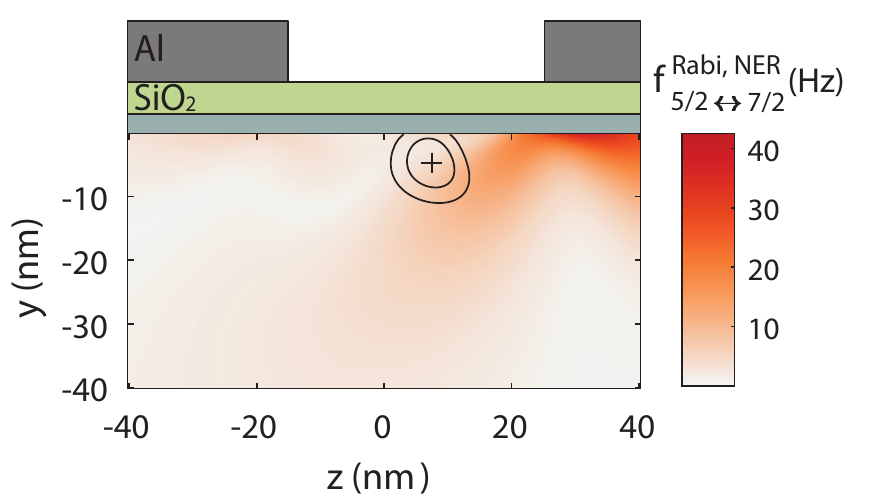}
  \caption{A possible piezo-electric drive is modelled by assuming a $\SI{3.5}{\nano\meter}$ thick, piezo-electric, quartz layer (blue) at the Si/SiO$_2$ interface, instead of a fully amorphous SiO$_2$ layer. 
  The experimental time-varying electric potential $\delta V_\mathrm{DFR}=\SI{20}{\milli \volt}$ is applied (gate on the right). 
  The resulting strain is calculated using the COMSOL model and subsequently converted into quadrupole interaction strength using $S_{11}$ and $S_{44}$ as found in \acrshort*{dft} calculations (see Sec.~\ref{subsubsec:gradient_elastic_tensor_calculations}.). 
  Shown is the $y-z$ cut-plane along $x=\SI{13}{\nano\meter}$ as indicated in Fig.~\ref{fig: COMSOLmodel}. 
  The contour lines correspond to the 68\% and 95\% confidence intervals for the donor position based on the triangulation procedure described in Sec.~\ref{sec: donor_triangulation}.
  The resulting mechanical \gls*{efg} driving strength on resonance, $f^{\rm Rabi, NER}_{5/2\leftrightarrow7/2}$, is two orders of magnitude smaller than the experimental value of $\SI{684}{\hertz}$. 
  This indicates that dynamical strain is an unlikely cause of NER in our device.}
  \label{fig: piezo_EFG}
\end{figure}

\subsubsection{Mechanical driving through SiO$_2$ piezoelectricity} \label{subsubsec:unlikely_source_of_ner_piezo}

Ref.~\citenum{lazovski2012detection} suggests that SiO$_2$ thin films thermally grown on Si can exhibit piezoelectric properties in the first few monolayers of the SiO$_2$ film. 
In the presence of a piezoelectric material, the time-dependent part of the applied electric potential will result in a periodic deformation of the sample.
Consequently, such a time-dependent strain component, which relates to a time-dependent \gls*{efg} (Eq.~\eqref{eq: Stensor}), results in a mechanical component to $\delta Q_{\alpha \beta}$ in Eq.~\eqref{eq: H_Q_time}. 
This would enable electrical driving of the nuclear spin.

Ref.~\citenum{lazovski2012detection} models the mechanical displacement by assuming that the first $\SI{3.5}{\nano\meter}$ of oxide consist of quartz. 
We therefore incorporate this quartz layer at the Si/SiO$_2$ interface in our COMSOL model.
As for the silicon, we need to adjust the orientation of the piezoelectric and stiffness matrices. 
The crystal structure of $\alpha$-quartz is trigonal trapezohedral.
Here, the [100]$_{\rm Quartz}$, [010]$_{\rm Quartz}$ and [001]$_{\rm Quartz}$ quartz crystal directions do not correspond to the Cartesian axes $x''$, $y''$ and $z''$ as for the silicon ([010], [001] and [100] respectively, see Sec.~\ref{sec: comsol_model}).
In the $x''-y''$ plane, the quartz unit cell has the shape of an equilateral parallelogram ($a=b=\SI{4.91}{\angstrom}$), where the angle between the crystal directions [100]$_{\rm Quartz}$ and [010]$_{\rm Quartz}$ is $\gamma=\SI{120}{\degree}$.
In this convention, the standard silicon Cartesian $x''$ and $y''$ axes correspond to the [010]$_{\rm Quartz}$ and [210]$_{\rm Quartz}$ crystal directions, respectively.
Perpendicular to these ($\alpha=\beta=\SI{90}{\degree}$) extends the [001]$_{\rm Quartz}$ direction of the unit cell with a length of $c=\SI{5.40}{\angstrom}$.
This is the exceptional three-fold rotation axis of $\alpha$-quartz.
Ref.~\citenum{carretero2013soft} found that the growth direction of $\alpha$-quartz on a (100) silicon substrate, using our crystal axis, is $[210]_{\rm Quartz}\parallel[001]_{\rm Silicon}$.
Hence in our model the exceptional [001]$_{\rm Quartz}$ axis will be in plane and aligned with the [100]$_{\rm Silicon}$ or [010]$_{\rm Silicon}$ directions.
We show the first orientation as an example.
We tested different quartz orientations, which all gave the same order of magnitude result.

A time-varying potential $\delta V_{\rm DFR}=\SI{20}{\milli \volt}$ applied to the DFR gate results in the driving strength shown in Fig.~\ref{fig: piezo_EFG}.
At the expected donor position, the estimated Rabi frequency on resonance is 5 to \SI{10}{\hertz}, two orders of magnitude smaller than the experimentally observed value $f^{\rm Rabi, NER}_{5/2\leftrightarrow7/2}=\SI{684}{\hertz}$.
We therefore conclude that piezoelectricity is an unlikely cause of \gls*{ner}.

\begin{figure}[ht]
  \centering
  \subfloat{\includegraphics[width=0.49\textwidth]{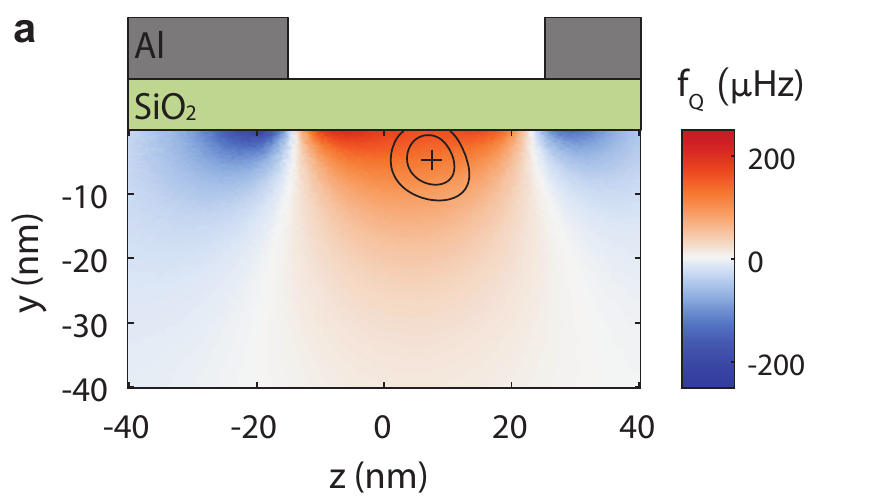}}
  \subfloat{\includegraphics[width=0.49\textwidth]{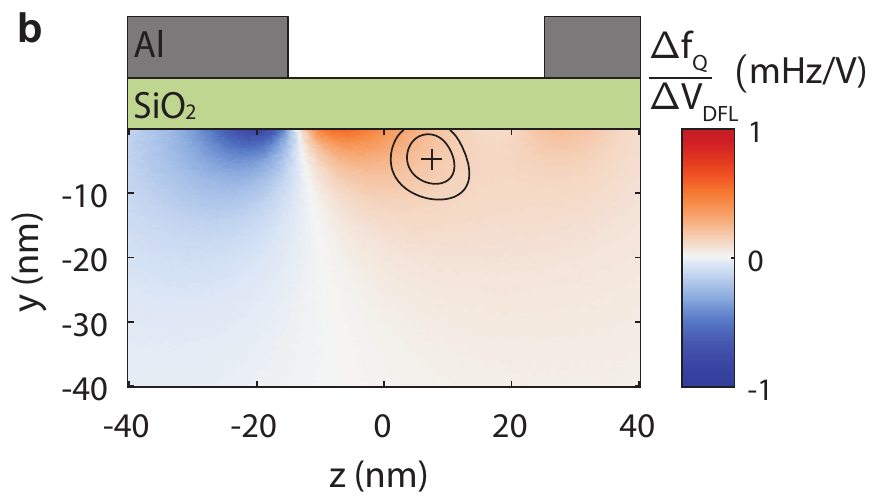}}\\
  \subfloat{\includegraphics[width=0.49\textwidth]{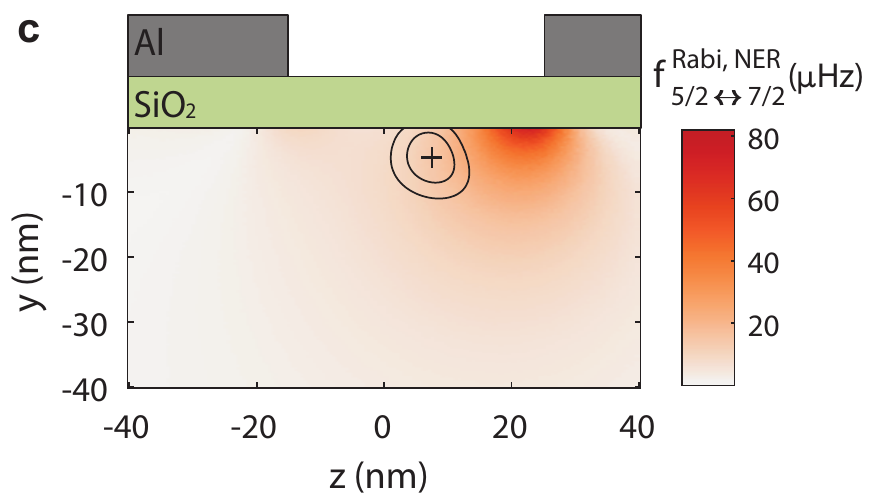}}
  \caption{Effect of the direct \acrshort*{efg} due to electric gate potentials creating an inhomogenous electric field. 
  This is in the absence of \acrshort*{efg} generation due to strain and \acrshort*{lqse}. 
  The $y-z$ cut-plane is shown along $x=\SI{13}{\nano\meter}$ as indicated in Fig.~\ref{fig: COMSOLmodel}. 
  The contour lines correspond to the 68\% and 95\% confidence intervals for the donor position based on the triangulation procedure described in Sec.~\ref{sec: donor_triangulation}.
  The resulting quadrupole splitting $f_Q$ (\textbf{a}) for the gate voltages given in Tab.~\ref{tab: sim_parameter}, the line shift per applied volt on gate DFL, $\Delta f_Q / \Delta V_{\textrm{DFL}}$ (\textbf{b}), and the \acrshort*{ner} driving strength for $\SI{20}{\milli\volt}$ driving amplitude on gate DFR, $f^{\textrm{Rabi, NER}}_{5/2 \leftrightarrow 7/2}$ (\textbf{c}), are more than 6 orders of magnitude smaller than the experimentally observed values, illustrating that this effect is entirely negligible.
  }
  \label{fig: intrinsic_EFG}
\end{figure}

\subsubsection{Direct effect of electric gate potentials} \label{subsubsec: direct_gate_effect}

The electric gate potentials will directly lead to an \gls*{efg} at the donor site, since the produced field is spatially inhomogeneous. 
To estimate the scale of this effect, the resulting quadrupole splitting is calculated in the absence of any other \gls*{efg} generating effects such as strain or the \gls*{lqse}. 
Fig.~\ref{fig: intrinsic_EFG} shows the resulting quadrupole splitting $f_Q$ (\textbf{a}), the spectral line shift per applied volt on gate DFL (\textbf{b}) and \gls*{ner} driving strength $f^{\rm Rabi, NER}_{5/2\leftrightarrow7/2}$ for $\SI{20}{\milli\volt}$ driving strength on the gate DFR (\textbf{c}), taking the \gls*{efg} directly generated by the gates at the donor site.
Their strengths are in the $\SI{100}{\micro\hertz}$, $\SI{1}{\milli\hertz\per\volt}$ and $\SI{1}{\micro\hertz}$ range, respectively, as opposed to the experimentally observed values of the order of $\SI{100}{\kilo\hertz}$, $\SI{10}{\kilo\hertz\per\volt}$ and $\SI{1}{\kilo\hertz}$.
Although some enhancement of this \gls*{efg} is expected at the \Sb nucleus due to the Sternheimer anti-shielding effect, this is not expected to surpass two orders of magnitude. 
As the direct effect of the electric gates results in an underestimation of the experimental observations by at least 6 orders of magnitude, we conclude this effect is insignificant. 
This illustrates that in our experiment a significant quadrupole interaction and \gls*{ner} strength requires a microscopic mechanism to be at play in direct vicinity of the donor site.

\section{Spin-mechanical coupling}

A fascinating prospect raised by our results is the possibility to coherently couple a single nuclear spin to a mechanical resonator. 
This would require that the dynamical strain caused by the zero-point fluctuations of the mechanical beam results in a nuclear Rabi frequency exceeding the inhomogenous linewidths of both the electron and the mechanical resonator. 
Fortunately, both systems can have exceptionally narrow linewidths, in the range of a few Hz. Below we provide an estimate of the zero-point strain in a silicon doubly-clamped mechanical oscillator, following the calculation presented in Ref.~\citenum{bennett2013phonon} (see also Ref.~\citenum{ovartchaiyapong2014dynamic} for a similar calculation in the case of a singly-clamped beam).

Consider a mechanical beam of length $L$, width $w$ and thickness $t$, clamped at both its extremities. We assume that the long axis of the beam coincides with the [110] crystallographic direction.
For Si, the Young's modulus is $E = 188$~GPa for stress along [110], and the mass density is $\rho = 2300$~kg/m$^3$. 
The moment of inertia of the beam is $I=w t^3/12$ and the wave number of the fundamental resonance mode is $k_0 = 4.73/L$. 
From this, one derives the (angular) frequency of the fundamental mode as:
\begin{equation} \label{eq:beamfreq}
    2\pi f_0 = k_0^2 \sqrt{\frac{E I}{\rho w t}} = 4.73^2 \frac{t}{L^2} \sqrt{\frac{E}{12 \rho}}
\end{equation}
In practice, one will need to choose $f_0$ to match the nuclear Larmor frequency $\gamma_n B_0$ (we neglect the small quadrupole shift $f_Q$ in this context). Therefore, $f_0$ and the resulting choice of beam geometry are not all free parameters. Choosing for example to fix the thickness $t$ of the beam (typically to the lowest value allowed by fabrication), we can derive the length $L$ necessary to yield the mechanical resonance that matches the nuclear precession frequency:
\begin{equation} \label{eq:beamlength}
    L=4.73 \sqrt{t} \sqrt{\frac{\sqrt{E/12\rho}}{2\pi \gamma_n B_0}}
\end{equation}
Next, we calculate the zero-point mechanical strain $\epsilon_{\rm zpf}$ in  such a beam. We assume that the nucleus is placed in the middle of the beam ($z=L/2$), as close as possible to the surface (where the strain is maximum), and find:
\begin{equation} \label{eq:epsilon}
    \epsilon_{\rm zpf} \approx 5 \sqrt{\frac{\hbar}{\sqrt{\rho E}}}\cdot w^{-1/2} L^{-3/2} \approx 1.23 \sqrt{\hbar} E^{-5/8} \rho^{1/8} w^{-1/2} t^{-3/4} (2\pi \gamma_n B_0)^{3/4}
\end{equation}
This equation shows that, at a constant frequency, $\epsilon_{\rm zpf}$ is maximized by making the beam as thin and narrow as possible, which will result in a minimal length too [Eq.~\eqref{eq:beamlength}]. It also shows that $\epsilon_{\rm zpf}$ depends on frequency (and thus on field) to the power of $3/4$.

Now we can use our models, benchmarked against the LQSE experimental data on the $^{123}$Sb nucleus, to estimate the spin-elastic coupling. In other words, we analyze a hypothetical Nuclear Acoustic Resonance (NAR) experiment \cite{sundfors1979nuclear} where a dynamical strain $\delta \epsilon$ at the nuclear site is caused by the zero-point strain $\epsilon_{\rm zpf}$ of the resonator. We assume that the length of the beam, the z-axis, coincides with the [110] crystallographic axis of Si. We apply a magnetic field along the [100] crystal axis. Considering the $m_I = 5/2 \leftrightarrow 7/2$ transition, the Rabi frequency is given by:
\begin{equation}
    f_{5/2\leftrightarrow 7/2}^{\rm Rabi,NAR} = \alpha_{5/2\leftrightarrow 7/2}\frac{eq_n}{2I(2I-1)h} S_{44} \sqrt{(\delta\epsilon_{xx}-\delta\epsilon_{zz})^2+2(\delta\epsilon_{yz}-\delta\epsilon_{xy})^2}
\end{equation}
where $\alpha_{5/2\leftrightarrow 7/2} = \sqrt{63}$. Taking the value $S_{44}=6.1\times 10^{22}$ V/m$^2$ extracted from the analysis in Sec.~\ref{subsubsec:gradient_elastic_tensor_calculations}, we obtain:
\begin{equation}
    f_{5/2\leftrightarrow 7/2}^{\rm Rabi,NAR} \approx 1.93 \times 10^8 \, \mathrm{Hz} \, \cdot \, \sqrt{(\delta\epsilon_{xx}-\delta\epsilon_{zz})^2+2(\delta\epsilon_{yz}-\delta\epsilon_{xy})^2}
\end{equation}
Identifying the dynamical strain component along the beam, $\delta \epsilon_{zz}$, with the zero-point strain of the mechanical beam, we find a simple linear relation:
\begin{equation} \label{eq:frabiNAR}
    f_{5/2\leftrightarrow 7/2}^{\rm Rabi,NAR} \approx 1.93 \times 10^8 \, \mathrm{Hz} \, \cdot \, \epsilon_{\rm zpf},
\end{equation}

In the context of cavity-QED, $f_{5/2\leftrightarrow 7/2}^{\rm Rabi,NAR}$ can be identified as the spin-mechanical vacuum Rabi splitting, $2g$. The interesting strong-coupling regime is achieved when $g$ exceeds both the phonon loss $\kappa$ and the qubit dephasing $\gamma$. In the notation and language of relevance to our system, this means $f_{5/2\leftrightarrow 7/2}^{\rm Rabi,NAR} \gg \Gamma_n, \Gamma_m$, where $\Gamma_n = 3.3$~Hz is the experimentally observed nuclear spin inhomogeneous linewidth, and $\Gamma_m = f_0 / Q_m$ is the resonator linewidth, determined by its mechanical quality factor $Q_m$. 

For the purpose of achieving strong coupling, the resonator design must be optimized considering both $\epsilon_{\rm zpf}$ and $\Gamma_m$. Typical Si resonators have a constant product $Q_m \cdot f_0 \sim 10^{13}$~Hz determined by material properties \cite{Ghaffari2013}. Therefore, $\Gamma_m \propto f_0^2$, whereas $g \propto f_0^{3/4}$ (\ref{eq:epsilon}), indicating that the best chance to achieve strong coupling is by reducing the operating frequency until the lower bound $\Gamma_m \sim \Gamma_n$ is reached ($\Gamma_n$ is usually independent of frequency). 

Taking for example $w=40$~nm, $t=20$~nm, and placing the $^{123}$Sb nuclear spin in a static field $B_0=1.5$~T, the resulting beam would need to have a resonance frequency $f_0 = 8.33$~MHz, a length $L=4.71$~$\mu$m, and would produce a zero-point strain $\epsilon_{\rm zpf}\approx 0.6\times 10^{-8}$, resulting in $f_{5/2\leftrightarrow 7/2}^{\rm Rabi,NAR} \approx 1.2$~Hz.

Assuming that the resonator has a quality factor $Q_m\approx 10^6$ at this frequency (consistent with $Q_m \cdot f_0 \sim 10^{13}$~Hz), i.e. $\Gamma_m = 8.3$~Hz, we thus see that the spin-cavity coupling $g$ is within less than a factor 10 of the dephasing rates.

Importantly, our experiment shows that the challenge in reaching strong coupling would not be set by the nuclear spin coherence, but rather by the zero-point strain and quality factor of the mechanical resonator, both of which can be significantly improved from the values used in the simple estimate above. 

Combining novel high-$Q_m$ resonators with  designs that explicitly maximize the zero-point strain, could bring the goal of coherent coupling between a single nuclear spin and a mechanical oscillator tantalizingly within reach. Achieving the single-phonon limit will be very challenging ($f_0 \sim 10$~MHz is equivalent to $\sim 0.5$~mK phonon energy), but interesting physics can already be explored in hot cavities \cite{schuetz2018quantum}.

\end{document}